\documentclass[twocolumn, trackchanges, onecolappendix]{aastex701}

\usepackage{amsmath}
\usepackage{physics}
\usepackage{enumerate}
\usepackage{booktabs}
\usepackage{multirow}
\usepackage{siunitx}

\usepackage[dvipsnames]{xcolor}

\begin{document}

\shorttitle{Close-In Brown Dwarfs' Eccentricity Distribution and Tidal Evolution}
\shortauthors{T. Ferreira \& M. Rice}

\title{On the Eccentricity Distribution and Tidal Evolution of Transiting Brown Dwarfs}

\correspondingauthor{Thiago Ferreira dos Santos}
\author[orcid=0000-0003-2059-470X,sname='Ferreira dos Santos']{Thiago Ferreira}
\affiliation{Department of Astronomy, Yale University, 219 Prospect Street, New Haven, CT 06511, USA}
\email[show]{thiago.dossantos@yale.edu}  

\author[orcid=0000-0002-7670-670X,sname='Rice']{Malena Rice}
\affiliation{Department of Astronomy, Yale University, 219 Prospect Street, New Haven, CT 06511, USA}
\email[]{malena.rice@yale.edu}  

\begin{abstract}

    Brown dwarfs on short-period orbits populate an intermediate regime between hot Jupiters and tight stellar binaries, lying at the intersection of possible evolutionary avenues. Their orbital eccentricities retain the dynamical imprint of both their formation pathways and any subsequent tidal evolution, providing a diagnostic for whether such objects formed in situ at small separations or were driven inward from higher-eccentricity orbits shaped by tidal dissipation. Using a hierarchical Bayesian framework, we characterise the orbital eccentricity distribution of transiting brown dwarfs. Short-period brown dwarfs ($P < 16$ days) are well represented by a Beta distribution with $\alpha < 1$ and $\beta > 1$, indicating a population concentrated at low eccentricities, whereas longer-period brown dwarfs ($P \geq 16$ days) display $\alpha,~\beta > 1$ and therefore occupy a more dynamically excited regime. This difference in eccentricity distributions likely reflects corresponding differences in the populations' eccentricity-damping timescales: close-in systems may evolve toward circular orbits on relatively short timescales, whilst wider companions experience negligible tidal processing over their lifetimes. Assuming that the full set of transiting brown dwarfs stems from a single primordial eccentricity distribution, {we constrain the typical brown dwarf tidal quality factor to $\mathcal{Q}_{\rm BD} = 10^{\rm 8.1\pm1.0}$ when neglecting the influence of tides raised on the host star, or $\mathcal{Q}_{\rm BD} = 10^{7.1\pm0.3}$ and $\mathcal{Q}_{\star} = 10^{6.0\pm0.1}$ when they are included.}
    
\end{abstract}


\section{Introduction}\label{sec:introduction}

Despite the thousands of brown dwarfs (BDs; $M \approx 13-{75}~M_{\rm Jup}$; {\citealt{2023A&A...671A.119C, 2024ApJ...975...59M}}) and exoplanets ($M \leq 13~M_{\rm Jup}$) discovered in recent decades\footnote{As of {May 2026}, nearly 6300 {confirmed} exoplanets are catalogued in the NASA Exoplanet Archive{'s Planetary Systems table} {\citep{2025PSJ.....6..186C}}, whilst the {\sc UltracoolSheet} catalogue {\citep{best_2025_15802304}} includes nearly 4000 spectroscopically confirmed ultracool dwarfs and BDs (M6-types and later, L-, T-, and Y-types).}, the dynamical boundaries demarcating these two classes remain indistinct. This arises, in part, from the small number of known transiting BDs---about 50 at the time of writing---that can be directly compared with the relatively well-characterised population of transiting exoplanets\footnote{One compiled catalogue of transiting BDs as of September 2023 is maintained at \url{https://theroncarmichael.com/browndwarfs}.}.

Delineating the boundary between planet-like and star-like evolutionary pathways requires a systematic examination of demographics alongside the primary mechanisms believed to govern the origins of BDs. Previous work \citep{2014MNRAS.439.2781M} has found that the subset of BD systems characterised by low mass ratios ($q \equiv M_{\rm BD}/M_{\star} < 0.01$) and BD masses below approximately 42.5 $M_{\rm Jup}$ exhibits properties broadly consistent with formation through gravitational instability within protoplanetary discs \citep{1997Sci...276.1836B, 2008MNRAS.389.1556B, 2009MNRAS.392..413S, 2016ARA&A..54..271K}. By contrast, systems with $q > 0.01$ and $M_{\rm BD} > 42.5~M_{\rm Jup}$ appear more consistent with stellar-like formation pathways involving the direct gravitational collapse of a molecular cloud core \citep{1969MNRAS.145..271L, 2002ARA&A..40..349T}. 

This result was derived from the population of wide-separation BDs---those most readily detectable via direct imaging and astrometric surveys. Trends observed in such systems may not extend to the short-period transiting BD population, however, whose formation mechanisms, dynamical evolution, and orbital characteristics remain comparatively understudied. If some BDs form in a {\it planet-like} manner, the existence of close-in, transiting BDs within this regime carries parallels to outstanding questions underlying hot Jupiter formation \citep{dawson2018}: in particular, {\it how do such massive bodies arrive on extremely close-in orbits?} 

{The potential formation channels for short-period BDs would imprint differing signatures onto the observed population of transiting BD orbital architectures. Disc-formed short-period BDs would emerge from the primordial circumstellar disc plane, suggesting a preference toward low orbital eccentricities and spin-orbit alignment \citep{rice2022b, rice2023TOI2202, radzom2024, wang2024, 2025Natur.644..356B}. Such BDs would not necessarily form in place, but may instead undergo significant dynamical interactions with the disc and with co-forming bodies, which may induce inward migration \citep{2008ASPC..398..235M, 2008A&A...482..315B, 2013ApJ...770..120B}. By contrast, short-period BDs originating from turbulent fragmentation of molecular clouds would follow stellar-like orbital distributions, including a high rate of eccentric and potentially spin-orbit misaligned orbits (e.g., \citealt{2012Natur.492..221R, 2014MNRAS.442..285B, 2022MNRAS.512.3383H}, and references therein).} 

{Beyond offering a new lens into BDs' formation and dynamical evolution pathways, close-in BDs are also subject to additional physical processes, such as tidal evolution, which can significantly modify their orbital eccentricities over time. BDs with sufficiently small periastron separations experience strong tidal interactions, which act to circularise their orbits on timescales determined by the efficiency of tidal damping in the BD---commonly parametrised by the tidal quality factor, $\mathcal{Q}$ \citep{2008CeMDA.101..171F, 2009ApJ...698L..42G, 2010A&A...514A..22H, 2016cole.book..169F}.}

In this work, we uniformly reanalyse the orbital eccentricities of the current transiting BD population to (1) model the population-level eccentricity distribution of short-period BDs and (2) place empirical constraints on their tidal properties, which cannot be easily measured in wider-orbit systems where tidal effects are minimal. We find that the distribution of transiting BD eccentricities is consistent with formation through a single dominant channel that excites moderate to high orbital eccentricities, in which the shortest-period BDs have undergone significant tidal circularisation, whilst longer-period systems more closely retain imprints of the primordial eccentricity distribution. Adopting this framework, we constrain the typical tidal quality factor of transiting BDs and their host stars at a population level. Throughout this work, we adopt the conventional tidal quality factor $\mathcal{Q}$ rather than the modified quantity $\mathcal{Q}^\prime$, with definitions discussed explicitly in $\S$\ref{sec:Q}, i.e., $\mathcal{Q}^\prime = (3/2\kappa_2)\mathcal{Q}$. 

This manuscript is organised as follows. In $\S$\ref{sec:re-fit}, we describe our re-fit of archival radial velocity data to compile a set of uniformly derived orbital eccentricities. In $\S$\ref{sec:BDs}, we describe the hierarchical model and its application to the sample of transiting BDs known to date. The role of tidal interactions in shaping the orbital evolution and eccentricity distribution of close-in BDs is explored in $\S$\ref{sec:tidal}, where we also derive the typical $\mathcal{Q}_{\rm BD}$ reported in this work. Our discussions are presented in $\S$\ref{sec:discussion} and conclusions in $\S$\ref{sec:conclusions}.

\section{Uniform Keplerian Re-Fitting}\label{sec:re-fit}

Orbital eccentricities in the literature are modelled and reported non-uniformly. Whilst some studies fix eccentricities to zero, others allow them to vary during the orbit-fitting process with uniform or normal priors on the eccentricity ($e$) and the argument of periastron ($\omega$), typically parametrised by $\sqrt{e_b}\cos\omega_b$ and $\sqrt{e_b}\sin\omega_b$. The choice of prior can significantly bias the inferred population-level distribution, as demonstrated in \citet{2023AJ....165...32N}, being particularly insidious in the regime of low signal-to-noise or sparse data and can lead to systematically inflated eccentricity estimates. 

To maximise the robustness of our population-level inferences, we therefore re-fit the available radial velocity data to derive orbital eccentricities in a uniform manner using the {\sc pyaneti} software \citep{2019MNRAS.482.1017B}, which employs an Affine Invariant Ensemble sampler \citep{2010CAMCS...5...65G} for Keplerian models. {We uniformly fit orbits for all transiting BD and very low-mass star systems compiled in the Cosmic Cookery catalog\footnote{\url{https://www.theroncarmichael.com/browndwarfs}}, originally published in \citet{2023MNRAS.519.5177C} and updated as of September 2023, which we supplemented with more recent discoveries published up to September 2024 \citep{2024MNRAS.530..318H, 2024MNRAS.533.2823H}. This yielded a total of 50 systems.} We considered 300 independent chains each with uniform distributions for the model parameters: orbital period $P_b$, mid-transit time $T_{0,b}$, combined eccentricity and argument of periastron ($\sqrt{e_b}\cos\omega_b$ and $\sqrt{e_b}\sin\omega_b$) to mitigate the Lucy–Sweeney bias \citep{1971AJ.....76..544L}, and Doppler semi-amplitude $K$. All radial velocity time series were retrieved from each object's original discovery paper, and the full set of phase-folded radial velocity fits is presented in Appendix \ref{app:keplerian_models}. 

\begin{figure}[h]
    \centering
    \includegraphics[width = \linewidth]{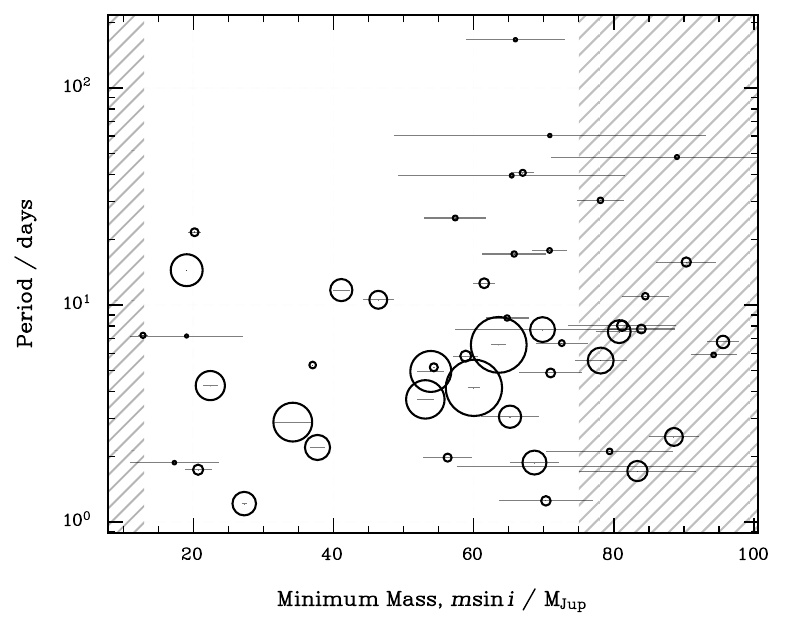}\\
    \includegraphics[width = \linewidth]{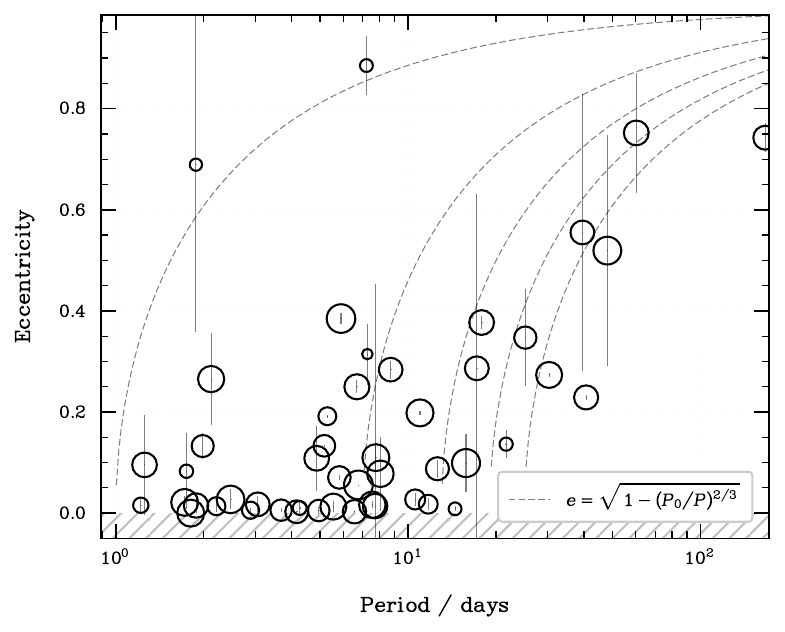}
    \caption{{Fundamental orbital properties of transiting BDs and low-mass stars inferred from our uniform re-fit of RV time-series in this work} (see $\S$\ref{sec:re-fit}). {While our full sample is shown here, only systems with derived $M_b\sin{i} < 75M_J$ are retained for the population-level eccentricity studies presented in this work.} \textit{Top panel:} Companion mass vs. orbital period. Marker sizes are scaled inversely with mean orbital eccentricity, such that higher-eccentricity systems are indicated by smaller circles. Hatched regions mark rough boundaries for deuterium burning \citep[$M_b <13~M_{\rm Jup}$;][]{2011ApJ...727...57S} and hydrogen burning \citep[$M_b > 75~M_{\rm Jup}$;][]{2023A&A...671A.119C, 2024ApJ...975...59M}. \textit{Bottom panel:} Orbital period vs. eccentricity. Marker sizes are scaled proportionally with BD mass. The hatched region marks $e < 0$, denoting unphysical parameter space. Dashed curves show angular momentum conservation tracks for initial orbital periods $P_0 = 1-30,~\delta P_0 = 6$ days, following $e = \sqrt{1 - (P_0/P)^{2/3}}$ (see \citealt{2016ApJ...829...34S}).}
    \label{fig:BDs_M_P_E}
\end{figure}

A complete list of targets and their uniformly derived orbital parameters is provided in Table \ref{tab:bdparams}, and Figure \ref{fig:BDs_M_P_E} shows the mass-period-eccentricity distribution of this re-analysed sample. The sample spans a broad range of host star types and orbital properties, with periods ranging from 0.55 to 166 days (with a mean of $\langle P_b \rangle = 13.7$ days), eccentricities up to 0.88 ($\langle e_b \rangle = 0.17$), and {companion minimum} masses from $12.8$ to $95.5~M_{\rm Jup}$ ($\langle {M_b\sin i} \rangle = 22.5~M_{\rm Jup}$), around stars with masses from $0.18$ to $2.38~M_\odot$ ($\langle M_{\star} \rangle = 0.5~M_\odot$) and effective temperatures from $3100$ to $10400$ K ($\langle T_{\rm eff} \rangle = 5810$ K). 

{We found that 14 of the modelled companions lie above the hydrogen-burning mass limit at roughly $75~M_{\rm Jup}$. Whilst these systems' derived parameters are reported in Table \ref{tab:bdparams} for completeness, we retain only the 36 transiting BD systems with $M_b\sin{i} < 75~M_{\rm Jup}$ for the remainder of this work.} {Because our sample consists of only transiting systems, the companions' orbital inclinations are close to edge-on ($i \approx 90^\circ$) such that their true masses differ only minimally from the RV-derived $M_b\sin i$ values (typically $<5\%$). Including literature inclination constraints in our mass cut-off does not, therefore, alter our sample.}

\section{The Eccentricity Distribution}\label{sec:BDs}

\subsection{Model Overview}\label{sec:genframework}

We apply a Hierarchical Bayesian Model (HBM) framework to model and contextualise the orbital eccentricity distribution for close-in BDs. {Hierarchy, in this case, is reflected in the model's multiple levels}. We first define the hyperparameters---also called the parameters of the prior distribution---and the hyper-priors, called the distribution of the hyperparameters\footnote{Consider, for example, a random variable $X$ with a normal distribution denoted by $\mathcal{N}(\mu, \sigma)$, where $\mu$ and $\sigma$ are the mean and standard deviation, respectively. Assume that $\mu$ follows a Uniform distribution between $\phi$ and $\xi$, making $\mu$ a hyperparameter, represented as $\mu\sim\mathcal{U}(\phi,\xi)$. This scenario is referred to as a two-level HBM. If $\phi$ and/or $\xi$ also follows another distribution $\mathcal{D}$ with hyperparameters $(\theta_1, \theta_2, \dots, \theta_n) \in \Theta$, such as $(\xi,\phi)\sim\mathcal{D}(\Theta)$, it results in a three-level HBM, and so forth.}. With a hierarchical structure, the model can capture variability across different systems, smoothing out information to improve parameter estimation. This approach is particularly helpful in cases with limited individual system data or biased populations, in that it allows for a more flexible and adaptive representation of underlying distributions without imposing rigid assumptions onto a dataset. For this reason, an HBM is preferred over a regular Bayesian model when inferring the eccentricity distribution of a sample (see, e.g., discussions in \citealt{2023AJ....166..112D}). 

In its general form (Equation 7 in \citealt{2010ApJ...725.2166H}), a posterior distribution of hyperparameters $\zeta\equiv(\alpha,\beta)\in\Theta$ given the data $\mathcal{D}$ follows\footnote{We seek the posterior of the global parameters $\Theta$ given $(\mathcal{D} = \{\mathcal{D}_1,\dots,\mathcal{D}_N\}$, where assuming conditional independence of datasets given $\Theta$, we have $\pi(\mathcal{D} \mid \Theta) = \prod_{k=1}^N \pi(\mathcal{D}_k \mid \Theta)$. If each $\mathcal{D}_k$ depends on a latent parameter $\zeta_k$ drawn from $\pi(\zeta_k \mid \Theta)$, then $\pi(\mathcal{D}_k \mid \Theta) = \int \pi(\zeta_k \mid \Theta) \, \pi(\mathcal{D}_k \mid \zeta_k) \,\mathrm{d}\zeta_k$. Substitutions yield Equation \ref{eq:pithetaD}.}

\begin{equation}
    \pi(\Theta \vert \mathcal{D}) \propto \pi(\Theta) \prod_{k=1}^{N} \int \pi(\zeta_k \vert \Theta) \pi(\mathcal{D}_k \vert \zeta_k) \dd{\zeta_k},\label{eq:pithetaD}
\end{equation}
where $\pi(\Theta)$ is the prior distribution of the model hyperparameters; $\zeta_k$ are the latent parameters for the $k-{\rm th}$ system, representing specific characteristics (e.g., eccentricity) that are inferred from the data; $\pi(\zeta_k\vert\Theta)$ is the likelihood of individual system parameters $\zeta_k$ given the hyperparameters $\Theta$; and $\pi(\mathcal{D}\vert\zeta_k)$ is the likelihood of the observed data $\mathcal{D}_k$ for the $k-{\rm th}$ system, given the latent parameters $\zeta_k$.

We followed the precedent outlined in several previous works fitting orbital eccentricity distributions \citep{2010ApJ...725.2166H, 2013MNRAS.434L..51K, 2019AJ....157...61V, 2020AJ....159...63B, 2023AJ....165...32N, 2023AJ....166...48D} by employing a Beta distribution $\mathcal{B}(\alpha, \beta)$. The Beta distribution has proven advantageous due to its simplicity---relying on only two parameters---and its notable flexibility in capturing a wide range of distribution shapes\footnote{For instance, a Uniform distribution is obtained when $\alpha = \beta = 1$, whilst a Normal distribution is approximated when both $\alpha$ and $\beta$ are sufficiently large, with the mean ($\mu$) calculated as $\alpha/(\alpha + \beta)$ and the standard deviation ($\sigma$) given by $\sqrt{\alpha\beta / [(\alpha + \beta)^2 (\alpha + \beta + 1)]}$.}. The distribution also aligns well with the range of potential orbital eccentricities for bounded companions, since $\mathcal{B}(e\vert\alpha,\beta)\in[0, 1]$. 

The qualitative behaviour of the Beta distribution is determined entirely by whether each shape parameter lies below, equal to, or above unity. When both fall below unity, the density is boundary-peaked; when both exceed unity, it becomes unimodal with mode $e^\ast = (\alpha - 1)/(\alpha + \beta - 2)$; and when $\alpha = \beta$, the distribution ranges from strongly U-shaped ($\alpha \ll 1$) to sharply concentrated around $e = 0.5$ for $\alpha \gg 1$, with the Uniform case recovered at $\alpha = \beta = 1$. Mixed cases can arise---in which one parameter lies below unity and the other above---yielding monotonic densities biased towards the endpoint associated with the smaller parameter. This behaviour also covers the edge cases in which exactly one parameter is equal to unity. 

The versatility of the Beta distribution enables it to effectively replicate the outcomes of various potential dynamical histories. {A Beta distribution can capture the behaviour of objects formed bottom-up within a disc, which are typically expected to maintain nearly circular orbits ($e\approx 0$) unless they are dynamically perturbed---e.g., by scattering phenomena that may naturally produce eccentric orbits \citep{1996Sci...274..954R, 2009ApJ...693L.113S}. It can also capture the broad range of orbital eccentricities expected from molecular cloud fragmentation, {such as the statistical equilibrium energy state represented by a thermal distribution $f(e) = 2e$ \citep{jeans1919origin, 1937AZh....14..207A}.}

A population-level eccentricity model following the Beta distribution can be expressed as:

\begin{equation}
    p(e\vert\alpha,\beta\in\Theta) = \frac{\Gamma(\alpha + \beta)}{\Gamma(\alpha)\Gamma(\beta)} e^{\alpha - 1}\left(1-e\right)^{\beta - 1}, 
    \label{eq:beta}
\end{equation}
where $(\alpha,\beta)\in\mathbb{R}^{+}$, and $\Gamma$ represents a Gamma function \citep{1980tisp.book.....G}

\begin{equation}
    \Gamma(x) = \int_0^\infty t^{x-1} e^{-t} \dd{t}, \quad \mbox{if } {\mathcal R}(x) > 0.
\end{equation}
For a population described by a Beta function, the empirical cumulative distribution function is given by the regularised incomplete Beta function \citep{1980tisp.book.....G}

\begin{equation}
    \mathcal{I}_{x}(\alpha,\beta) = \frac{B(x;\alpha,\beta)}{B(\alpha,\beta)} = \frac{1}{B(\alpha,\beta)}\int_{\frac{1-x}{x}}^{+\infty} \frac{\xi^{\beta-1}}{(1+\xi)^{\alpha + \beta}} d\xi.
    \label{eq:incomplete_beta}
\end{equation}

For the models presented in this work, the Markov hyperparameter spaces were explored using an adaptive Hamiltonian Monte Carlo (HMC) No-U-Turn sampler (NUTS; \citealt{2011arXiv1111.4246H}), with sampling conducted with $10^4$ iterations over 2500 tuning steps and a target acceptance rate of 0.99 within the {\sc PyMC} {\tt v3.11.5} software\footnote{\url{https://github.com/pymc-devs/pymc}} \citep{2016ascl.soft10016S}. Model convergence was assessed with the Gelman-Rubin statistic $\hat{R} \leq 1$ \citep{1992StaSc...7..457G}. 

\subsection{Population-Level Inference}\label{sec:population_level_inference}

To constrain the typical BD tidal quality factor (see $\S$\ref{sec:tidal}), we examine differences between the short-period (SP) and long-period (LP) transiting BD populations, dividing our sample into two parts. We used a non-parametric Kolmogorov-Smirnov (KS; \citealt{an1933sulla, smirnov1939estimate}) test to identify a period threshold between eccentricity regimes, motivated by the expectation that shorter-period BDs should have systematically more circular orbits than longer-period BDs due to tidal circularisation. We sampled thresholds from 1 to 40 days in 1-day increments, dividing our sample at each threshold into short- and long-period subsets. For each division, we applied a one-sided KS test with the alternative hypothesis that the sample distribution is shifted toward larger values compared to the reference, quantifying whether the short-period transiting BD population had systematically lower eccentricities than long-period companions. The threshold yielding the minimum $p-$value was selected as the optimal boundary between eccentricity regimes.

A period threshold of $P_{\rm GP, threshold} = 16$ days was found for the BD population with a KS statistic of $D = 0.76$ and a $p-$value of $10^{-5}$, leading to a rejection of the null hypothesis that the two samples are drawn from the same underlying distribution. The corresponding cumulative eccentricity distributions for the full sample and for the sub-samples above and below the period threshold are shown in Figure \ref{fig:BDs_ECDF}. The significance of the KS statistic for the $P_{\rm GP, threshold} = 16$ day division suggests that the eccentricity distribution substantially differs between the shortest-period and slightly longer-period transiting BDs, as expected for a population that is influenced by tidal evolution. 

{We note that, though other measures of orbital separation (such as $a/R_*$ for semimajor axis $a$ and stellar radius $R_*$) more directly trace tidal evolution whilst incorporating the influence of stellar type, orbital period is a direct observable from transit photometry and is more uniformly reported across heterogeneous archival samples. We therefore adopt orbital period as our dividing factor, rather than $a/R_*$, to ensure consistency across the full dataset and facilitate comparison with other recent studies \citep[e.g.][]{2026A&A...709A.130S}. To search for possible systematic biases in host star properties introduced by our choice of period-based threshold, we performed a KS test comparing the masses of the SP and LP transiting BD host stars. We found that the two populations do not show significant differences in their host star mass distributions (KS = 0.24, $p = 0.32$), indicating no immediately evident systematic differences in the stellar samples.}

\begin{figure}[t]
    \centering
    \includegraphics[width = \linewidth]{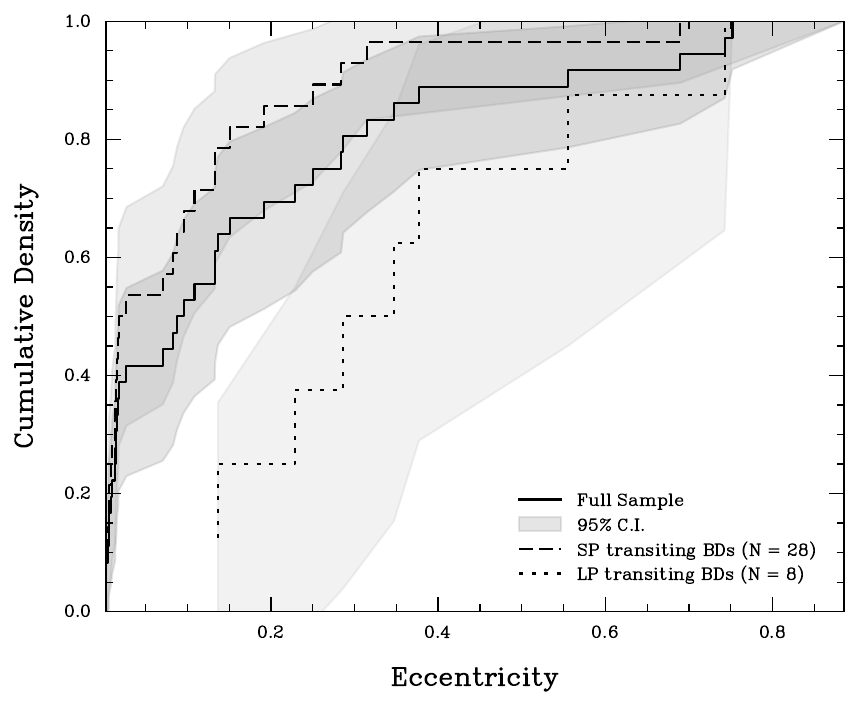}
    \caption{Empirical cumulative distribution function with shaded $95\%$ confidence intervals for the short-period ($P < 16$ days; dashed) and long-period ($P \geq 16$ days; dotted) BD populations, delineated by a period threshold determined via a KS test and  $p-$value analysis (see $\S$\ref{sec:population_level_inference}). The full sample CDF is represented as a solid black line.}
    \label{fig:BDs_ECDF}
\end{figure}

Within the HBM framework, for both the short- and long-period populations we set the hyperparameters $\alpha$ and $\beta$ of the Beta distributions using Normal priors $(\alpha, \beta) \sim \mathcal{N}(\mu_k, \sigma_k)$, with $k = \alpha, \beta$ and $\mu, \sigma$ corresponding to the mean and standard deviation of the distribution. To ensure positive values and improve sampling efficiency, we define the location and scale hyperparameters via natural-logarithm-transformed Uniform priors: $\ln \mu_k, \ln \sigma_k \sim \mathcal{U}(\ln 0.001, \ln 1000)$, motivated by recent studies for improved hierarchical modelling by \cite{2023AJ....165...32N}.

The top left panel of Figure \ref{fig:BDs_Beta} illustrates the resulting Beta distributions for the short- and long-period transiting BD populations, alongside the eccentricity distribution for the full sample (see summarised Table \ref{tab:beta_summary}). To provide further context for these findings, we applied the same HBM approach to two different comparison samples---namely, Jovian planets from the {\it California Legacy Survey} and wide binaries from the \emph{Gaia} catalogue---with results presented in Appendix \ref{app:jovianplanets} and \ref{app:binaries}. Figure \ref{fig:BDs_Beta} also shows a comparison of eccentricity distributions determined from previous work for various populations of BDs, giant exoplanets, and stellar binaries. 

\begin{figure*}[t]
    \centering
    \includegraphics[width = \linewidth]{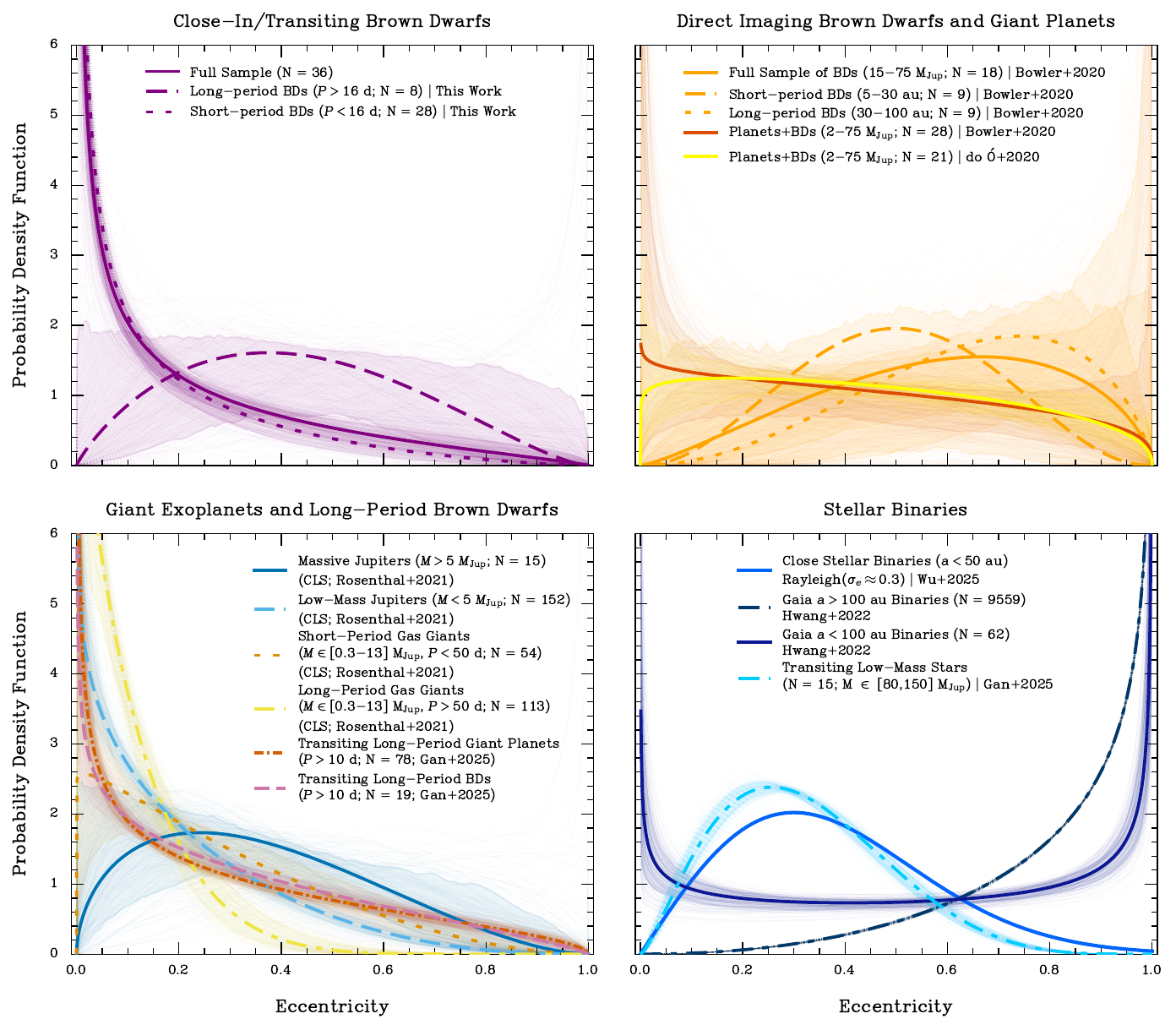}
    \caption{Eccentricity distributions for transiting BDs (top left panel; this work), divided into short- and long-period sub-populations, alongside the eccentricity distributions of directly imaged BDs and gas giant planets (top right panel; \citealt{2020AJ....159...63B, 2023AJ....166...48D}); Jovian planets from the {\it California Legacy Survey}, together with long-period giant planets and transiting BDs analysed in \cite{2025arXiv250709461G} (bottom left panel; \citealt{2021ApJS..255....8R}; see Appendix \ref{app:jovianplanets}); and stellar binaries (bottom right panel; \citealt{2022MNRAS.512.3383H, 2025ApJ...982L..34W, 2025arXiv250709461G}; see Appendix \ref{app:binaries}). {Except for the distribution of close stellar binaries from \cite{2025ApJ...982L..34W}, which is depicted with a Rayleigh distribution with scale parameter $\sigma_e = 0.3$, each panel shows 200 posterior samples of a Beta distribution fitted to eccentricity measurements from each distinct population, following the process presented in $\S$\ref{sec:genframework}.} Shaded regions for each Beta distribution represent the 68\% credible intervals ($1\sigma$) derived via Monte Carlo sampling.}
    \label{fig:BDs_Beta}
\end{figure*}

\begin{table*}[t]
\centering
\renewcommand{\arraystretch}{1.3}
\begin{tabular}{@{} l c c c @{}}
\toprule
{\bf Population} & {\bf Mean}         & {\bf Beta Parameters}  & {\bf Note} \\
                 & {\bf Eccentricity} & {\bf $(\alpha,\beta)$} &                  \\
\midrule
\midrule

{Short-period BDs (SP BDs)$^{(1)}$}

    & {$0.15 \pm 0.05$}
    & {$(0.426^{+0.100}_{-0.084},\ 2.354^{+0.807}_{-0.658})$} 
    & Skewed toward low eccentricity ($\alpha < 1$, $\beta > 1$). \\

{Long-period BDs (LP BDs)$^{(1)}$}

    & {$0.43 \pm 0.26$} 
    & {$(1.879^{+1.088}_{-0.753},\ 2.470^{+1.480}_{-1.018})$} 
    & Broad distribution, roughly symmetric ($\alpha, \beta > 1$). \\

{Full sample BDs (FS BDs)$^{(1)}$}
    
    & {$0.20 \pm 0.05$} 
    & {$(0.469^{+0.082}_{-0.087},\ 1.815^{+0.414}_{-0.457})$}
    & Mildly skewed toward circular orbits. \\

\midrule
\midrule

Low-mass Jovians (LM GPs)$^{(2)}$ 
    
    & $0.20 \pm 0.02$ 
    & $(0.844^{+0.087}_{-0.081},\ 3.313^{+0.421}_{-0.388})$ 
    & Tendency for circular orbits ($\alpha < 1$, $\beta > 1$). \\

Massive Jovians (M GPs)$^{(2)}$ 
    
    & $0.36 \pm 0.14$ 
    & $(1.545^{+0.545}_{-0.432},\ 2.716^{+1.013}_{-0.804})$ 
    & Broad distribution ($\alpha, \beta > 1$). \\

\midrule

Close binaries ($a < 100$ au)$^{(3)}$ 
    
    & $0.55 \pm 0.11$ 
    & $(0.704^{+0.123}_{-0.110},\ 0.573^{+0.095}_{-0.085})$ 
    & Bimodal; preference for extremes ($\alpha, \beta < 1$). \\

Wide binaries ($a > 100$ au)$^{(3)}$ 
    
    & $0.82 \pm 0.01$ 
    & $(3.047^{+0.049}_{-0.047},\ 0.675^{+0.008}_{-0.008})$ 
    & Skewed toward high eccentricity ($\alpha > 1$, $\beta < 1$). \\

\bottomrule
\end{tabular}
\caption{Summary of eccentricity distributions, statistical means, and fitted Beta function hyperparameters for the various populations analysed. {\it Notes:} $^{(1)}$This work with uniform re-fit (see $\S$\ref{sec:re-fit}); $^{(2)}$Derived from the {\it California Legacy Survey} sample (\citealt{2021ApJS..255....8R}; see Appendix~\ref{app:jovianplanets}); $^{(3)}$Based on the {\it Gaia} wide binaries catalogue (\citealt{2022MNRAS.512.3383H}; see Appendix \ref{app:binaries}).}
\label{tab:beta_summary}
\end{table*}

The short-period transiting BD population ($P < 16$ days) is well fit using Beta-regression parameters with $\alpha < 1$ and $\beta > 1$, indicating a population-level preference for nearly circular orbits (average eccentricity $\langle e \rangle = 0.15$; see also Figure \ref{fig:BDs_ECDF}). This is expected for a population with such close-in orbits that have been sculpted by tides and orbital circularisation through eccentricity damping. We examine the implications of this modelled distribution for BD tidal quality factors and circularisation timescales in $\S$\ref{sec:tidal}. 

By contrast, the eccentricity distribution for long-period transiting BDs ($P \geq 16$ days) is characterised by hyperparameters $\alpha > 1$ and a similar $\beta$ to the short-period population, but with a broad peak around $\langle e \rangle\approx 0.4$. Since tidal damping is inefficient at long orbital periods, the observed eccentricities are able to persist over Gyr timescales commensurate with the systems' ages. These moderate eccentricities may reflect binary-star-like formation with primordially elevated eccentricities and/or past dynamical excitation---for example, via von Zeipel-Lidov-Kozai (vZLK) oscillations (\citealt{1910AN....183..345V, 1962P&SS....9..719L, 1962AJ.....67..591K}; see also \citealt{2019MEEP....7....1I}), scattering \citep{1996Sci...274..954R}, or secular chaos \citep{2011ApJ...739...31L, 2011ApJ...735..109W}. 

\subsection{Comparing Eccentricity Distributions of Transiting Brown Dwarfs, Binary Stars, and Giant Planets}

To provide further context for the transiting BD eccentricity distributions, several benchmark comparison samples are shown in Figure \ref{fig:BDs_Beta}. We also compare the populations' eccentricity distributions through empirical cumulative distribution functions in Figure \ref{fig:ks_populations}. 

{We compare the SP and LP transiting BD populations against the eccentricity distributions of giant planets from the {\it California Legacy Survey} (CLS; \citealt{2021ApJS..255....8R}); transiting low-mass stars ($M_{\rm stars} \in [80,150]~M_{\rm Jup}$) from \cite{2025arXiv250709461G}, characterised by $\mathcal{B}(2.42, 5.12)$; and close stellar binaries from \cite{2025ApJ...982L..34W}, described by a Rayleigh distribution with $\sigma_e = 0.3$. The LP BD eccentricity distribution is indistinguishable from that of both transiting low-mass stars from \citet{2025arXiv250709461G} ($D = 0.28$, $p = 0.45$) and close stellar binaries from \citet{2025ApJ...982L..34W} ($D = 0.20$, $p = 0.81$), and it is marginally inconsistent with the eccentricity distribution of long-period giant planets from CLS \citep[$D = 0.40$, $p = 0.12$;][]{2021ApJS..255....8R}. By contrast, SP transiting BDs have a distinct eccentricity distribution that is strongly inconsistent with both stellar populations: $D = 0.65$, $p < 10^{-12}$ when comparing with the low-mass stars and close stellar binaries. The SP BDs show marginal consistency only with short-period giant planets from CLS ($D = 0.26$, $p = 0.13$).} 

{These results indicate that the orbital eccentricity distribution of LP BDs is comparable to that of close stellar binary systems.} By contrast, SP BDs display orbital characteristics more akin to those of short-period giant planets, which may result from comparable tidal circularisation processes operating for both system types -- irrespective of whether they originated with comparable orbital eccentricity distributions.

\begin{figure}[h]
    \centering
    \includegraphics[width = \linewidth]{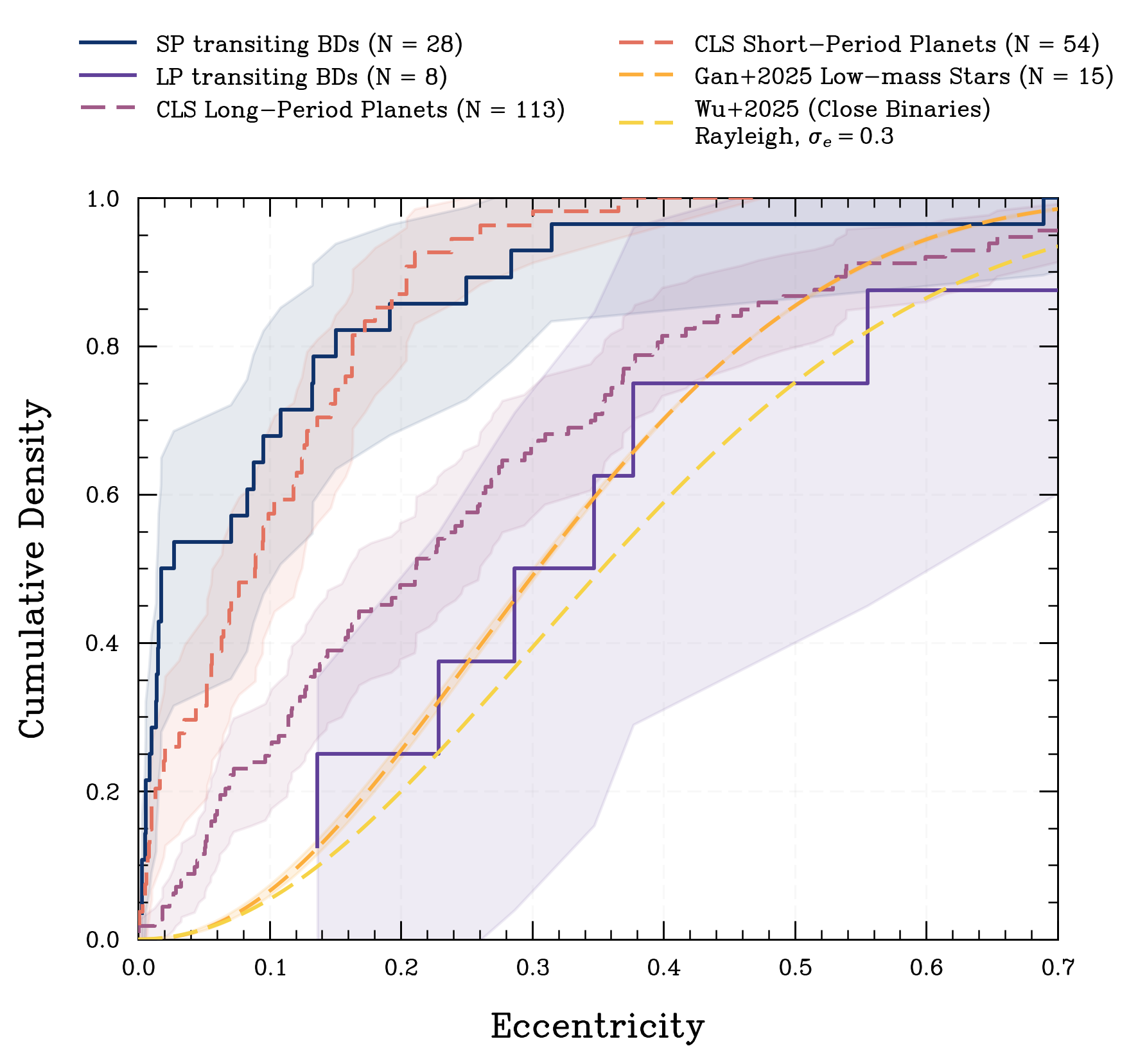}
    \caption{Empirical cumulative distribution functions comparing short-period (SP, $P < 16$ days) and long-period (LP, $P \geq 16$ days) transiting BDs (solid lines) with various comparison samples (dashed lines): short-period ($N = 54$) and long-period ($N = 113$) giant planets from the {\it California Legacy Survey}, eccentricity distributions for transiting low-mass stars ($M_\star \in [80,150]~M_{\rm Jup}$; $N = 15$) from \cite{2025arXiv250709461G}, and close stellar binaries (Rayleigh distribution, $\sigma_e = 0.3$) from \cite{2025ApJ...982L..34W}. Shaded regions represent 95\% confidence intervals.}
    \label{fig:ks_populations}
\end{figure}

\section{Tidal Evolution of Close-In Brown Dwarfs}\label{sec:tidal}

The long-term dynamical evolution of transiting BDs is shaped by the cumulative impact of tidal dissipation, which progressively attenuates eccentricity and circularises orbits---a process that is particularly relevant for systems in which the BD resides at small orbital separations. The efficacy of this mechanism is linked to the internal stratification and dissipative properties of the BD. In this section, we consider how tidal dissipation sculpts eccentricity across the population. Specifically, we demonstrate that the low orbital eccentricities of short-period transiting BDs can be reproduced as the tidally processed relics of the longer-period transiting BD eccentricity distribution. 

Under this assumption, we derive limits on the typical transiting BD tidal quality factor $\mathcal{Q}_{\rm BD}$, as well as the host stars' typical tidal quality factor $\mathcal{Q}_\star$, using two separate equilibrium tides frameworks. Neither tidal prescription considered in this work should be regarded as a complete physical description of BD tides, since both neglect dynamical tides, inertial-wave dissipation, resonance locking, and structural feedback from tidal heating. $\mathcal{Q}_{\rm BD}$ represents an effective ensemble-averaged dissipation efficiency for substellar companions, and $\mathcal{Q}_\star$ characterises the mean tidal response of the stellar hosts. Inferred values should therefore be interpreted as population-level effective parameters.

\subsection{\citet{2008Icar..193..637W} Tidal Formalism $-$ Model A}\label{subsubsec:model1}

We first consider the tidal framework of \citet{2008Icar..193..637W}, which assumes a synchronously rotating body on an eccentric orbit around the host star. In this framing, we consider only the tidal distortion of the BD, which is assumed to be incompressible with a low orbital eccentricity. The secular, orbit-averaged eccentricity evolution is written as

\begin{equation}
    \frac{de}{dt} \equiv \dot{e} = \left[\frac{21 \kappa_2 G M_\star^2 \Omega R_{\rm BD}^5 \zeta(e)}{2 \mathcal{Q}_{\rm BD} a^6} \right] \frac{a (1 - e^2)}{G M_\star M_{\rm BD} e},\label{eq:dote}
\end{equation}
where $\kappa_2$ is the Love number, $\mathcal{Q_{\rm BD}}$ is the effective tidal quality factor, and $M_\star$, $R_\star$, $M_{\rm BD}$, $a$, and $e$ denote the stellar mass and radius and the BD mass, semi-major axis, and eccentricity, respectively. The auxiliary eccentricity functions are

\begin{equation}
    \Omega = \frac{1}{P} \frac{f_1(e)}{f_2(e) \beta^3}
\end{equation}

\begin{equation}
    \zeta(e) = \frac{2}{7} \left[ \frac{f_0(e)}{\beta^{15}} - 2 \frac{f_1(e)}{\beta^{12}} + \frac{f_2(e)}{\beta^9} \right],
\end{equation}
with

\begin{align}
    f_0(e) &= 1 + \frac{31}{2} e^2 + \frac{255}{8} e^4 + \frac{185}{16} e^6 + \frac{25}{64} e^8, \\
    f_1(e) &= 1 + \frac{15}{2} e^2 + \frac{45}{8} e^4 + \frac{5}{16} e^6, \\
    f_2(e) &= 1 + 3 e^2 + \frac{3}{8} e^4, \\
    \beta &= \sqrt{1 - e^2}. \label{eq:aux}
\end{align}

\subsection{\citet{2008ApJ...678.1396J} Tidal Formalism $-$ Model B}\label{subsubsec:model2}

We also consider the tidal formalism of \cite{2008ApJ...678.1396J}, which considers the coupled tidal dissipation within both the BD and the host star. In our case, these are characterised by constant quality factors $\mathcal{Q}_{\rm BD}$ and $\mathcal{Q}_\star$, respectively. The secular evolution equations are given by

\begin{eqnarray}
    \frac{de}{dt}\equiv\dot{e}&=&-e\left[\frac{63}{4}\frac{\left(GM_\star^3\right)^{1/2}R_{\rm BD}^5}{\mathcal{Q}_{\rm BD}M_{\rm BD}}\right. \\ &+& \left.\frac{171}{16}\frac{\left(G/M_\star\right)^{1/2}R_\star^5 M_{\rm BD}}{\mathcal{Q}_\star}\right] a^{-13/2}
\end{eqnarray}

and 

\begin{eqnarray}
    \frac{da}{dt}\equiv\dot{a}&=&-a\left[\frac{63}{2}\frac{\left(GM_\star^3\right)^{1/2}R_{\rm BD}}{\mathcal{Q}_{\rm BD} M_{\rm BD}} e^2\right. \\ &+& \left.\frac{9}{2} \frac{\left(G/M_\star\right)^{1/2}R_\star^5 M_{\rm BD}}{\mathcal{Q}_\star}\right] a^{-13/2}.
\end{eqnarray}

\subsection{Empirical Tidal Efficiency $\mathcal{Q}$ Constraints}\label{sec:Q}

We model the tidal evolution of the short-period transiting BD population by following both frameworks introduced in Sections \ref{subsubsec:model1} and \ref{subsubsec:model2}. We demonstrate that the eccentricity distribution of the SP population can be reproduced through the cumulative tidal processing of shorter-period {orbital eccentricity} analogues to the LP population. 

In our forward-modelling procedure, BD masses and orbital periods are sampled from the observed SP population. For each realisation, we generate $5000$ systems by bootstrapping from the SP BD population of $P < 16$ days. Measurement uncertainties are incorporated by drawing BD masses and periods from Gaussian distributions centred on the reported mean values, with standard deviations set by the associated uncertainties. Stellar masses and radii are similarly drawn from the observed stellar host population. Initial eccentricities are drawn from a Beta distribution, {$\mathcal{B}(1.879,~2.470)$}, consistent with the observed LP transiting BD population (see Table \ref{tab:beta_summary}). 

Since the drawn orbital periods correspond to the currently observed, rather than initial, evolutionary states of each system, we compute initial orbital periods following angular momentum conservation: $P_{\rm final} = P_{\rm initial} (1-e_0^2)^{3/2}$ \citep{2016ApJ...829...34S}, which accounts for the orbital contraction that accompanies circularisation. {This prescription is agnostic to the mechanism for the BD's eccentricity excitation, which may have resulted in the prior ejection of neighbouring companions from the system. We assume that after any such instability, the BD's orbital circularisation conserves angular momentum. Hence, the BD's angular momentum budget is assumed to be only minimally affected by inner companions that may be disrupted during the BD's orbital circularisation process.} {We note that the fractional difference in the BD's angular momentum from a putative $\sim3~M_{\rm Jup}$ companion ejected during circularisation can be estimated as $\Delta L_{\rm BD}/L_{\rm BD} \sim M_p/M_{\rm BD} \lesssim 10\%$ for typical BDs of $M_{\rm BD}\approx30~M_{\rm Jup}$, corresponding to a maximum difference of up to a factor of $\approx$1.4 in $P_{\rm final}$ from unaccounted-for ejected planetary companions.}

The population is evolved to a fixed age of 6.5 Gyr---the median age of this work's transiting BD population with available ages from the \emph{Gaia} DR3 FLAME module---using $\log_{10}\mathcal{Q}$ values ranging from $1$ to $10$ in increments of $\delta \log_{10} \mathcal{Q} = 0.1$ for Model A. For Model B, which incorporates both the stellar and planetary tidal influence, we adopt a two-dimensional grid of tidal quality factors, sampling $\mathcal{Q}_\star$ and $\mathcal{Q}_{\rm BD}$ independently over the range $10^{6}$ to $10^{9}$ in steps of $\delta \log_{10} \mathcal{Q}_{\rm BD\vert\star} = 0.1$ dex.

To quantify the goodness of fit for each evolved population and for both models, we compare models with the data via a Kullback-Leibler (KL) divergence test:

\begin{equation} 
    \mathcal{D}_{\rm KL}(R \parallel S) \;=\; \int_0^1 R(e)\, \log_{10} \!\left[\frac{R(e)}{S(e)}\right] \dd{e},
\end{equation} 
which measures the information lost when a candidate distribution $S(e)$ is used to approximate a reference distribution $R(e)$. Here, $R(e)$ represents the observed eccentricity distribution and $S(e)$ the simulated one. A value of $\mathcal{D}_{\rm KL} = 0$ corresponds to a perfect match, whilst larger values indicate increasing statistical dissimilarity. Unlike traditional metrics such as the KS statistic, which probe differences in cumulative distributions, the KL divergence is sensitive to discrepancies in the full probability density, particularly in the tails of the distribution. 

For Beta distributions, $\mathcal{D}_{\rm KL}$ admits a closed-form solution

\begin{align}
    \mathcal{D}_{\rm KL}&\!\left(\mathcal{B}(\alpha_p, \beta_p) \,\big\|\, \mathcal{B}(\alpha_q, \beta_q)\right) 
    = \ln\frac{\mathcal{B}(\alpha_q, \beta_q)}{\mathcal{B}(\alpha_p, \beta_p)} \notag \\
    & + (\alpha_p - \alpha_q)\,\big[\psi(\alpha_p) - \psi(\alpha_p + \beta_p)\big] \notag \\
    & + (\beta_p - \beta_q)\,\big[\psi(\beta_p) - \psi(\alpha_p + \beta_p)\big],
\end{align}
where $\mathcal{B}(\alpha,\beta)=\Gamma(\alpha)\Gamma(\beta)/\Gamma(\alpha+\beta)$ is the Beta function and $\psi(x)=\tfrac{d}{dx}\ln\Gamma(x)$ is the Digamma function---enabling efficient evaluation across parameter space.

We infer the best-fit tidal quality factor $\log_{10}\mathcal{Q}$ by minimising the KL divergence between the evolved eccentricity distribution and the observed SP BD distribution, modelled as {$\mathcal{B}\left(0.426,\ 2.354\right)$}. To derive a posterior constraint, we treat the KL divergence as an effective misfit statistic and construct a likelihood $\ln \mathcal{L}(\mathcal{Q}) = -\langle \mathcal{D}_{\rm KL}(\mathcal{Q}) \rangle$, where the average is taken over all realisations at each grid point. Sampling of $\log_{10}\mathcal{Q}$ is performed via a Metropolis-Hastings MCMC with a Gaussian proposal distribution of width 0.1 dex and a Uniform prior over $1 \le \log_{10}\mathcal{Q} \le 10$. Each proposed $\mathcal{Q}$ is mapped to the nearest grid entry for likelihood evaluation and accepted according to the standard Metropolis ratio. Chains are evolved for $10^6$ steps, yielding {$\mathcal{Q}_{\rm BD} = 10^{\rm 8.1\pm1.0}$ across the population with model A, and a joint constraint $\mathcal{Q}_{\rm BD} = 10^{7.1\pm0.3}$ and $\mathcal{Q}_{\star} = 10^{6.0\pm0.1}$ with model B.}

This $\mathcal{Q}_{\rm BD}$ inference was initially performed at a single, fixed evolutionary age (6.5 Gyr) across the full population. To quantify the influence of age, which, in practice, varies across the population, we repeated the KL-based inference over a grid of $1$, $5$, $8$, and $13$ Gyr. {The resulting values show a monotonic increase in $\mathcal{Q}_{\rm BD}$ with age, from $10^{7.6\pm1.2}$ at 1 Gyr to $10^{8.3\pm0.9}$ at 13 Gyr with model A.}

Across the full range of BD masses, we find an excellent fit for the tidal quality factor, {$\langle\mathcal{Q_{\rm BD}}\rangle\simeq10^{7-9}$}, through forward modelling of the observed SP distribution. {The KL divergence values for all ages are $\mathcal{D}_{\rm KL,~A} \sim 0.08$ and $\mathcal{D}_{\rm KL,~B} \sim 0.3$ for models A and B, respectively. These relatively small values indicate that both tidally evolved populations reproduce the observed SP BD eccentricity distribution well, with model A providing the closer agreement.} 

\begin{figure*}[t]
    \centering
    \includegraphics[width = 0.49\linewidth]{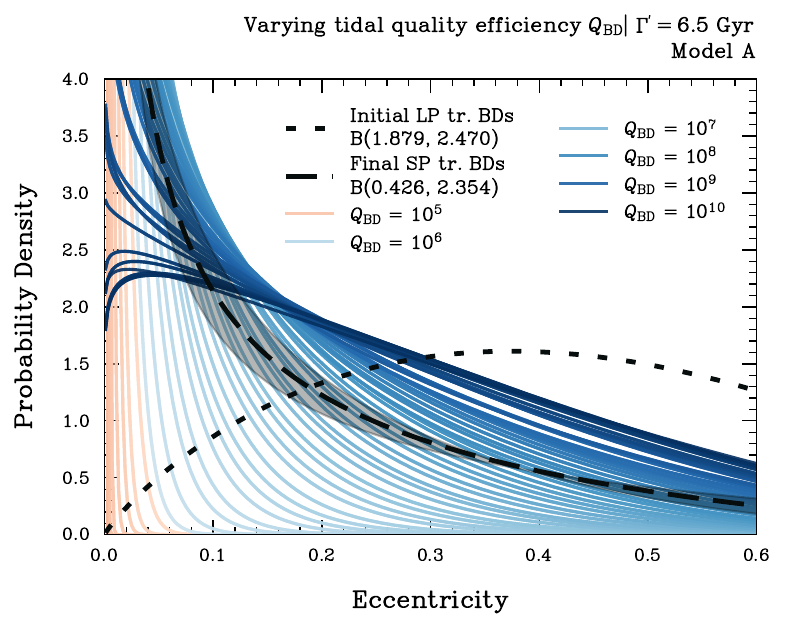}
    \includegraphics[width = 0.49\linewidth]{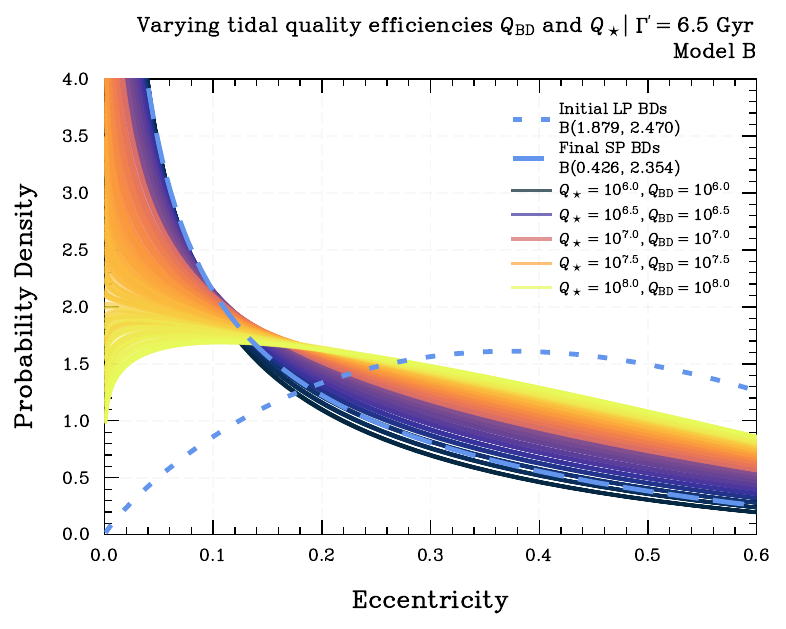}
    \caption{Eccentricity distributions for transiting BDs after 6.5 Gyr of tidal evolution following the \cite{2008Icar..193..637W} model (first panel; model A) and the \cite{2008ApJ...678.1396J} model (second panel; model B). The coloured curves show the evolved distributions for different BD tidal quality factors $\mathcal{Q}_{\rm BD}$ ranging from $10^{1}$ to $10^{10}$ (first panel) and grid combinations of $\mathcal{Q}_\star = 10^{6}-10^{9}$ and $\mathcal{Q}_{\rm BD} = 10^{6}-10^{9}$ (second panel). The initial LP transiting BD distribution (dotted line) and observed SP distribution (dashed line with shaded uncertainties) are shown for comparison (see Table \ref{tab:beta_summary}). Higher $\mathcal{Q}_{\rm BD}$ and/or $\mathcal{Q}_\star$ values indicate weaker tidal dissipation, preserving more eccentric orbits.}
    \label{fig:tidalevolution_mass}
\end{figure*}

\subsection{Intuition from Simplified Populations: Eccentricity Evolution Under Tidal Sculpting}\label{sec:synthetic}

To provide further pedagogical intuition for how the BD eccentricity distribution is shaped by tides, we also evolve synthetic populations assuming a range of unimodal BD masses and $Q$ values in the \citet{2008Icar..193..637W} tidal framework, using Equations \ref{eq:dote}$-$\ref{eq:aux}. Initial eccentricities are drawn from a Beta distribution, {$\mathcal{B}(1.879, 2.470)$}, consistent with the observed long-period transiting BD population (see Table \ref{tab:beta_summary}). We again follow the systems' evolution up to 6.5 Gyr and assume a solar-mass/solar-radius host for all systems. Our test grid spans $\mathcal{Q} = 10^{1}-10^{10}$ in logarithmic steps of $\Delta\log_{10}\mathcal{Q} = 0.1$ and BD masses spanning $M_{\rm BD} = 20-90~M_{\rm Jup}$ in $10~M_{\rm Jup}$ increments. Initial orbital periods were randomly sampled from a continuous Uniform distribution on the interval $[1,16]$ days. 

{Figure \ref{fig:tidalmodelQs} presents the evolved synthetic distributions for varying $\mathcal{Q}$ and BD mass. At low $\mathcal{Q}$, more massive BDs circularise more slowly, consistent with the $\dot{e} \propto M_{\rm BD}^{-1}$ scaling in Equation \ref{eq:dote}. At higher $\mathcal{Q}$, however, this trend reverses: more massive BDs occupy slightly larger semi-major axes at a given period (since these are drawn uniformly), and the steep $a^{-5}$ dependence in Equation \ref{eq:dote} reduces tidal efficiency preferentially for these systems, such that lower-mass BDs become the more circularised population above $\mathcal{Q}\approx10^{8.5}$.}

\begin{figure*}[t]
    \centering
    \includegraphics[width = 0.32\linewidth]{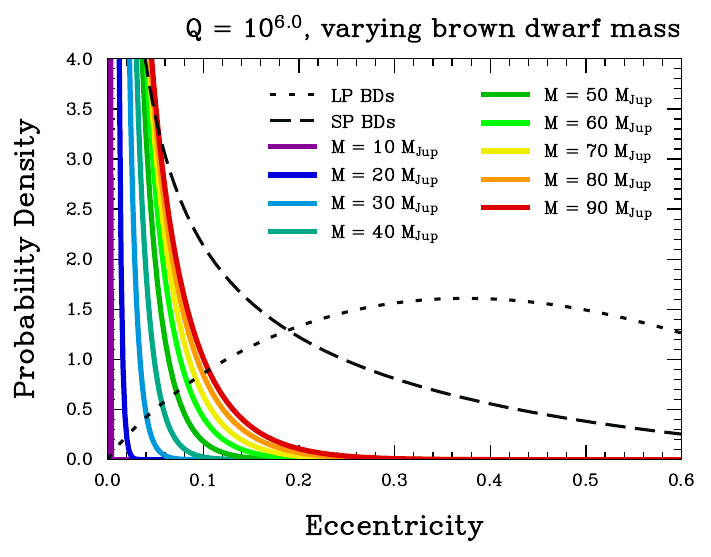}
    \includegraphics[width = 0.32\linewidth]{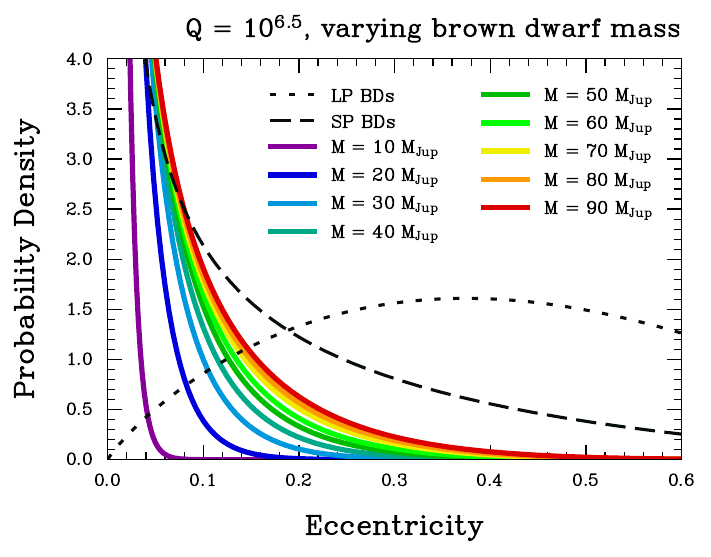}
    \includegraphics[width = 0.32\linewidth]{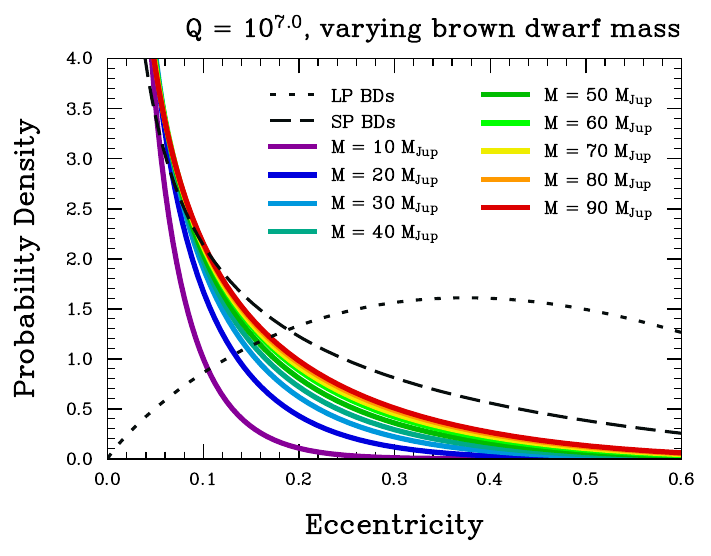}\\
    \includegraphics[width = 0.32\linewidth]{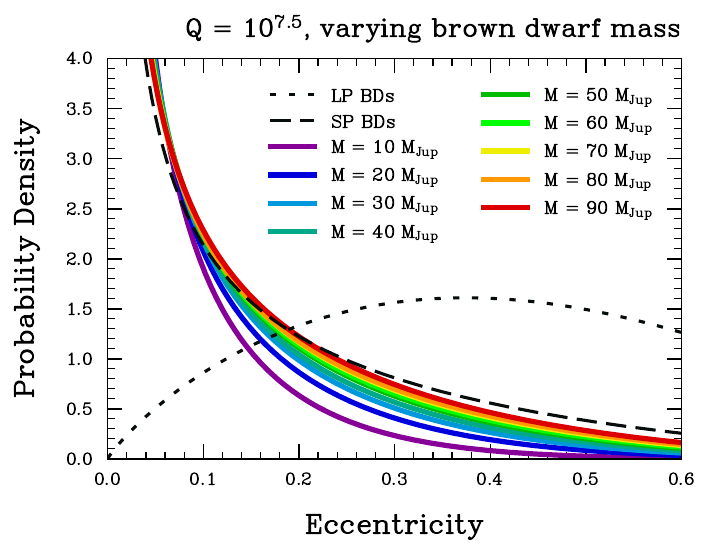}
    \includegraphics[width = 0.32\linewidth]{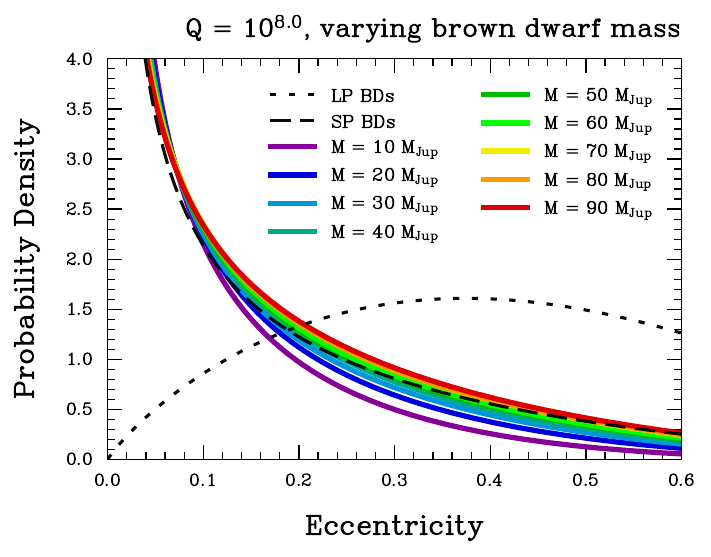}
    \includegraphics[width = 0.32\linewidth]{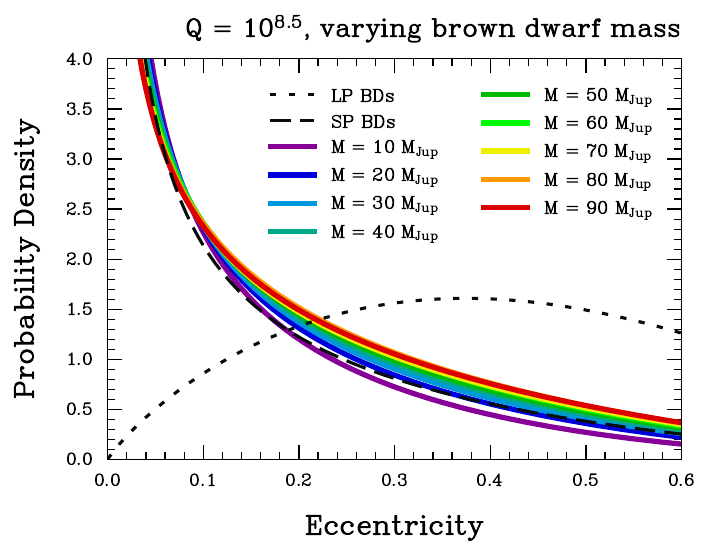}\\
    \includegraphics[width = 0.32\linewidth]{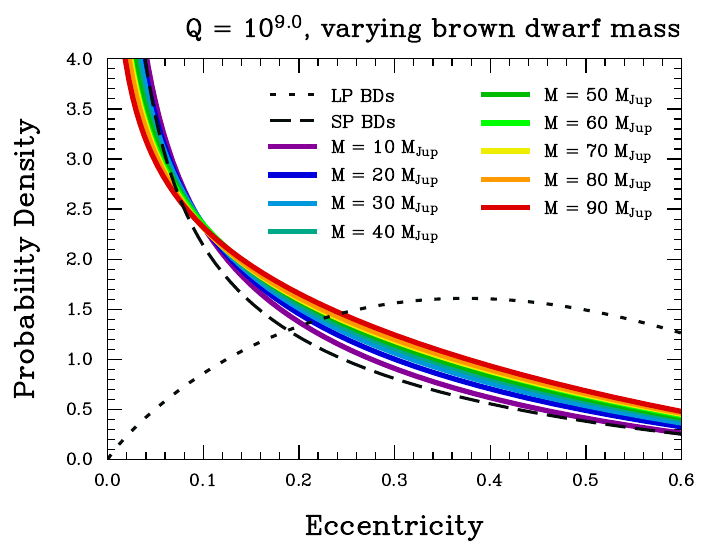}
    \includegraphics[width = 0.32\linewidth]{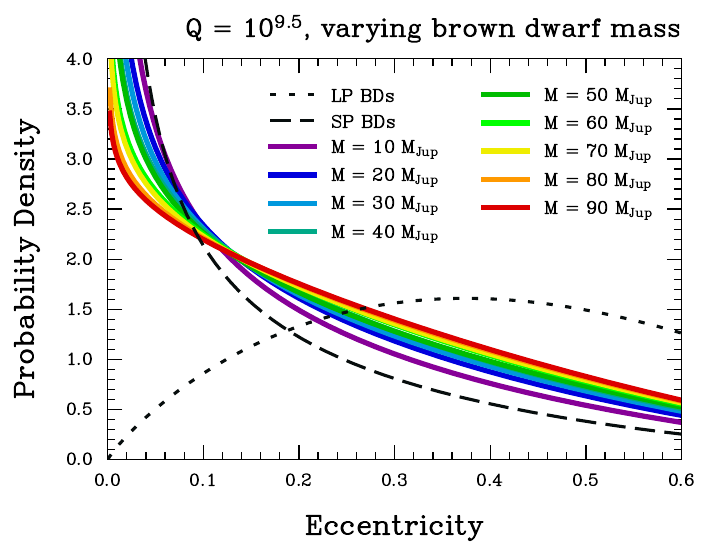}
    \includegraphics[width = 0.32\linewidth]{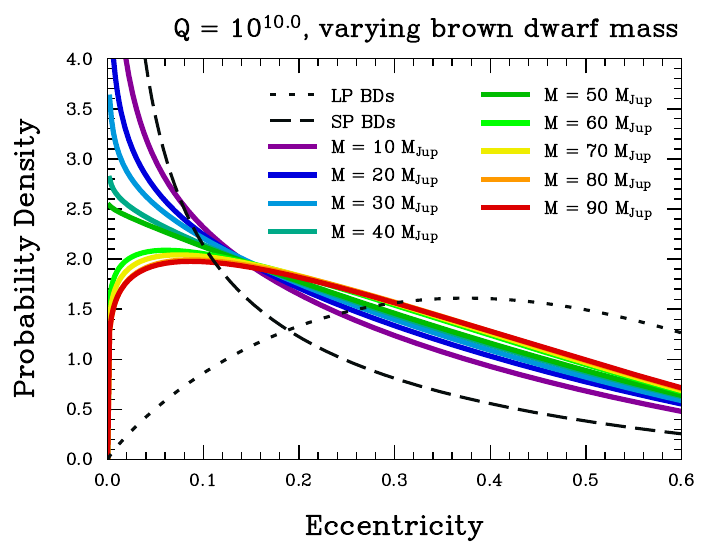}
    \caption{Eccentricity distributions for BDs as a function of mass for tidal quality factors of $\log_{10}(\mathcal{Q}) = 6-10$ with steps of $\delta\log_{10}(\mathcal{Q}) = 0.5$, after 6.5 Gyr of tidal evolution under the \citet{2008Icar..193..637W} framework (Model A; $\S$\ref{subsubsec:model1}). Each population begins, before tidal evolution, with an eccentricity distribution set to match the LP BD Beta distribution. The present-day observed LP and SP BD Beta distributions are indicated with dotted and dashed black lines, respectively. Increasing $\mathcal{Q}$ reduces the efficiency of tidal circularisation, leaving the population at relatively highly excited orbits.}
    \label{fig:tidalmodelQs}
\end{figure*}

\section{Discussion}\label{sec:discussion}

Measuring tidal quality factors is widely recognised as a challenging task. Relevant tidal mechanisms include equilibrium tides damped by turbulent convection \citep{1966AnAp...29..489Z,1977A&A....57..383Z,1977Icar...30..301G} and dynamical tides mediated by inertial waves, whose excitation is restricted to forcing frequencies $|\omega| \leq 2\Omega$ \citep{2007ApJ...661.1180O,2014ARA&A..52..171O}. The efficiency of these processes varies strongly with rotation rate, age, and thermal stratification, such that related uncertainties in the modelling and parameter dependence can naturally lead to effective tidal quality factors spanning several orders of magnitude \citep{2015A&A...580L...3M,2020MNRAS.498.2270B}. The inferred tidal quality factor $\mathcal{Q}_{\rm BD}$ presented in this work should, therefore, be understood as an effective, frequency-averaged parameter.

Previous studies have generally provided modest tidal constraints for BDs: for instance, \cite{2010A&A...514A..22H} derived a lower limit of $\log_{10}\mathcal{Q}_{\rm BD} \gtrsim 4.5$ for a BD in the eclipsing binary 2MASS J05352184-0546085, and \cite{2018AJ....156..168B} inferred $\log_{10}\mathcal{Q}_{\rm BD} \gtrsim 4.15$ for CWW 89A b, both based on tidal synchronisation arguments. In comparison, hot Jupiters (HJ) are typically inferred to have $\log_{10}\mathcal{Q}_{\rm HJ} \sim 5-6$, whilst estimates for Jupiter itself yield $\log_{10}\mathcal{Q}_{\rm J} \sim 6$, albeit with substantial uncertainty \citep{1979Natur.279..767Y, 1980LPI....11..871P}. Stellar tidal quality factors are expected to vary significantly across stellar types, with estimates spanning $\log_{10}\mathcal{Q}_{\star} \sim 6-9$ depending on mass, evolutionary stage, and tidal frequency (e.g., \citealt{2011ApJ...731...67P, 2023MNRAS.524.5575P, 2024AN....34530132I}). Indeed, for certain forcing regimes, extremely inefficient dissipation corresponding to $\log_{10}\mathcal{Q}_{\star} \rightarrow \infty$ cannot be formally excluded (e.g., $\S$2.2 in \citealt{2014ARA&A..52..171O}). 

Our inference of {{$\log_{10}\mathcal{Q}_{\rm BD} \approx 7-9$}---with $\mathcal{Q}_{\rm BD}$} approximately two orders of magnitude higher than typical values for gas giant planets---implies that tidal dissipation is substantially less efficient in BDs than in gas giant planets. Relative to hot Jupiters---which are rapidly sculpted by efficient tidal dissipation on Gyr timescales \citep{2009ApJ...692L...9L, 2010ApJ...725.1995M}---transiting BDs more effectively preserve orbital eccentricities and architectural signatures established during their formation and early dynamical evolution. Hence, commonly adopted Jupiter-like $\mathcal{Q}$ values may overestimate dissipation efficiency in the BD regime. 

The dissipative behaviour of BDs in our sample is instead more comparable to that of low-mass stars. The weak dissipation of BDs may arise from short convective turnover times relative to tidal forcing periods, which limits the efficiency of turbulent dissipation \citep{1977A&A....57..383Z, 2014ARA&A..52..171O}; high internal densities, which tend to modify the propagation and damping of inertial waves \citep{2003A&A...402..701B}; and/or the onset of partial electron degeneracy at higher masses, which stiffens the interior and reduces tidal deformation (though the quantitative impact on energy dissipation remains uncertain; see, e.g., \citealt{2000ApJ...542..464C, 2018AJ....156..149B}). 

{The structural properties of BDs also set the magnetic dynamo strength \citep{macdonald2009structural, 2010ApJ...713.1249M,2010A&A...522A..13R} and may therefore link tidal dissipation efficiency to other observable properties of brown dwarfs. Field brown dwarfs have been found with extremely rapid rotation \citep{kao2016auroral, kao2018strongest,tannock2021weather}, coupled with strong magnetic fields that may offer insight into their internal structures. However, these extreme rotators are not necessarily representative of our sample: wide-orbiting brown dwarfs bound to a stellar companion have been found to spin more slowly than field brown dwarfs \citep{hsu2026distinct}, and spin is regulated by the natal circumsubstellar disk properties alongside internal structure. Tidal studies have shown that frequency-averaged dissipation is surprisingly insensitive to magnetic field strength \citep{lin2018tidal,astoul2019does}, such that derived $Q$, magnetic field strength, and implications for brown dwarfs’ structural properties, while related, may not be directly translatable.} 

Recent observational studies further strengthen the picture presented in this work. For instance, examining the orbital eccentricities of 70 substellar companions spanning the planet-BD mass range ($0.8<M/M_{\rm Jup}<80$) and the orbital separation range $1< a/\mathrm{AU} < 10$, \citet{2026AJ....171...67G} suggest that the transition toward bottom-up (``planet-like'') formation processes may largely occur at masses below the lower limit of our sample ($M < 13~M_{\rm Jup}$). Leveraging an Epps-Singleton test, \citet{2026A&A...709A.130S} also concluded that the $P>16$ day transiting BD population follows an eccentricity distribution more comparable to that of low-mass stars than to that of giant planets.

Our analysis is limited, however, by sample size: {only 36 transiting BDs} are known, with fewer than a dozen LP systems, and higher-mass objects dominate the sample (see Figure \ref{fig:BDs_M_P_E}). Consequently, our population-level trends overall reflect the properties of relatively high-mass BDs. It is possible that the lowest-mass BDs may form through a separate avenue and thus may have distinct interior structures and correspondingly distinct $\mathcal{Q}$ from the higher-mass population. Dividing our sample into sub-samples at $42.5\, M_{\rm Jup}$ following \citet{2014MNRAS.439.2781M}, we find no significant difference in the eccentricity distribution of the low- vs. high-mass regimes. However, the absence of an observable difference may merely be a reflection of our limited sample size. 

\section{Conclusions}\label{sec:conclusions}

In this work, we present a population-level characterisation of the orbital eccentricities of transiting BDs, which we uniformly refit across a sample of 36 systems. Our conclusions are as follows:

\begin{itemize}

    \item Beta distribution fits indicate that the long-period ($P \geq 16$ days) transiting BD population is skewed toward higher eccentricities {($\mathcal{B}_{\rm LP~BDs}(1.879^{+1.088}_{-0.753},~2.470^{+1.480}_{-1.018})$), whilst the short-period BD population ($P < 16$ days) is skewed toward lower eccentricities ($\mathcal{B}_{\rm SP~BDs}(0.426^{+0.100}_{-0.084},~2.354^{+0.807}_{-0.658})$).} This is consistent with expectations in the case that tidal interactions have circularised the orbits of the shortest-period BDs (see Table \ref{tab:beta_summary}).

    \item As shown in Figure \ref{fig:ks_populations}, the eccentricity distribution for LP BDs is consistent with previously characterised eccentricity distributions for close stellar binaries. This supports a star-like formation scenario dominating our sample, which is largely comprised of relatively high-mass BDs $M_{\rm BD}>42.5M_{\rm Jup}$. 

    \item Assuming that short-period ($P<16$ day) BDs share a primordial eccentricity distribution with the $P \geq 16$ day long-period transiting BD population, {we constrain the typical tidal quality factor of transiting BDs to {$\mathcal{Q}_{\rm BD} = 10^{8.1\pm1.0}$} when neglecting the influence of tides raised on the host star, and $\mathcal{Q}_{\rm BD} = 10^{7.1\pm0.3}$ and $\mathcal{Q}_{\star} = 10^{6.0\pm0.1}$ when including this effect} (see Figure \ref{fig:tidalevolution_mass}). Our results are consistent {within 1$\sigma$} in both cases, and each is indicative of a star-like tidal $\mathcal{Q}$ value for the BD population.

\end{itemize} 

Our results suggest that BDs dissipate tidal energy less efficiently than gas giant planets and that they more closely resemble low-mass stars in this respect. Such weak tidal coupling allows BDs to act as dynamical fossils, retaining orbital characteristics set during formation and early epochs of evolution (see \cite{2026A&A...709A.130S} for similar findings regarding long-period BDs). Hence, even relatively short-period BDs may efficiently retain the orbital configurations imprinted from their formation histories. Forthcoming missions such as the PLAnetary Transits and Oscillations of stars (PLATO; \citealt{2025ExA....59...26R}) and the Roman Space Telescope \citep{2021AJ....161...84M, 2023arXiv230612363H} have the potential to substantially expand the census of transiting BDs, enabling more precise eccentricity mapping, mass-dependent analyses, and tighter constraints on the systems' dynamical origins.

\begin{acknowledgments} 
    {We thank the referee for helpful comments that strengthened this analysis. We also thank Sebastian Pineda, Dino Chih-Chun Hsu, Daniella Bardalez Gagliuffi, Joel Ong, Yaguang Li, Tiger Lu, and Jason Wang for helpful conversations that refined the interpretation presented in this work.} TF acknowledges support from the Yale Graduate School of Arts and Sciences. MR acknowledges support from Heising-Simons Foundation Grants \#2021-2802 and \#2023-4478. {This research has made use of the NASA Exoplanet Archive (\url{https://doi.org/10.26133/NEA12}), which is operated by the California Institute of Technology, under contract with the National Aeronautics and Space Administration under the Exoplanet Exploration Program.}
\end{acknowledgments}

\software{{\sc numpy} \citep{2020Natur.585..357H}, {\sc matplotlib} \citep{2024zndo..14249941T}, {\sc pandas} \citep{2022zndo...3509134T}, {\sc smplotlib} \citep{jiaxuan_li_2023_8126529}, {\sc astropy} \citep{2022ApJ...935..167A}, {\sc pyaneti} \citep{2019MNRAS.482.1017B}, {\sc statsmodels} \citep{2023zndo....593847P}, {\sc corner} \citep{2016JOSS....1...24F}, {\sc arviz} \citep{2019JOSS....4.1143K}, {\sc scipy} \citep{2020NatMe..17..261V}, {\sc seaborn} \citep{2020zndo...3767070W}, {\sc scikit\mbox{-}learn} \citep{2011JMLR...12.2825P}, {\sc pymc3} \citep{2016ascl.soft10016S}.}

\facilities{\emph{Gaia}, Keck I (HIRES), NOT (FIES), Harlan J. Smith (Tull), ESO 3.6m (HARPS), Euler 1.2m (CORALIE), Struve 2.1m (SANDIFORD), OHP 1.93m (SOPHIE), TLS 2m (Alfred Jensch), VLT (UVES, FLAMES, ESPRESSO), ESO 2.2m (FEROS), SMARTS 1.5m (CHIRON), KELT (North, South), Tillinghast 1.5m (TRES), NGTS, CFHT (SPIRou), ANU 2.3m, LCOGT (NRES), Mount Abu 1.2m (PARAS), Ondrejov 2m (Perek), Tautenburg 2m (Alfred Jensch), TNG 3.58m (HARPS-N).}

\bibliography{bib}

@ARTICLE{jeans1919origin,
       author = {{Jeans}, J.~H.},
        title = "{The origin of binary systems}",
      journal = {\mnras},
         year = 1919,
        month = apr,
       volume = {79},
        pages = {408},
          doi = {10.1093/mnras/79.6.408},
       adsurl = {https://ui.adsabs.harvard.edu/abs/1919MNRAS..79..408J}
}

@ARTICLE{dawson2018,
       author = {{Dawson}, Rebekah I. and {Johnson}, John Asher},
        title = "{Origins of Hot Jupiters}",
      journal = {\araa},
         year = 2018,
        month = sep,
       volume = {56},
        pages = {175-221},
          doi = {10.1146/annurev-astro-081817-051853},
archivePrefix = {arXiv},
       eprint = {1801.06117},
 primaryClass = {astro-ph.EP},
       adsurl = {https://ui.adsabs.harvard.edu/abs/2018ARA&A..56..175D}
}

@ARTICLE{2020AJ....159...63B,
       author = {{Bowler}, Brendan P. and {Blunt}, Sarah C. and {Nielsen}, Eric L.},
        title = "{Population-level Eccentricity Distributions of Imaged Exoplanets and Brown Dwarf Companions: Dynamical Evidence for Distinct Formation Channels}",
      journal = {\aj},
         year = 2020,
        month = feb,
       volume = {159},
       number = {2},
          eid = {63},
        pages = {63},
          doi = {10.3847/1538-3881/ab5b11},
archivePrefix = {arXiv},
       eprint = {1911.10569},
 primaryClass = {astro-ph.EP},
       adsurl = {https://ui.adsabs.harvard.edu/abs/2020AJ....159...63B}
}

@ARTICLE{2014MNRAS.439.2781M,
       author = {{Ma}, Bo and {Ge}, Jian},
        title = "{Statistical properties of brown dwarf companions: implications for different formation mechanisms}",
      journal = {\mnras},
         year = 2014,
        month = apr,
       volume = {439},
       number = {3},
        pages = {2781-2789},
          doi = {10.1093/mnras/stu134},
archivePrefix = {arXiv},
       eprint = {1303.6442},
 primaryClass = {astro-ph.EP},
       adsurl = {https://ui.adsabs.harvard.edu/abs/2014MNRAS.439.2781M}
}

@ARTICLE{rice2022b,
       author = {{Rice}, Malena and {Wang}, Songhu and {Wang}, Xian-Yu and {Stef{\'a}nsson}, Gu{\dj}mundur and {Isaacson}, Howard and {Howard}, Andrew W. and {Logsdon}, Sarah E. and {Schweiker}, Heidi and {Dai}, Fei and {Brinkman}, Casey and {Giacalone}, Steven and {Holcomb}, Rae},
        title = "{A Tendency Toward Alignment in Single-star Warm-Jupiter Systems}",
      journal = {\aj},
         year = 2022,
        month = sep,
       volume = {164},
       number = {3},
          eid = {104},
        pages = {104},
          doi = {10.3847/1538-3881/ac8153},
archivePrefix = {arXiv},
       eprint = {2207.06511},
 primaryClass = {astro-ph.EP},
       adsurl = {https://ui.adsabs.harvard.edu/abs/2022AJ....164..104R}
}

@ARTICLE{wang2024,
       author = {{Wang}, Xian-Yu and {Rice}, Malena and {Wang}, Songhu and {Kanodia}, Shubham and {Dai}, Fei and {Logsdon}, Sarah E. and {Schweiker}, Heidi and {Teske}, Johanna K. and {Butler}, R. Paul and {Crane}, Jeffrey D. and {Shectman}, Stephen and {Quinn}, Samuel N. and {Kostov}, Veselin and {Osborn}, Hugh P. and {Goeke}, Robert F. and {Eastman}, Jason D. and {Shporer}, Avi and {Rapetti}, David and {Collins}, Karen A. and {Watkins}, Cristilyn N. and {Relles}, Howard M. and {Ricker}, George R. and {Seager}, Sara and {Winn}, Joshua N. and {Jenkins}, Jon M.},
        title = "{Single-star Warm-Jupiter Systems Tend to Be Aligned, Even around Hot Stellar Hosts: No T $_{eff}${\textendash}{\ensuremath{\lambda}} Dependency}",
      journal = {\apjl},
         year = 2024,
        month = sep,
       volume = {973},
       number = {1},
          eid = {L21},
        pages = {L21},
          doi = {10.3847/2041-8213/ad7469},
archivePrefix = {arXiv},
       eprint = {2408.10038},
 primaryClass = {astro-ph.EP},
       adsurl = {https://ui.adsabs.harvard.edu/abs/2024ApJ...973L..21W}
}

@ARTICLE{blunt2026evidence,
       author = {{Blunt}, Sarah and {Wang}, Jason and {Murray-Clay}, Ruth and {Macintosh}, Bruce and {Rubenzahl}, Ryan A. and {Fulton}, B.~J.},
        title = "{Evidence for a Peak at $\sim$0.3 in the Eccentricity Distribution of Typical Super-Jovian Exoplanets}",
      journal = {arXiv e-prints},
         year = 2026,
        month = jan,
          eid = {arXiv:2601.18877},
        pages = {arXiv:2601.18877},
          doi = {10.48550/arXiv.2601.18877},
archivePrefix = {arXiv},
       eprint = {2601.18877},
 primaryClass = {astro-ph.EP},
       adsurl = {https://ui.adsabs.harvard.edu/abs/2026arXiv260118877B}
}

@ARTICLE{rice2023TOI2202,
       author = {{Rice}, Malena and {Wang}, Xian-Yu and {Wang}, Songhu and {Shporer}, Avi and {Barkaoui}, Khalid and {Brahm}, Rafael and {Collins}, Karen A. and {Jord{\'a}n}, Andr{\'e}s and {Lowson}, Nataliea and {Butler}, R. Paul and {Crane}, Jeffrey D. and {Shectman}, Stephen and {Teske}, Johanna K. and {Osip}, David and {Collins}, Kevin I. and {Murgas}, Felipe and {Boyle}, Gavin and {Pozuelos}, Francisco J. and {Timmermans}, Mathilde and {Jehin}, Emmanuel and {Gillon}, Micha{\"e}l},
        title = "{Evidence for Low-level Dynamical Excitation in Near-resonant Exoplanet Systems}",
      journal = {\aj},
         year = 2023,
        month = dec,
       volume = {166},
       number = {6},
          eid = {266},
        pages = {266},
          doi = {10.3847/1538-3881/ad09de},
archivePrefix = {arXiv},
       eprint = {2311.02478},
 primaryClass = {astro-ph.EP},
       adsurl = {https://ui.adsabs.harvard.edu/abs/2023AJ....166..266R}
}

@ARTICLE{radzom2024,
       author = {{Radzom}, Brandon T. and {Dong}, Jiayin and {Rice}, Malena and {Wang}, Xian-Yu and {Yee}, Samuel W. and {Fairnington}, Tyler R. and {Petrovich}, Cristobal and {Wang}, Songhu},
        title = "{Evidence for Primordial Alignment: Insights from Stellar Obliquity Measurements for Compact Sub-Saturn Systems}",
      journal = {\aj},
         year = 2024,
        month = sep,
       volume = {168},
       number = {3},
          eid = {116},
        pages = {116},
          doi = {10.3847/1538-3881/ad61d8},
archivePrefix = {arXiv},
       eprint = {2404.06504},
 primaryClass = {astro-ph.EP},
       adsurl = {https://ui.adsabs.harvard.edu/abs/2024AJ....168..116R}
}

@ARTICLE{2025Natur.644..356B,
       author = {{Biddle}, Lauren I. and {Bowler}, Brendan P. and {Morgan}, Marvin and {Tran}, Quang H. and {Wu}, Ya-Lin},
        title = "{One-third of Sun-like stars are born with misaligned planet-forming disks}",
      journal = {\nat},
         year = 2025,
        month = aug,
       volume = {644},
       number = {8076},
        pages = {356-361},
          doi = {10.1038/s41586-025-09324-0},
archivePrefix = {arXiv},
       eprint = {2508.06488},
 primaryClass = {astro-ph.EP},
       adsurl = {https://ui.adsabs.harvard.edu/abs/2025Natur.644..356B}
}

@ARTICLE{2019MEEP....7....1I,
       author = {{Ito}, Takashi and {Ohtsuka}, Katsuhito},
        title = "{The Lidov-Kozai Oscillation and Hugo von Zeipel}",
      journal = {Monographs on Environment, Earth and Planets},
         year = 2019,
        month = nov,
       volume = {7},
       number = {1},
        pages = {1-113},
          doi = {10.5047/meep.2019.00701.0001},
archivePrefix = {arXiv},
       eprint = {1911.03984},
 primaryClass = {astro-ph.EP},
       adsurl = {https://ui.adsabs.harvard.edu/abs/2019MEEP....7....1I}
}

@ARTICLE{1962AJ.....67..591K,
       author = {{Kozai}, Yoshihide},
        title = "{Secular perturbations of asteroids with high inclination and eccentricity}",
      journal = {\aj},
         year = 1962,
        month = nov,
       volume = {67},
        pages = {591-598},
          doi = {10.1086/108790},
       adsurl = {https://ui.adsabs.harvard.edu/abs/1962AJ.....67..591K}
}

@ARTICLE{1962P&SS....9..719L,
       author = {{Lidov}, M.~L.},
        title = "{The evolution of orbits of artificial satellites of planets under the action of gravitational perturbations of external bodies}",
      journal = {\planss},
         year = 1962,
        month = oct,
       volume = {9},
       number = {10},
        pages = {719-759},
          doi = {10.1016/0032-0633(62)90129-0},
       adsurl = {https://ui.adsabs.harvard.edu/abs/1962P&SS....9..719L}
}

@ARTICLE{2019AJ....157...61V,
       author = {{Van Eylen}, Vincent and {Albrecht}, Simon and {Huang}, Xu and {MacDonald}, Mariah G. and {Dawson}, Rebekah I. and {Cai}, Maxwell X. and {Foreman-Mackey}, Daniel and {Lundkvist}, Mia S. and {Silva Aguirre}, Victor and {Snellen}, Ignas and {Winn}, Joshua N.},
        title = "{The Orbital Eccentricity of Small Planet Systems}",
      journal = {\aj},
         year = 2019,
        month = feb,
       volume = {157},
       number = {2},
          eid = {61},
        pages = {61},
          doi = {10.3847/1538-3881/aaf22f},
archivePrefix = {arXiv},
       eprint = {1807.00549},
 primaryClass = {astro-ph.EP},
       adsurl = {https://ui.adsabs.harvard.edu/abs/2019AJ....157...61V}
}

@INPROCEEDINGS{2008ASPC..398..235M,
       author = {{Mordasini}, C. and {Alibert}, Y. and {Benz}, W. and {Naef}, D.},
        title = "{Giant Planet Formation by Core Accretion}",
    booktitle = {Extreme Solar Systems},
         year = 2008,
       editor = {{Fischer}, D. and {Rasio}, F.~A. and {Thorsett}, S.~E. and {Wolszczan}, A.},
       series = {Astronomical Society of the Pacific Conference Series},
       volume = {398},
        month = jan,
        pages = {235},
          doi = {10.48550/arXiv.0710.5667},
archivePrefix = {arXiv},
       eprint = {0710.5667},
 primaryClass = {astro-ph},
       adsurl = {https://ui.adsabs.harvard.edu/abs/2008ASPC..398..235M}
}

@ARTICLE{2008A&A...482..315B,
       author = {{Baraffe}, I. and {Chabrier}, G. and {Barman}, T.},
        title = "{Structure and evolution of super-Earth to super-Jupiter exoplanets. I. Heavy element enrichment in the interior}",
      journal = {\aap},
         year = 2008,
        month = apr,
       volume = {482},
       number = {1},
        pages = {315-332},
          doi = {10.1051/0004-6361:20079321},
archivePrefix = {arXiv},
       eprint = {0802.1810},
 primaryClass = {astro-ph},
       adsurl = {https://ui.adsabs.harvard.edu/abs/2008A&A...482..315B}
}

@ARTICLE{2013ApJ...770..120B,
       author = {{Bodenheimer}, Peter and {D'Angelo}, Gennaro and {Lissauer}, Jack J. and {Fortney}, Jonathan J. and {Saumon}, Didier},
        title = "{Deuterium Burning in Massive Giant Planets and Low-mass Brown Dwarfs Formed by Core-nucleated Accretion}",
      journal = {\apj},
         year = 2013,
        month = jun,
       volume = {770},
       number = {2},
          eid = {120},
        pages = {120},
          doi = {10.1088/0004-637X/770/2/120},
archivePrefix = {arXiv},
       eprint = {1305.0980},
 primaryClass = {astro-ph.EP},
       adsurl = {https://ui.adsabs.harvard.edu/abs/2013ApJ...770..120B}
}

@ARTICLE{2002ARA&A..40..349T,
       author = {{Tohline}, Joel E.},
        title = "{The Origin of Binary Stars}",
      journal = {\araa},
         year = 2002,
        month = jan,
       volume = {40},
        pages = {349-385},
          doi = {10.1146/annurev.astro.40.060401.093810},
       adsurl = {https://ui.adsabs.harvard.edu/abs/2002ARA&A..40..349T}
}

@ARTICLE{2011arXiv1111.4246H,
       author = {{Hoffman}, Matthew D. and {Gelman}, Andrew},
        title = "{The No-U-Turn Sampler: Adaptively Setting Path Lengths in Hamiltonian Monte Carlo}",
      journal = {arXiv e-prints},
         year = 2011,
        month = nov,
          eid = {arXiv:1111.4246},
        pages = {arXiv:1111.4246},
          doi = {10.48550/arXiv.1111.4246},
archivePrefix = {arXiv},
       eprint = {1111.4246},
 primaryClass = {stat.CO},
       adsurl = {https://ui.adsabs.harvard.edu/abs/2011arXiv1111.4246H}
}

@software{2016ascl.soft10016S,
       author = {{Salvatier}, John and {Wiecki{\^a}}, Thomas V. and {Fonnesbeck}, Christopher},
        title = "{PyMC3: Python probabilistic programming framework}",
 howpublished = {Astrophysics Source Code Library, record ascl:1610.016},
         year = 2016,
        month = oct,
          eid = {ascl:1610.016},
       adsurl = {https://ui.adsabs.harvard.edu/abs/2016ascl.soft10016S}
}

@ARTICLE{1992StaSc...7..457G,
       author = {{Gelman}, Andrew and {Rubin}, Donald B.},
        title = "{Inference from Iterative Simulation Using Multiple Sequences}",
      journal = {Statistical Science},
         year = 1992,
        month = jan,
       volume = {7},
        pages = {457-472},
          doi = {10.1214/ss/1177011136},
       adsurl = {https://ui.adsabs.harvard.edu/abs/1992StaSc...7..457G}
}

@ARTICLE{2010ApJ...725.2166H,
       author = {{Hogg}, David W. and {Myers}, Adam D. and {Bovy}, Jo},
        title = "{Inferring the Eccentricity Distribution}",
      journal = {\apj},
         year = 2010,
        month = dec,
       volume = {725},
       number = {2},
        pages = {2166-2175},
          doi = {10.1088/0004-637X/725/2/2166},
archivePrefix = {arXiv},
       eprint = {1008.4146},
 primaryClass = {astro-ph.SR},
       adsurl = {https://ui.adsabs.harvard.edu/abs/2010ApJ...725.2166H}
}

@ARTICLE{2013MNRAS.434L..51K,
       author = {{Kipping}, D.~M.},
        title = "{Parametrizing the exoplanet eccentricity distribution with the beta  distribution.}",
      journal = {\mnras},
         year = 2013,
        month = jul,
       volume = {434},
        pages = {L51-L55},
          doi = {10.1093/mnrasl/slt075},
archivePrefix = {arXiv},
       eprint = {1306.4982},
 primaryClass = {astro-ph.EP},
       adsurl = {https://ui.adsabs.harvard.edu/abs/2013MNRAS.434L..51K}
}

@ARTICLE{2023AJ....166...48D,
       author = {{Do {\'O}}, Clarissa R. and {O'Neil}, Kelly K. and {Konopacky}, Quinn M. and {Do}, Tuan and {Martinez}, Gregory D. and {Ruffio}, Jean-Baptiste and {Ghez}, Andrea M.},
        title = "{The Orbital Eccentricities of Directly Imaged Companions Using Observable-based Priors: Implications for Population-level Distributions}",
      journal = {\aj},
         year = 2023,
        month = aug,
       volume = {166},
       number = {2},
          eid = {48},
        pages = {48},
          doi = {10.3847/1538-3881/acdc9a},
archivePrefix = {arXiv},
       eprint = {2306.04080},
 primaryClass = {astro-ph.EP},
       adsurl = {https://ui.adsabs.harvard.edu/abs/2023AJ....166...48D}
}

@ARTICLE{1996Sci...274..954R,
       author = {{Rasio}, Frederic A. and {Ford}, Eric B.},
        title = "{Dynamical instabilities and the formation of extrasolar planetary systems}",
      journal = {Science},
         year = 1996,
        month = nov,
       volume = {274},
        pages = {954-956},
          doi = {10.1126/science.274.5289.954},
       adsurl = {https://ui.adsabs.harvard.edu/abs/1996Sci...274..954R}
}

@ARTICLE{2009ApJ...693L.113S,
       author = {{Scharf}, Caleb and {Menou}, Kristen},
        title = "{Long-Period Exoplanets From Dynamical Relaxation}",
      journal = {\apjl},
         year = 2009,
        month = mar,
       volume = {693},
       number = {2},
        pages = {L113-L117},
          doi = {10.1088/0004-637X/693/2/L113},
archivePrefix = {arXiv},
       eprint = {0811.1981},
 primaryClass = {astro-ph},
       adsurl = {https://ui.adsabs.harvard.edu/abs/2009ApJ...693L.113S}
}

@ARTICLE{1937AZh....14..207A,
       author = {{Ambartsumian}, V.~A.},
        title = "{On the Statistics of Double Stars}",
      journal = {\azh},
         year = 1937,
        month = jan,
       volume = {14},
        pages = {207-219},
       adsurl = {https://ui.adsabs.harvard.edu/abs/1937AZh....14..207A}
}

@ARTICLE{2003A&A...402..701B,
       author = {{Baraffe}, I. and {Chabrier}, G. and {Barman}, T.~S. and {Allard}, F. and {Hauschildt}, P.~H.},
        title = "{Evolutionary models for cool brown dwarfs and extrasolar giant planets. The case of HD 209458}",
      journal = {\aap},
         year = 2003,
        month = may,
       volume = {402},
        pages = {701-712},
          doi = {10.1051/0004-6361:20030252},
archivePrefix = {arXiv},
       eprint = {astro-ph/0302293},
 primaryClass = {astro-ph},
       adsurl = {https://ui.adsabs.harvard.edu/abs/2003A&A...402..701B}
}

@ARTICLE{2023MNRAS.519.5177C,
       author = {{Carmichael}, Theron W.},
        title = "{Improved radius determinations for the transiting brown dwarf population in the era of Gaia and TESS}",
      journal = {\mnras},
         year = 2023,
        month = mar,
       volume = {519},
       number = {4},
        pages = {5177-5190},
          doi = {10.1093/mnras/stac3720},
archivePrefix = {arXiv},
       eprint = {2212.02502},
 primaryClass = {astro-ph.SR},
       adsurl = {https://ui.adsabs.harvard.edu/abs/2023MNRAS.519.5177C}
}

@ARTICLE{2023AJ....166..112D,
       author = {{Dong}, Jiayin and {Foreman-Mackey}, Daniel},
        title = "{A Hierarchical Bayesian Framework for Inferring the Stellar Obliquity Distribution}",
      journal = {\aj},
         year = 2023,
        month = sep,
       volume = {166},
       number = {3},
          eid = {112},
        pages = {112},
          doi = {10.3847/1538-3881/ace105},
archivePrefix = {arXiv},
       eprint = {2305.14220},
 primaryClass = {astro-ph.EP},
       adsurl = {https://ui.adsabs.harvard.edu/abs/2023AJ....166..112D}
}

@BOOK{1980tisp.book.....G,
       author = {{Gradshteyn}, I.~S. and {Ryzhik}, I.~M.},
        title = "{Table of integrals, series and products}",
         year = 1980,
       adsurl = {https://ui.adsabs.harvard.edu/abs/1980tisp.book.....G}
}

@ARTICLE{2023AJ....166..225S,
       author = {{Schmidt}, Stephen P. and {Schlaufman}, Kevin C. and {Ding}, Keyi and {Grunblatt}, Samuel K. and {Carmichael}, Theron and {Bieryla}, Allyson and {Rodriguez}, Joseph E. and {Schulte}, Jack and {Vowell}, Noah and {Zhou}, George and {Quinn}, Samuel N. and {Yee}, Samuel W. and {Winn}, Joshua N. and {Hartman}, Joel D. and {Latham}, David W. and {Caldwell}, Douglas A. and {Fausnaugh}, M.~M. and {Hedges}, Christina and {Jenkins}, Jon M. and {Osborn}, Hugh P. and {Seager}, S.},
        title = "{Verification of Gaia Data Release 3 Single-lined Spectroscopic Binary Solutions With Three Transiting Low-mass Secondaries}",
      journal = {\aj},
         year = 2023,
        month = dec,
       volume = {166},
       number = {6},
          eid = {225},
        pages = {225},
          doi = {10.3847/1538-3881/ad0135},
archivePrefix = {arXiv},
       eprint = {2310.07936},
 primaryClass = {astro-ph.SR},
       adsurl = {https://ui.adsabs.harvard.edu/abs/2023AJ....166..225S}
}

@article{smirnov1939estimate,
  title={Estimate of deviation between empirical distribution functions in two independent samples},
  author={Smirnov, Nikolai V},
  journal={Bulletin Moscow University},
  volume={2},
  number={2},
  pages={3--16},
  year={1939}
}

@article{an1933sulla,
  title={Sulla determinazione empirica di una legge didistribuzione},
  author={Kolmogorov, Andrei},
  journal={Giorn Dell'inst Ital Degli Att},
  volume={4},
  pages={89--91},
  year={1933}
}

@misc{jiaxuan_li_2023_8126529,
  author       = {Jiaxuan Li},
  title        = {AstroJacobLi/smplotlib: v0.0.9},
  month        = jul,
  year         = 2023,
  publisher    = {Zenodo},
  version      = {v0.0.9},
  doi          = {10.5281/zenodo.8126529},
  url          = {https://doi.org/10.5281/zenodo.8126529}
}

@software{2022zndo...3509134T,
       author = {{The pandas development Team}},
        title = "{pandas-dev/pandas: Pandas}",
         year = 2024,
        month = apr,
          eid = {10.5281/zenodo.3509134},
          doi = {10.5281/zenodo.3509134},
      version = {v2.2.2},
    publisher = {Zenodo},
       adsurl = {https://ui.adsabs.harvard.edu/abs/2022zndo...3509134T}
}

@ARTICLE{2020NatMe..17..261V,
       author = {{Virtanen}, Pauli and {Gommers}, Ralf and {Oliphant}, Travis E. and {Haberland}, Matt and {Reddy}, Tyler and {Cournapeau}, David and {Burovski}, Evgeni and {Peterson}, Pearu and {Weckesser}, Warren and {Bright}, Jonathan and {van der Walt}, St{\'e}fan J. and {Brett}, Matthew and {Wilson}, Joshua and {Millman}, K. Jarrod and {Mayorov}, Nikolay and {Nelson}, Andrew R.~J. and {Jones}, Eric and {Kern}, Robert and {Larson}, Eric and {Carey}, C.~J. and {Polat}, {\.I}lhan and {Feng}, Yu and {Moore}, Eric W. and {VanderPlas}, Jake and {Laxalde}, Denis and {Perktold}, Josef and {Cimrman}, Robert and {Henriksen}, Ian and {Quintero}, E.~A. and {Harris}, Charles R. and {Archibald}, Anne M. and {Ribeiro}, Ant{\^o}nio H. and {Pedregosa}, Fabian and {van Mulbregt}, Paul and {SciPy 1. 0 Contributors}},
        title = "{SciPy 1.0: fundamental algorithms for scientific computing in Python}",
      journal = {Nature Methods},
         year = 2020,
        month = feb,
       volume = {17},
        pages = {261-272},
          doi = {10.1038/s41592-019-0686-2},
archivePrefix = {arXiv},
       eprint = {1907.10121},
 primaryClass = {cs.MS},
       adsurl = {https://ui.adsabs.harvard.edu/abs/2020NatMe..17..261V}
}

@ARTICLE{1969MNRAS.145..271L,
       author = {{Larson}, Richard B.},
        title = "{Numerical calculations of the dynamics of collapsing proto-star}",
      journal = {\mnras},
         year = 1969,
        month = jan,
       volume = {145},
        pages = {271},
          doi = {10.1093/mnras/145.3.271},
       adsurl = {https://ui.adsabs.harvard.edu/abs/1969MNRAS.145..271L}
}

@ARTICLE{2010CAMCS...5...65G,
       author = {{Goodman}, Jonathan and {Weare}, Jonathan},
        title = "{Ensemble samplers with affine invariance}",
      journal = {Communications in Applied Mathematics and Computational Science},
         year = 2010,
        month = jan,
       volume = {5},
       number = {1},
        pages = {65-80},
          doi = {10.2140/camcos.2010.5.65},
       adsurl = {https://ui.adsabs.harvard.edu/abs/2010CAMCS...5...65G}
}

@ARTICLE{2019MNRAS.482.1017B,
       author = {{Barrag{\'a}n}, O. and {Gandolfi}, D. and {Antoniciello}, G.},
        title = "{PYANETI: a fast and powerful software suite for multiplanet radial velocity and transit fitting}",
      journal = {\mnras},
         year = 2019,
        month = jan,
       volume = {482},
       number = {1},
        pages = {1017-1030},
          doi = {10.1093/mnras/sty2472},
archivePrefix = {arXiv},
       eprint = {1809.04609},
 primaryClass = {astro-ph.EP},
       adsurl = {https://ui.adsabs.harvard.edu/abs/2019MNRAS.482.1017B}
}

@ARTICLE{1971AJ.....76..544L,
       author = {{Lucy}, L.~B. and {Sweeney}, M.~A.},
        title = "{Spectroscopic binaries with circular orbits.}",
      journal = {\aj},
         year = 1971,
        month = aug,
       volume = {76},
        pages = {544-556},
          doi = {10.1086/111159},
       adsurl = {https://ui.adsabs.harvard.edu/abs/1971AJ.....76..544L}
}

@ARTICLE{2010A&A...514A..22H,
       author = {{Heller}, R. and {Jackson}, B. and {Barnes}, R. and {Greenberg}, R. and {Homeier}, D.},
        title = "{Tidal effects on brown dwarfs: application to the eclipsing binary 2MASS J05352184-0546085. The anomalous temperature reversal in the context of tidal heating}",
      journal = {\aap},
         year = 2010,
        month = may,
       volume = {514},
          eid = {A22},
        pages = {A22},
          doi = {10.1051/0004-6361/200912826},
archivePrefix = {arXiv},
       eprint = {1002.1246},
 primaryClass = {astro-ph.SR},
       adsurl = {https://ui.adsabs.harvard.edu/abs/2010A&A...514A..22H}
}

@ARTICLE{1997Sci...276.1836B,
       author = {{Boss}, A.~P.},
        title = "{Giant planet formation by gravitational instability.}",
      journal = {Science},
         year = 1997,
        month = jan,
       volume = {276},
        pages = {1836-1839},
          doi = {10.1126/science.276.5320.1836},
       adsurl = {https://ui.adsabs.harvard.edu/abs/1997Sci...276.1836B}
}

@ARTICLE{2008Icar..193..637W,
       author = {{Wisdom}, Jack},
        title = "{Tidal dissipation at arbitrary eccentricity and obliquity}",
      journal = {\icarus},
         year = 2008,
        month = feb,
       volume = {193},
       number = {2},
        pages = {637-640},
          doi = {10.1016/j.icarus.2007.09.002},
       adsurl = {https://ui.adsabs.harvard.edu/abs/2008Icar..193..637W}
}

@ARTICLE{2023AJ....165...32N,
       author = {{Nagpal}, Vighnesh and {Blunt}, Sarah and {Bowler}, Brendan P. and {Dupuy}, Trent J. and {Nielsen}, Eric L. and {Wang}, Jason J.},
        title = "{The Impact of Bayesian Hyperpriors on the Population-level Eccentricity Distribution of Imaged Planets}",
      journal = {\aj},
         year = 2023,
        month = feb,
       volume = {165},
       number = {2},
          eid = {32},
        pages = {32},
          doi = {10.3847/1538-3881/ac9fd2},
archivePrefix = {arXiv},
       eprint = {2211.02121},
 primaryClass = {astro-ph.EP},
       adsurl = {https://ui.adsabs.harvard.edu/abs/2023AJ....165...32N}
}

@ARTICLE{2008MNRAS.389.1556B,
       author = {{Bonnell}, Ian A. and {Clark}, Paul and {Bate}, Matthew R.},
        title = "{Gravitational fragmentation and the formation of brown dwarfs in stellar clusters}",
      journal = {\mnras},
         year = 2008,
        month = oct,
       volume = {389},
       number = {4},
        pages = {1556-1562},
          doi = {10.1111/j.1365-2966.2008.13679.x},
archivePrefix = {arXiv},
       eprint = {0807.0460},
 primaryClass = {astro-ph},
       adsurl = {https://ui.adsabs.harvard.edu/abs/2008MNRAS.389.1556B}
}

@ARTICLE{2016ARA&A..54..271K,
       author = {{Kratter}, Kaitlin and {Lodato}, Giuseppe},
        title = "{Gravitational Instabilities in Circumstellar Disks}",
      journal = {\araa},
         year = 2016,
        month = sep,
       volume = {54},
        pages = {271-311},
          doi = {10.1146/annurev-astro-081915-023307},
archivePrefix = {arXiv},
       eprint = {1603.01280},
 primaryClass = {astro-ph.SR},
       adsurl = {https://ui.adsabs.harvard.edu/abs/2016ARA&A..54..271K}
}

@ARTICLE{2009MNRAS.392..413S,
       author = {{Stamatellos}, Dimitris and {Whitworth}, Anthony P.},
        title = "{The properties of brown dwarfs and low-mass hydrogen-burning stars formed by disc fragmentation}",
      journal = {\mnras},
         year = 2009,
        month = jan,
       volume = {392},
       number = {1},
        pages = {413-427},
          doi = {10.1111/j.1365-2966.2008.14069.x},
archivePrefix = {arXiv},
       eprint = {0810.1687},
 primaryClass = {astro-ph},
       adsurl = {https://ui.adsabs.harvard.edu/abs/2009MNRAS.392..413S}
}

@ARTICLE{2024MNRAS.533.2823H,
       author = {{Henderson}, Beth A. and {Casewell}, Sarah L. and {Jord{\'a}n}, Andr{\'e}s and {Brahm}, Rafael and {Henning}, Thomas and {Gill}, Samuel and {Mayorga}, L.~C. and {Ziegler}, Carl and {Stassun}, Keivan G. and {Goad}, Michael R. and {Acton}, Jack and {Alves}, Douglas R. and {Anderson}, David R. and {Apergis}, Ioannis and {Armstrong}, David J. and {Bayliss}, Daniel and {Burleigh}, Matthew R. and {Dragomir}, Diana and {Gillen}, Edward and {G{\"u}nther}, Maximilian N. and {Hedges}, Christina and {Hesse}, Katharine M. and {Hobson}, Melissa J. and {Jenkins}, James S. and {Jenkins}, Jon M. and {Kendall}, Alicia and {Lendl}, Monika and {Lund}, Michael B. and {McCormac}, James and {Moyano}, Maximiliano and {Osborn}, Ares and {Pinto}, Marcelo Tala and {Ramsay}, Gavin and {Rapetti}, David and {Saha}, Suman and {Seager}, Sara and {Trifonov}, Trifon and {Udry}, St{\'e}phane and {Vines}, Jose I. and {West}, Richard G. and {Wheatley}, Peter J. and {Winn}, Joshua N. and {Zivave}, Tafadzwa},
        title = "{TOI-2490b - the most eccentric brown dwarf transiting in the brown dwarf desert}",
      journal = {\mnras},
         year = 2024,
        month = sep,
       volume = {533},
       number = {3},
        pages = {2823-2842},
          doi = {10.1093/mnras/stae1940},
archivePrefix = {arXiv},
       eprint = {2408.04475},
 primaryClass = {astro-ph.EP},
       adsurl = {https://ui.adsabs.harvard.edu/abs/2024MNRAS.533.2823H}
}

@ARTICLE{2024MNRAS.530..318H,
       author = {{Henderson}, Beth A. and {Casewell}, Sarah L. and {Goad}, Michael R. and {Acton}, Jack S. and {G{\"u}nther}, Maximilian N. and {Nielsen}, Louise D. and {Burleigh}, Matthew R. and {Belardi}, Claudia and {Tilbrook}, Rosanna H. and {Turner}, Oliver and {Howell}, Steve B. and {Clark}, Catherine A. and {Littlefield}, Colin and {Barkaoui}, Khalid and {Alves}, Douglas R. and {Anderson}, David R. and {Bayliss}, Daniel and {Bouchy}, Francois and {Bryant}, Edward M. and {Dransfield}, George and {Ducrot}, Elsa and {Eigm{\"u}ller}, Philipp and {Gill}, Samuel and {Gillen}, Edward and {Gillon}, Micha{\"e}l and {Hawthorn}, Faith and {Hooton}, Matthew J. and {Jackman}, James A.~G. and {Jehin}, Emmanuel and {Jenkins}, James S. and {Kendall}, Alicia and {Lendl}, Monika and {McCormac}, James and {Moyano}, Maximiliano and {Pedersen}, Peter Pihlmann and {Pozuelos}, Francisco J. and {Ramsay}, Gavin and {Sefako}, Ramotholo R. and {Timmermans}, Mathilde and {Triaud}, Amaury H.~M.~J. and {Udry}, Stephane and {Vines}, Jose I. and {Watson}, Christopher A. and {West}, Richard G. and {Wheatley}, Peter J. and {Z{\'u}{\~n}iga-Fern{\'a}ndez}, Sebasti{\'a}n},
        title = "{NGTS-28Ab: a short period transiting brown dwarf}",
      journal = {\mnras},
         year = 2024,
        month = may,
       volume = {530},
       number = {1},
        pages = {318-339},
          doi = {10.1093/mnras/stae508},
archivePrefix = {arXiv},
       eprint = {2402.09943},
 primaryClass = {astro-ph.EP},
       adsurl = {https://ui.adsabs.harvard.edu/abs/2024MNRAS.530..318H}
}

@ARTICLE{2025ApJ...982L..34W,
       author = {{Wu}, Yanqin and {Hadden}, Sam and {Dewberry}, Janosz and {El-Badry}, Kareem and {Matzner}, Christopher D.},
        title = "{Eccentricities of Close Stellar Binaries}",
      journal = {\apjl},
         year = 2025,
        month = mar,
       volume = {982},
       number = {1},
          eid = {L34},
        pages = {L34},
          doi = {10.3847/2041-8213/adb751},
archivePrefix = {arXiv},
       eprint = {2411.09905},
 primaryClass = {astro-ph.SR},
       adsurl = {https://ui.adsabs.harvard.edu/abs/2025ApJ...982L..34W}
}

@ARTICLE{2021ApJS..255....8R,
       author = {{Rosenthal}, Lee J. and {Fulton}, Benjamin J. and {Hirsch}, Lea A. and {Isaacson}, Howard T. and {Howard}, Andrew W. and {Dedrick}, Cayla M. and {Sherstyuk}, Ilya A. and {Blunt}, Sarah C. and {Petigura}, Erik A. and {Knutson}, Heather A. and {Behmard}, Aida and {Chontos}, Ashley and {Crepp}, Justin R. and {Crossfield}, Ian J.~M. and {Dalba}, Paul A. and {Fischer}, Debra A. and {Henry}, Gregory W. and {Kane}, Stephen R. and {Kosiarek}, Molly and {Marcy}, Geoffrey W. and {Rubenzahl}, Ryan A. and {Weiss}, Lauren M. and {Wright}, Jason T.},
        title = "{The California Legacy Survey. I. A Catalog of 178 Planets from Precision Radial Velocity Monitoring of 719 Nearby Stars over Three Decades}",
      journal = {\apjs},
         year = 2021,
        month = jul,
       volume = {255},
       number = {1},
          eid = {8},
        pages = {8},
          doi = {10.3847/1538-4365/abe23c},
archivePrefix = {arXiv},
       eprint = {2105.11583},
 primaryClass = {astro-ph.EP},
       adsurl = {https://ui.adsabs.harvard.edu/abs/2021ApJS..255....8R}
}

@software{2024zndo..14249941T,
       author = {{The Matplotlib Development Team}},
        title = "{Matplotlib: Visualization with Python}",
         year = 2024,
        month = nov,
          eid = {10.5281/zenodo.14249941},
          doi = {10.5281/zenodo.14249941},
      version = {v3.9.3},
    publisher = {Zenodo},
       adsurl = {https://ui.adsabs.harvard.edu/abs/2024zndo..14249941T}
}

@ARTICLE{2020Natur.585..357H,
       author = {{Harris}, Charles R. and {Millman}, K. Jarrod and {van der Walt}, St{\'e}fan J. and {Gommers}, Ralf and {Virtanen}, Pauli and {Cournapeau}, David and {Wieser}, Eric and {Taylor}, Julian and {Berg}, Sebastian and {Smith}, Nathaniel J. and {Kern}, Robert and {Picus}, Matti and {Hoyer}, Stephan and {van Kerkwijk}, Marten H. and {Brett}, Matthew and {Haldane}, Allan and {del R{\'\i}o}, Jaime Fern{\'a}ndez and {Wiebe}, Mark and {Peterson}, Pearu and {G{\'e}rard-Marchant}, Pierre and {Sheppard}, Kevin and {Reddy}, Tyler and {Weckesser}, Warren and {Abbasi}, Hameer and {Gohlke}, Christoph and {Oliphant}, Travis E.},
        title = "{Array programming with NumPy}",
      journal = {\nat},
         year = 2020,
        month = sep,
       volume = {585},
       number = {7825},
        pages = {357-362},
          doi = {10.1038/s41586-020-2649-2},
archivePrefix = {arXiv},
       eprint = {2006.10256},
 primaryClass = {cs.MS},
       adsurl = {https://ui.adsabs.harvard.edu/abs/2020Natur.585..357H}
}

@ARTICLE{2022ApJ...935..167A,
       author = {{Astropy Collaboration} and {Price-Whelan}, Adrian M. and {Lim}, Pey Lian and {Earl}, Nicholas and {Starkman}, Nathaniel and {Bradley}, Larry and {Shupe}, David L. and {Patil}, Aarya A. and {Corrales}, Lia and {Brasseur}, C.~E. and {N{\"o}the}, Maximilian and {Donath}, Axel and {Tollerud}, Erik and {Morris}, Brett M. and {Ginsburg}, Adam and {Vaher}, Eero and {Weaver}, Benjamin A. and {Tocknell}, James and {Jamieson}, William and {van Kerkwijk}, Marten H. and {Robitaille}, Thomas P. and {Merry}, Bruce and {Bachetti}, Matteo and {G{\"u}nther}, H. Moritz and {Aldcroft}, Thomas L. and {Alvarado-Montes}, Jaime A. and {Archibald}, Anne M. and {B{\'o}di}, Attila and {Bapat}, Shreyas and {Barentsen}, Geert and {Baz{\'a}n}, Juanjo and {Biswas}, Manish and {Boquien}, M{\'e}d{\'e}ric and {Burke}, D.~J. and {Cara}, Daria and {Cara}, Mihai and {Conroy}, Kyle E. and {Conseil}, Simon and {Craig}, Matthew W. and {Cross}, Robert M. and {Cruz}, Kelle L. and {D'Eugenio}, Francesco and {Dencheva}, Nadia and {Devillepoix}, Hadrien A.~R. and {Dietrich}, J{\"o}rg P. and {Eigenbrot}, Arthur Davis and {Erben}, Thomas and {Ferreira}, Leonardo and {Foreman-Mackey}, Daniel and {Fox}, Ryan and {Freij}, Nabil and {Garg}, Suyog and {Geda}, Robel and {Glattly}, Lauren and {Gondhalekar}, Yash and {Gordon}, Karl D. and {Grant}, David and {Greenfield}, Perry and {Groener}, Austen M. and {Guest}, Steve and {Gurovich}, Sebastian and {Handberg}, Rasmus and {Hart}, Akeem and {Hatfield-Dodds}, Zac and {Homeier}, Derek and {Hosseinzadeh}, Griffin and {Jenness}, Tim and {Jones}, Craig K. and {Joseph}, Prajwel and {Kalmbach}, J. Bryce and {Karamehmetoglu}, Emir and {Ka{\l}uszy{\'n}ski}, Miko{\l}aj and {Kelley}, Michael S.~P. and {Kern}, Nicholas and {Kerzendorf}, Wolfgang E. and {Koch}, Eric W. and {Kulumani}, Shankar and {Lee}, Antony and {Ly}, Chun and {Ma}, Zhiyuan and {MacBride}, Conor and {Maljaars}, Jakob M. and {Muna}, Demitri and {Murphy}, N.~A. and {Norman}, Henrik and {O'Steen}, Richard and {Oman}, Kyle A. and {Pacifici}, Camilla and {Pascual}, Sergio and {Pascual-Granado}, J. and {Patil}, Rohit R. and {Perren}, Gabriel I. and {Pickering}, Timothy E. and {Rastogi}, Tanuj and {Roulston}, Benjamin R. and {Ryan}, Daniel F. and {Rykoff}, Eli S. and {Sabater}, Jose and {Sakurikar}, Parikshit and {Salgado}, Jes{\'u}s and {Sanghi}, Aniket and {Saunders}, Nicholas and {Savchenko}, Volodymyr and {Schwardt}, Ludwig and {Seifert-Eckert}, Michael and {Shih}, Albert Y. and {Jain}, Anany Shrey and {Shukla}, Gyanendra and {Sick}, Jonathan and {Simpson}, Chris and {Singanamalla}, Sudheesh and {Singer}, Leo P. and {Singhal}, Jaladh and {Sinha}, Manodeep and {Sip{\H{o}}cz}, Brigitta M. and {Spitler}, Lee R. and {Stansby}, David and {Streicher}, Ole and {{\v{S}}umak}, Jani and {Swinbank}, John D. and {Taranu}, Dan S. and {Tewary}, Nikita and {Tremblay}, Grant R. and {de Val-Borro}, Miguel and {Van Kooten}, Samuel J. and {Vasovi{\'c}}, Zlatan and {Verma}, Shresth and {de Miranda Cardoso}, Jos{\'e} Vin{\'\i}cius and {Williams}, Peter K.~G. and {Wilson}, Tom J. and {Winkel}, Benjamin and {Wood-Vasey}, W.~M. and {Xue}, Rui and {Yoachim}, Peter and {Zhang}, Chen and {Zonca}, Andrea and {Astropy Project Contributors}},
        title = "{The Astropy Project: Sustaining and Growing a Community-oriented Open-source Project and the Latest Major Release (v5.0) of the Core Package}",
      journal = {\apj},
         year = 2022,
        month = aug,
       volume = {935},
       number = {2},
          eid = {167},
        pages = {167},
          doi = {10.3847/1538-4357/ac7c74},
archivePrefix = {arXiv},
       eprint = {2206.14220},
 primaryClass = {astro-ph.IM},
       adsurl = {https://ui.adsabs.harvard.edu/abs/2022ApJ...935..167A}
}

@ARTICLE{2016JOSS....1...24F,
       author = {{Foreman-Mackey}, Daniel},
        title = "{corner.py: Scatterplot matrices in Python}",
      journal = {The Journal of Open Source Software},
         year = 2016,
        month = jun,
       volume = {1},
        pages = {24},
          doi = {10.21105/joss.00024},
       adsurl = {https://ui.adsabs.harvard.edu/abs/2016JOSS....1...24F}
}

@software{2023zndo....593847P,
       author = {{Perktold}, Josef and {Seabold}, Skipper and {Sheppard}, Kevin and {ChadFulton} and {Shedden}, Kerby and {jbrockmendel} and {j-grana6} and {Quackenbush}, Peter and {Arel-Bundock}, Vincent and {McKinney}, Wes and {Langmore}, Ian and {Baker}, Bart and {Gommers}, Ralf and {yogabonito} and {s-scherrer} and {Zhurko}, Yauhen and {Brett}, Matthew and {Giampieri}, Enrico and {yl565} and {Millman}, Jarrod and {Hobson}, Paul and {Vincent} and {Roy}, Pamphile and {Augspurger}, Tom and {tvanzyl} and {alexbrc} and {Hartley}, Tyler and {Perez}, Fernando and {Tamiya}, Yuji and {Halchenko}, Yaroslav},
        title = "{statsmodels/statsmodels: Release 0.14.2}",
         year = 2024,
        month = apr,
          eid = {10.5281/zenodo.593847},
          doi = {10.5281/zenodo.593847},
      version = {v0.14.2},
    publisher = {Zenodo},
       adsurl = {https://ui.adsabs.harvard.edu/abs/2023zndo....593847P}
}

@ARTICLE{2019JOSS....4.1143K,
       author = {{Kumar}, Ravin and {Carroll}, Colin and {Hartikainen}, Ari and {Martin}, Osvaldo},
        title = "{ArviZ a unified library for exploratory analysis of Bayesian models in Python}",
      journal = {The Journal of Open Source Software},
         year = 2019,
        month = jan,
       volume = {4},
       number = {33},
          eid = {1143},
        pages = {1143},
          doi = {10.21105/joss.01143},
       adsurl = {https://ui.adsabs.harvard.edu/abs/2019JOSS....4.1143K}
}

@software{2020zndo...3767070W,
       author = {{Waskom}, Michael and {Botvinnik}, Olga and {Ostblom}, Joel and {Gelbart}, Maoz and {Lukauskas}, Saulius and {Hobson}, Paul and {Gemperline}, David C and {Augspurger}, Tom and {Halchenko}, Yaroslav and {Cole}, John B. and {Warmenhoven}, Jordi and {De Ruiter}, Julian and {Pye}, Cameron and {Hoyer}, Stephan and {Vanderplas}, Jake and {Villalba}, Santi and {Kunter}, Gero and {Quintero}, Eric and {Bachant}, Pete and {Martin}, Marcel and {Meyer}, Kyle and {Swain}, Corban and {Miles}, Alistair and {Brunner}, Thomas and {O'Kane}, Drew and {Yarkoni}, Tal and {Williams}, Mike Lee and {Evans}, Constantine and {Fitzgerald}, Clark and {Brian}},
        title = "{mwaskom/seaborn: v0.10.1 (April 2020)}",
         year = 2020,
        month = apr,
          eid = {10.5281/zenodo.3767070},
          doi = {10.5281/zenodo.3767070},
      version = {v0.10.1},
    publisher = {Zenodo},
       adsurl = {https://ui.adsabs.harvard.edu/abs/2020zndo...3767070W}
}

@ARTICLE{2011JMLR...12.2825P,
       author = {{Pedregosa}, Fabian and {Varoquaux}, Ga{\"e}l and {Gramfort}, Alexandre and {Michel}, Vincent and {Thirion}, Bertrand and {Grisel}, Olivier and {Blondel}, Mathieu and {M{\"u}ller}, Andreas and {Nothman}, Joel and {Louppe}, Gilles and {Prettenhofer}, Peter and {Weiss}, Ron and {Dubourg}, Vincent and {Vanderplas}, Jake and {Passos}, Alexandre and {Cournapeau}, David and {Brucher}, Matthieu and {Perrot}, Matthieu and {Duchesnay}, {\'E}douard},
        title = "{Scikit-learn: Machine Learning in Python}",
      journal = {Journal of Machine Learning Research},
         year = 2011,
        month = oct,
       volume = {12},
        pages = {2825-2830},
          doi = {10.48550/arXiv.1201.0490},
archivePrefix = {arXiv},
       eprint = {1201.0490},
 primaryClass = {cs.LG},
       adsurl = {https://ui.adsabs.harvard.edu/abs/2011JMLR...12.2825P}
}

@ARTICLE{2019MNRAS.489.5146J,
       author = {{Jackman}, James A.~G. and {Wheatley}, Peter J. and {Bayliss}, Dan and {Gill}, Samuel and {Hodgkin}, Simon T. and {Burleigh}, Matthew R. and {Braker}, Ian P. and {G{\"u}nther}, Maximilian N. and {Louden}, Tom and {Turner}, Oliver and {Anderson}, David R. and {Belardi}, Claudia and {Bouchy}, Fran{\c{c}}ois and {Briegal}, Joshua T. and {Bryant}, Edward M. and {Cabrera}, Juan and {Casewell}, Sarah L. and {Chaushev}, Alexander and {Costes}, Jean C. and {Csizmadia}, Szilard and {Eigm{\"u}ller}, Philipp and {Erikson}, Anders and {G{\"a}nsicke}, Boris T. and {Gillen}, Edward and {Goad}, Michael R. and {Jenkins}, James S. and {McCormac}, James and {Moyano}, Maximiliano and {Nielsen}, Louise D. and {Pollacco}, Don and {Poppenhaeger}, Katja and {Queloz}, Didier and {Rauer}, Heike and {Raynard}, Liam and {Smith}, Alexis M.~S. and {Udry}, St{\'e}phane and {Vines}, Jose I. and {Watson}, Christopher A. and {West}, Richard G.},
        title = "{NGTS-7Ab: an ultrashort-period brown dwarf transiting a tidally locked and active M dwarf}",
      journal = {\mnras},
         year = 2019,
        month = nov,
       volume = {489},
       number = {4},
        pages = {5146-5164},
          doi = {10.1093/mnras/stz2496},
archivePrefix = {arXiv},
       eprint = {1906.08219},
 primaryClass = {astro-ph.SR},
       adsurl = {https://ui.adsabs.harvard.edu/abs/2019MNRAS.489.5146J}
}

@ARTICLE{2017ApJ...849...11G,
       author = {{Gillen}, Edward and {Hillenbrand}, Lynne A. and {David}, Trevor J. and {Aigrain}, Suzanne and {Rebull}, Luisa and {Stauffer}, John and {Cody}, Ann Marie and {Queloz}, Didier},
        title = "{New Low-mass Eclipsing Binary Systems in Praesepe Discovered by K2}",
      journal = {\apj},
         year = 2017,
        month = nov,
       volume = {849},
       number = {1},
          eid = {11},
        pages = {11},
          doi = {10.3847/1538-4357/aa84b3},
archivePrefix = {arXiv},
       eprint = {1706.03084},
 primaryClass = {astro-ph.SR},
       adsurl = {https://ui.adsabs.harvard.edu/abs/2017ApJ...849...11G}
}

@ARTICLE{2008A&A...491..889D,
       author = {{Deleuil}, M. and {Deeg}, H.~J. and {Alonso}, R. and {Bouchy}, F. and {Rouan}, D. and {Auvergne}, M. and {Baglin}, A. and {Aigrain}, S. and {Almenara}, J.~M. and {Barbieri}, M. and {Barge}, P. and {Bruntt}, H. and {Bord{\'e}}, P. and {Collier Cameron}, A. and {Csizmadia}, Sz. and {de La Reza}, R. and {Dvorak}, R. and {Erikson}, A. and {Fridlund}, M. and {Gandolfi}, D. and {Gillon}, M. and {Guenther}, E. and {Guillot}, T. and {Hatzes}, A. and {H{\'e}brard}, G. and {Jorda}, L. and {Lammer}, H. and {L{\'e}ger}, A. and {Llebaria}, A. and {Loeillet}, B. and {Mayor}, M. and {Mazeh}, T. and {Moutou}, C. and {Ollivier}, M. and {P{\"a}tzold}, M. and {Pont}, F. and {Queloz}, D. and {Rauer}, H. and {Schneider}, J. and {Shporer}, A. and {Wuchterl}, G. and {Zucker}, S.},
        title = "{Transiting exoplanets from the CoRoT space mission . VI. CoRoT-Exo-3b: the first secure inhabitant of the brown-dwarf desert}",
      journal = {\aap},
         year = 2008,
        month = dec,
       volume = {491},
       number = {3},
        pages = {889-897},
          doi = {10.1051/0004-6361:200810625},
archivePrefix = {arXiv},
       eprint = {0810.0919},
 primaryClass = {astro-ph},
       adsurl = {https://ui.adsabs.harvard.edu/abs/2008A&A...491..889D}
}

@ARTICLE{2011A&A...525A..68B,
       author = {{Bouchy}, F. and {Deleuil}, M. and {Guillot}, T. and {Aigrain}, S. and {Carone}, L. and {Cochran}, W.~D. and {Almenara}, J.~M. and {Alonso}, R. and {Auvergne}, M. and {Baglin}, A. and {Barge}, P. and {Bonomo}, A.~S. and {Bord{\'e}}, P. and {Csizmadia}, Sz. and {de Bondt}, K. and {Deeg}, H.~J. and {D{\'\i}az}, R.~F. and {Dvorak}, R. and {Endl}, M. and {Erikson}, A. and {Ferraz-Mello}, S. and {Fridlund}, M. and {Gandolfi}, D. and {Gazzano}, J.~C. and {Gibson}, N. and {Gillon}, M. and {Guenther}, E. and {Hatzes}, A. and {Havel}, M. and {H{\'e}brard}, G. and {Jorda}, L. and {L{\'e}ger}, A. and {Lovis}, C. and {Llebaria}, A. and {Lammer}, H. and {MacQueen}, P.~J. and {Mazeh}, T. and {Moutou}, C. and {Ofir}, A. and {Ollivier}, M. and {Parviainen}, H. and {P{\"a}tzold}, M. and {Queloz}, D. and {Rauer}, H. and {Rouan}, D. and {Santerne}, A. and {Schneider}, J. and {Tingley}, B. and {Wuchterl}, G.},
        title = "{Transiting exoplanets from the CoRoT space mission. XV. CoRoT-15b: a brown-dwarf transiting companion}",
      journal = {\aap},
         year = 2011,
        month = jan,
       volume = {525},
          eid = {A68},
        pages = {A68},
          doi = {10.1051/0004-6361/201015276},
archivePrefix = {arXiv},
       eprint = {1010.0179},
 primaryClass = {astro-ph.EP},
       adsurl = {https://ui.adsabs.harvard.edu/abs/2011A&A...525A..68B}
}

@ARTICLE{2015A&A...584A..13C,
       author = {{Csizmadia}, Sz. and {Hatzes}, A. and {Gandolfi}, D. and {Deleuil}, M. and {Bouchy}, F. and {Fridlund}, M. and {Szabados}, L. and {Parviainen}, H. and {Cabrera}, J. and {Aigrain}, S. and {Alonso}, R. and {Almenara}, J. -M. and {Baglin}, A. and {Bord{\'e}}, P. and {Bonomo}, A.~S. and {Deeg}, H.~J. and {D{\'\i}az}, R.~F. and {Erikson}, A. and {Ferraz-Mello}, S. and {Tadeu dos Santos}, M. and {Guenther}, E.~W. and {Guillot}, T. and {Grziwa}, S. and {H{\'e}brard}, G. and {Klagyivik}, P. and {Ollivier}, M. and {P{\"a}tzold}, M. and {Rauer}, H. and {Rouan}, D. and {Santerne}, A. and {Schneider}, J. and {Mazeh}, T. and {Wuchterl}, G. and {Carpano}, S. and {Ofir}, A.},
        title = "{Transiting exoplanets from the CoRoT space mission{\ensuremath{\star}}. XXVIII. CoRoT-33b, an object in the brown dwarf desert with 2:3 commensurability with its host star}",
      journal = {\aap},
         year = 2015,
        month = dec,
       volume = {584},
          eid = {A13},
        pages = {A13},
          doi = {10.1051/0004-6361/201526763},
archivePrefix = {arXiv},
       eprint = {1508.05763},
 primaryClass = {astro-ph.EP},
       adsurl = {https://ui.adsabs.harvard.edu/abs/2015A&A...584A..13C}
}

@ARTICLE{2022MNRAS.516..636S,
       author = {{Sebastian}, D. and {Guenther}, E.~W. and {Deleuil}, M. and {Dorsch}, M. and {Heber}, U. and {Heuser}, C. and {Gandolfi}, D. and {Grziwa}, S. and {Deeg}, H.~J. and {Alonso}, R. and {Bouchy}, F. and {Csizmadia}, Sz and {Cusano}, F. and {Fridlund}, M. and {Geier}, S. and {Irrgang}, A. and {Korth}, J. and {Nespral}, D. and {Rauer}, H. and {Tal-Or}, L. and {CoRoT-team}},
        title = "{Sub-stellar companions of intermediate-mass stars with CoRoT: CoRoT-34b, CoRoT-35b, and CoRoT-36b}",
      journal = {\mnras},
         year = 2022,
        month = oct,
       volume = {516},
       number = {1},
        pages = {636-655},
          doi = {10.1093/mnras/stac2131},
archivePrefix = {arXiv},
       eprint = {2207.08742},
 primaryClass = {astro-ph.EP},
       adsurl = {https://ui.adsabs.harvard.edu/abs/2022MNRAS.516..636S}
}

@ARTICLE{2018AJ....156..168B,
       author = {{Beatty}, Thomas G. and {Morley}, Caroline V. and {Curtis}, Jason L. and {Burrows}, Adam and {Davenport}, James R.~A. and {Montet}, Benjamin T.},
        title = "{A Significant Overluminosity in the Transiting Brown Dwarf CWW 89Ab}",
      journal = {\aj},
         year = 2018,
        month = oct,
       volume = {156},
       number = {4},
          eid = {168},
        pages = {168},
          doi = {10.3847/1538-3881/aad697},
archivePrefix = {arXiv},
       eprint = {1807.11500},
 primaryClass = {astro-ph.SR},
       adsurl = {https://ui.adsabs.harvard.edu/abs/2018AJ....156..168B}
}

@ARTICLE{2017A&A...604L...6V,
       author = {{von Boetticher}, Alexander and {Triaud}, Amaury H.~M.~J. and {Queloz}, Didier and {Gill}, Sam and {Lendl}, Monika and {Delrez}, Laetitia and {Anderson}, David R. and {Collier Cameron}, Andrew and {Faedi}, Francesca and {Gillon}, Micha{\"e}l and {G{\'o}mez Maqueo Chew}, Yilen and {Hebb}, Leslie and {Hellier}, Coel and {Jehin}, Emmanu{\"e}l and {Maxted}, Pierre F.~L. and {Martin}, David V. and {Pepe}, Francesco and {Pollacco}, Don and {S{\'e}gransan}, Damien and {Smalley}, Barry and {Udry}, St{\'e}phane and {West}, Richard},
        title = "{The EBLM project. III. A Saturn-size low-mass star at the hydrogen-burning limit}",
      journal = {\aap},
         year = 2017,
        month = aug,
       volume = {604},
          eid = {L6},
        pages = {L6},
          doi = {10.1051/0004-6361/201731107},
archivePrefix = {arXiv},
       eprint = {1706.08781},
 primaryClass = {astro-ph.SR},
       adsurl = {https://ui.adsabs.harvard.edu/abs/2017A&A...604L...6V}
}

@ARTICLE{2017AJ....153...15B,
       author = {{Bayliss}, D. and {Hojjatpanah}, S. and {Santerne}, A. and {Dragomir}, D. and {Zhou}, G. and {Shporer}, A. and {Col{\'o}n}, K.~D. and {Almenara}, J. and {Armstrong}, D.~J. and {Barrado}, D. and {Barros}, S.~C.~C. and {Bento}, J. and {Boisse}, I. and {Bouchy}, F. and {Brown}, D.~J.~A. and {Brown}, T. and {Cameron}, A. and {Cochran}, W.~D. and {Demangeon}, O. and {Deleuil}, M. and {D{\'\i}az}, R.~F. and {Fulton}, B. and {Horne}, K. and {H{\'e}brard}, G. and {Lillo-Box}, J. and {Lovis}, C. and {Mawet}, D. and {Ngo}, H. and {Osborn}, H. and {Palle}, E. and {Petigura}, E. and {Pollacco}, D. and {Santos}, N. and {Sefako}, R. and {Siverd}, R. and {Sousa}, S.~G. and {Tsantaki}, M.},
        title = "{EPIC 201702477b: A Transiting Brown Dwarf from K2 in a 41 day Orbit}",
      journal = {\aj},
         year = 2017,
        month = jan,
       volume = {153},
       number = {1},
          eid = {15},
        pages = {15},
          doi = {10.3847/1538-3881/153/1/15},
archivePrefix = {arXiv},
       eprint = {1606.04047},
 primaryClass = {astro-ph.EP},
       adsurl = {https://ui.adsabs.harvard.edu/abs/2017AJ....153...15B}
}

@ARTICLE{2021MNRAS.505.4956B,
       author = {{Benni}, P. and {Burdanov}, A.~Y. and {Krushinsky}, V.~V. and {Bonfanti}, A. and {H{\'e}brard}, G. and {Almenara}, J.~M. and {Dalal}, S. and {Demangeon}, O.~D.~S. and {Tsantaki}, M. and {Pepper}, J. and {Stassun}, K.~G. and {Vanderburg}, A. and {Belinski}, A. and {Kashaev}, F. and {Barkaoui}, K. and {Kim}, T. and {Kang}, W. and {Antonyuk}, K. and {Dyachenko}, V.~V. and {Rastegaev}, D.~A. and {Beskakotov}, A. and {Mitrofanova}, A.~A. and {Pozuelos}, F.~J. and {Kuznetsov}, E.~D. and {Popov}, A. and {Kiefer}, F. and {Wilson}, P.~A. and {Ricker}, G. and {Vanderspek}, R. and {Latham}, D.~W. and {Seager}, S. and {Jenkins}, J.~M. and {Sokov}, E. and {Sokova}, I. and {Marchini}, A. and {Papini}, R. and {Salvaggio}, F. and {Banfi}, M. and {Ba{\c{s}}t{\"u}rk}, {\"O}. and {Torun}, {\c{S}}. and {Yal{\c{c}}{\i}nkaya}, S. and {Ivanov}, K. and {Valyavin}, G. and {Jehin}, E. and {Gillon}, M. and {Pak{\v{s}}tien{\.{e}}}, E. and {Hentunen}, V. -P. and {Shadick}, S. and {Bretton}, M. and {W{\"u}nsche}, A. and {Garlitz}, J. and {Jongen}, Y. and {Molina}, D. and {Girardin}, E. and {Grau Horta}, F. and {Naves}, R. and {Benkhaldoun}, Z. and {Joner}, M.~D. and {Spencer}, M. and {Bieryla}, A. and {Stevens}, D.~J. and {Jensen}, E.~L.~N. and {Collins}, K.~A. and {Charbonneau}, D. and {Quintana}, E.~V. and {Mullally}, S.~E. and {Henze}, C.~E.},
        title = "{Discovery of a young low-mass brown dwarf transiting a fast-rotating F-type star by the Galactic Plane eXoplanet (GPX) survey}",
      journal = {\mnras},
         year = 2021,
        month = aug,
       volume = {505},
       number = {4},
        pages = {4956-4967},
          doi = {10.1093/mnras/stab1567},
archivePrefix = {arXiv},
       eprint = {2009.11899},
 primaryClass = {astro-ph.SR},
       adsurl = {https://ui.adsabs.harvard.edu/abs/2021MNRAS.505.4956B}
}

@ARTICLE{2019AJ....157...31Z,
       author = {{Zhou}, G. and {Bakos}, G. {\'A}. and {Bayliss}, D. and {Bento}, J. and {Bhatti}, W. and {Brahm}, R. and {Csubry}, Z. and {Espinoza}, N. and {Hartman}, J.~D. and {Henning}, T. and {Jord{\'a}n}, A. and {Mancini}, L. and {Penev}, K. and {Rabus}, M. and {Sarkis}, P. and {Suc}, V. and {de Val-Borro}, M. and {Rodriguez}, J.~E. and {Osip}, D. and {Kedziora-Chudczer}, L. and {Bailey}, J. and {Tinney}, C.~G. and {Durkan}, S. and {L{\'a}z{\'a}r}, J. and {Papp}, I. and {S{\'a}ri}, P.},
        title = "{HATS-70b: A 13 MJ Brown Dwarf Transiting an A Star}",
      journal = {\aj},
         year = 2019,
        month = jan,
       volume = {157},
       number = {1},
          eid = {31},
        pages = {31},
          doi = {10.3847/1538-3881/aaf1bb},
archivePrefix = {arXiv},
       eprint = {1811.06925},
 primaryClass = {astro-ph.EP},
       adsurl = {https://ui.adsabs.harvard.edu/abs/2019AJ....157...31Z}
}

@ARTICLE{2023AJ....165..268V,
       author = {{Vowell}, Noah and {Rodriguez}, Joseph E. and {Quinn}, Samuel N. and {Zhou}, George and {Vanderburg}, Andrew and {Mann}, Andrew W. and {Hooton}, Matthew J. and {Stassun}, Keivan G. and {Howard}, Saburo and {Bieryla}, Allyson and {Latham}, David W. and {Howell}, Steve B. and {Guillot}, Tristan and {Ziegler}, Carl and {Collins}, Karen A. and {Carmichael}, Theron W. and {Jenkins}, Jon M. and {Shporer}, Avi and {ABE}, Lyu and {Bendjoya}, Philippe and {Bush}, Jonathan L. and {Buttu}, Marco and {Collins}, Kevin I. and {Eastman}, Jason D. and {Fields}, Matthew J. and {Gasparetto}, Thomas and {G{\"u}nther}, Maximilian N. and {Kostov}, Veselin B. and {Kraus}, Adam L. and {Lester}, Kathryn V. and {Levine}, Alan M. and {Littlefield}, Colin and {Marie-Sainte}, Wenceslas and {M{\'e}karnia}, Djamel and {Osborn}, Hugh P. and {Rapetti}, David and {Ricker}, George R. and {Seager}, S. and {Sefako}, Ramotholo and {Srdoc}, Gregor and {Suarez}, Olga and {Torres}, Guillermo and {Triaud}, Amaury H.~M.~J. and {Vanderspek}, R. and {Winn}, Joshua N.},
        title = "{HIP 33609 b: An Eccentric Brown Dwarf Transiting a V = 7.3 Rapidly Rotating B Star}",
      journal = {\aj},
         year = 2023,
        month = jun,
       volume = {165},
       number = {6},
          eid = {268},
        pages = {268},
          doi = {10.3847/1538-3881/acd197},
archivePrefix = {arXiv},
       eprint = {2301.09663},
 primaryClass = {astro-ph.EP},
       adsurl = {https://ui.adsabs.harvard.edu/abs/2023AJ....165..268V}
}

@ARTICLE{2013A&A...549A..18T,
       author = {{Triaud}, A.~H.~M.~J. and {Hebb}, L. and {Anderson}, D.~R. and {Cargile}, P. and {Collier Cameron}, A. and {Doyle}, A.~P. and {Faedi}, F. and {Gillon}, M. and {Gomez Maqueo Chew}, Y. and {Hellier}, C. and {Jehin}, E. and {Maxted}, P. and {Naef}, D. and {Pepe}, F. and {Pollacco}, D. and {Queloz}, D. and {S{\'e}gransan}, D. and {Smalley}, B. and {Stassun}, K. and {Udry}, S. and {West}, R.~G.},
        title = "{The EBLM project. I. Physical and orbital parameters, including spin-orbit angles, of two low-mass eclipsing binaries on opposite sides of the brown dwarf limit}",
      journal = {\aap},
         year = 2013,
        month = jan,
       volume = {549},
          eid = {A18},
        pages = {A18},
          doi = {10.1051/0004-6361/201219643},
archivePrefix = {arXiv},
       eprint = {1208.4940},
 primaryClass = {astro-ph.SR},
       adsurl = {https://ui.adsabs.harvard.edu/abs/2013A&A...549A..18T}
}

@ARTICLE{2012ApJ...761..123S,
       author = {{Siverd}, Robert J. and {Beatty}, Thomas G. and {Pepper}, Joshua and {Eastman}, Jason D. and {Collins}, Karen and {Bieryla}, Allyson and {Latham}, David W. and {Buchhave}, Lars A. and {Jensen}, Eric L.~N. and {Crepp}, Justin R. and {Street}, Rachel and {Stassun}, Keivan G. and {Gaudi}, B. Scott and {Berlind}, Perry and {Calkins}, Michael L. and {DePoy}, D.~L. and {Esquerdo}, Gilbert A. and {Fulton}, Benjamin J. and {F{\H{u}}r{\'e}sz}, G{\'a}bor and {Geary}, John C. and {Gould}, Andrew and {Hebb}, Leslie and {Kielkopf}, John F. and {Marshall}, Jennifer L. and {Pogge}, Richard and {Stanek}, K.~Z. and {Stefanik}, Robert P. and {Szentgyorgyi}, Andrew H. and {Trueblood}, Mark and {Trueblood}, Patricia and {Stutz}, Amelia M. and {van Saders}, Jennifer L.},
        title = "{KELT-1b: A Strongly Irradiated, Highly Inflated, Short Period, 27 Jupiter-mass Companion Transiting a Mid-F Star}",
      journal = {\apj},
         year = 2012,
        month = dec,
       volume = {761},
       number = {2},
          eid = {123},
        pages = {123},
          doi = {10.1088/0004-637X/761/2/123},
archivePrefix = {arXiv},
       eprint = {1206.1635},
 primaryClass = {astro-ph.EP},
       adsurl = {https://ui.adsabs.harvard.edu/abs/2012ApJ...761..123S}
}

@ARTICLE{2013A&A...558L...6M,
       author = {{Moutou}, C. and {Bonomo}, A.~S. and {Bruno}, G. and {Montagnier}, G. and {Bouchy}, F. and {Almenara}, J.~M. and {Barros}, S.~C.~C. and {Deleuil}, M. and {D{\'\i}az}, R.~F. and {H{\'e}brard}, G. and {Santerne}, A.},
        title = "{SOPHIE velocimetry of Kepler transit candidates. IX. KOI-415 b: a long-period, eccentric transiting brown dwarf to an evolved Sun}",
      journal = {\aap},
         year = 2013,
        month = oct,
       volume = {558},
          eid = {L6},
        pages = {L6},
          doi = {10.1051/0004-6361/201322201},
archivePrefix = {arXiv},
       eprint = {1309.0905},
 primaryClass = {astro-ph.EP},
       adsurl = {https://ui.adsabs.harvard.edu/abs/2013A&A...558L...6M}
}

@ARTICLE{2019AJ....158...38C,
       author = {{Carmichael}, Theron W. and {Latham}, David W. and {Vanderburg}, Andrew M.},
        title = "{New Substellar Discoveries from Kepler  and K2: Is There a Brown Dwarf Desert?}",
      journal = {\aj},
         year = 2019,
        month = jul,
       volume = {158},
       number = {1},
          eid = {38},
        pages = {38},
          doi = {10.3847/1538-3881/ab245e},
archivePrefix = {arXiv},
       eprint = {1903.03118},
 primaryClass = {astro-ph.EP},
       adsurl = {https://ui.adsabs.harvard.edu/abs/2019AJ....158...38C}
}

@ARTICLE{2011ApJ...730...79J,
       author = {{Johnson}, John Asher and {Apps}, Kevin and {Gazak}, J. Zachary and {Crepp}, Justin R. and {Crossfield}, Ian J. and {Howard}, Andrew W. and {Marcy}, Geoff W. and {Morton}, Timothy D. and {Chubak}, Carly and {Isaacson}, Howard},
        title = "{LHS 6343 C: A Transiting Field Brown Dwarf Discovered by the Kepler Mission}",
      journal = {\apj},
         year = 2011,
        month = apr,
       volume = {730},
       number = {2},
          eid = {79},
        pages = {79},
          doi = {10.1088/0004-637X/730/2/79},
archivePrefix = {arXiv},
       eprint = {1008.4141},
 primaryClass = {astro-ph.EP},
       adsurl = {https://ui.adsabs.harvard.edu/abs/2011ApJ...730...79J}
}

@ARTICLE{2018AJ....156..140I,
       author = {{Irwin}, Jonathan M. and {Charbonneau}, David and {Esquerdo}, Gilbert A. and {Latham}, David W. and {Winters}, Jennifer G. and {Dittmann}, Jason A. and {Newton}, Elisabeth R. and {Berta-Thompson}, Zachory K. and {Berlind}, Perry and {Calkins}, Michael L.},
        title = "{Four New Eclipsing Mid M-dwarf Systems from the New Luyten Two Tenths Catalog}",
      journal = {\aj},
         year = 2018,
        month = oct,
       volume = {156},
       number = {4},
          eid = {140},
        pages = {140},
          doi = {10.3847/1538-3881/aad9a3},
archivePrefix = {arXiv},
       eprint = {1808.03243},
 primaryClass = {astro-ph.SR},
       adsurl = {https://ui.adsabs.harvard.edu/abs/2018AJ....156..140I}
}

@ARTICLE{2021MNRAS.505.2741A,
       author = {{Acton}, Jack S. and {Goad}, Michael R. and {Burleigh}, Matthew R. and {Casewell}, Sarah L. and {Breytenbach}, Hannes and {Nielsen}, Louise D. and {Smith}, Gareth and {Anderson}, David R. and {Battley}, Matthew P. and {Bayliss}, Daniel and {Bouchy}, Fran{\c{c}}ois and {Bryant}, Edward M. and {Csizmadia}, Szil{\'a}rd and {Eigm{\"u}ller}, Philipp and {Gill}, Samuel and {Gillen}, Edward and {Grieves}, Nolan and {G{\"u}nther}, Maximilian N. and {Henderson}, Beth A. and {Hodgkin}, Simon T. and {Jackman}, James A.~G. and {Jenkins}, James S. and {Lendl}, Monika and {McCormac}, James and {Moyano}, Maximiliano and {Nelson}, Richard P. and {Sefako}, Ramotholo R. and {Smith}, Alexis M.~S. and {Stalport}, Manu and {Thomas}, Jessymol K. and {Tilbrook}, Rosanna H. and {Udry}, St{\'e}phane and {West}, Richard G. and {Wheatley}, Peter J. and {Worters}, Hannah L. and {Vines}, Jose I. and {Alves}, Douglas R.},
        title = "{NGTS-19b: a high-mass transiting brown dwarf in a 17-d eccentric orbit}",
      journal = {\mnras},
         year = 2021,
        month = aug,
       volume = {505},
       number = {2},
        pages = {2741-2752},
          doi = {10.1093/mnras/stab1459},
archivePrefix = {arXiv},
       eprint = {2105.08574},
 primaryClass = {astro-ph.EP},
       adsurl = {https://ui.adsabs.harvard.edu/abs/2021MNRAS.505.2741A}
}

@ARTICLE{2010ApJ...718.1353I,
       author = {{Irwin}, Jonathan and {Buchhave}, Lars and {Berta}, Zachory K. and {Charbonneau}, David and {Latham}, David W. and {Burke}, Christopher J. and {Esquerdo}, Gilbert A. and {Everett}, Mark E. and {Holman}, Matthew J. and {Nutzman}, Philip and {Berlind}, Perry and {Calkins}, Michael L. and {Falco}, Emilio E. and {Winn}, Joshua N. and {Johnson}, John A. and {Gazak}, J. Zachary},
        title = "{NLTT 41135: A Field M Dwarf + Brown Dwarf Eclipsing Binary in a Triple System, Discovered by the MEarth Observatory}",
      journal = {\apj},
         year = 2010,
        month = aug,
       volume = {718},
       number = {2},
        pages = {1353-1366},
          doi = {10.1088/0004-637X/718/2/1353},
archivePrefix = {arXiv},
       eprint = {1006.1793},
 primaryClass = {astro-ph.SR},
       adsurl = {https://ui.adsabs.harvard.edu/abs/2010ApJ...718.1353I}
}

@ARTICLE{2006A&A...447.1035P,
       author = {{Pont}, F. and {Moutou}, C. and {Bouchy}, F. and {Behrend}, R. and {Mayor}, M. and {Udry}, S. and {Queloz}, D. and {Santos}, N. and {Melo}, C.},
        title = "{Radius and mass of a transiting M dwarf near the hydrogen-burning limit. OGLE-TR-123}",
      journal = {\aap},
         year = 2006,
        month = mar,
       volume = {447},
       number = {3},
        pages = {1035-1039},
          doi = {10.1051/0004-6361:20053692},
       adsurl = {https://ui.adsabs.harvard.edu/abs/2006A&A...447.1035P}
}

@ARTICLE{2021A&A...650A..55P,
       author = {{Palle}, E. and {Luque}, R. and {Zapatero Osorio}, M.~R. and {Parviainen}, H. and {Ikoma}, M. and {Tabernero}, H.~M. and {Zechmeister}, M. and {Mustill}, A.~J. and {Bejar}, V.~S.~J. and {Narita}, N. and {Murgas}, F.},
        title = "{ESPRESSO mass determination of TOI-263b: an extreme inhabitant of the brown dwarf desert}",
      journal = {\aap},
         year = 2021,
        month = jun,
       volume = {650},
          eid = {A55},
        pages = {A55},
          doi = {10.1051/0004-6361/202039937},
archivePrefix = {arXiv},
       eprint = {2103.11150},
 primaryClass = {astro-ph.EP},
       adsurl = {https://ui.adsabs.harvard.edu/abs/2021A&A...650A..55P}
}

@ARTICLE{2020AJ....159..151S,
       author = {{{\v{S}}ubjak}, J{\'a}n and {Sharma}, Rishikesh and {Carmichael}, Theron W. and {Johnson}, Marshall C. and {Gonzales}, Erica J. and {Matthews}, Elisabeth and {Boffin}, Henri M.~J. and {Brahm}, Rafael and {Chaturvedi}, Priyanka and {Chakraborty}, Abhijit and {Ciardi}, David R. and {Collins}, Karen A. and {Esposito}, Massimiliano and {Fridlund}, Malcolm and {Gan}, Tianjun and {Gandolfi}, Davide and {Garc{\'\i}a}, Rafael A. and {Guenther}, Eike and {Hatzes}, Artie and {Latham}, David W. and {Mathis}, St{\'e}phane and {Mathur}, Savita and {Persson}, Carina M. and {Relles}, Howard M. and {Schlieder}, Joshua E. and {Barclay}, Thomas and {Dressing}, Courtney D. and {Crossfield}, Ian and {Howard}, Andrew W. and {Rodler}, Florian and {Zhou}, George and {Quinn}, Samuel N. and {Esquerdo}, Gilbert A. and {Calkins}, Michael L. and {Berlind}, Perry and {Stassun}, Keivan G. and {Bla{\v{z}}ek}, Martin and {Skarka}, Marek and {{\v{S}}pokov{\'a}}, Magdalena and {{\v{Z}}{\'a}k}, Ji{\v{r}}{\'\i} and {Albrecht}, Simon and {Sobrino}, Roi Alonso and {Beck}, Paul and {Cabrera}, Juan and {Carleo}, Ilaria and {Cochran}, William D. and {Csizmadia}, Szilard and {Dai}, Fei and {Deeg}, Hans J. and {de Leon}, Jerome P. and {Eigm{\"u}ller}, Philipp and {Endl}, Michael and {Erikson}, Anders and {Fukui}, Akihiko and {Georgieva}, Iskra and {Gonz{\'a}lez-Cuesta}, Luc{\'\i}a and {Grziwa}, Sascha and {Hidalgo}, Diego and {Hirano}, Teruyuki and {Hjorth}, Maria and {Knudstrup}, Emil and {Korth}, Judith and {Lam}, Kristine W.~F. and {Livingston}, John H. and {Lund}, Mikkel N. and {Luque}, Rafael and {Montanes Rodr{\'\i}guez}, Pilar and {Murgas}, Felipe and {Narita}, Norio and {Nespral}, David and {Niraula}, Prajwal and {Nowak}, Grzegorz and {Pall{\'e}}, Enric and {P{\"a}tzold}, Martin and {Prieto-Arranz}, Jorge and {Rauer}, Heike and {Redfield}, Seth and {Ribas}, Ignasi and {Smith}, Alexis M.~S. and {Van Eylen}, Vincent and {Kab{\'a}th}, Petr},
        title = "{TOI-503: The First Known Brown-dwarf Am-star Binary from the TESS Mission}",
      journal = {\aj},
         year = 2020,
        month = apr,
       volume = {159},
       number = {4},
          eid = {151},
        pages = {151},
          doi = {10.3847/1538-3881/ab7245},
archivePrefix = {arXiv},
       eprint = {1909.07984},
 primaryClass = {astro-ph.SR},
       adsurl = {https://ui.adsabs.harvard.edu/abs/2020AJ....159..151S}
}

@ARTICLE{2020AJ....160...53C,
       author = {{Carmichael}, Theron W. and {Quinn}, Samuel N. and {Mustill}, Alexander J. and {Huang}, Chelsea and {Zhou}, George and {Persson}, Carina M. and {Nielsen}, Louise D. and {Collins}, Karen A. and {Ziegler}, Carl and {Collins}, Kevin I. and {Rodriguez}, Joseph E. and {Shporer}, Avi and {Brahm}, Rafael and {Mann}, Andrew W. and {Bouchy}, Francois and {Fridlund}, Malcolm and {Stassun}, Keivan G. and {Hellier}, Coel and {Seidel}, Julia V. and {Stalport}, Manu and {Udry}, Stephane and {Pepe}, Francesco and {Ireland}, Michael and {{\v{Z}}erjal}, Maru{\v{s}}a and {Brice{\~n}o}, C{\'e}sar and {Law}, Nicholas and {Jord{\'a}n}, Andr{\'e}s and {Espinoza}, N{\'e}stor and {Henning}, Thomas and {Sarkis}, Paula and {Latham}, David W.},
        title = "{Two Intermediate-mass Transiting Brown Dwarfs from the TESS Mission}",
      journal = {\aj},
         year = 2020,
        month = jul,
       volume = {160},
       number = {1},
          eid = {53},
        pages = {53},
          doi = {10.3847/1538-3881/ab9b84},
archivePrefix = {arXiv},
       eprint = {2002.01943},
 primaryClass = {astro-ph.SR},
       adsurl = {https://ui.adsabs.harvard.edu/abs/2020AJ....160...53C}
}

@ARTICLE{2021A&A...652A.127G,
       author = {{Grieves}, Nolan and {Bouchy}, Fran{\c{c}}ois and {Lendl}, Monika and {Carmichael}, Theron and {Mireles}, Ismael and {Shporer}, Avi and {McLeod}, Kim K. and {Collins}, Karen A. and {Brahm}, Rafael and {Stassun}, Keivan G. and {Gill}, Sam and {Bouma}, Luke G. and {Guillot}, Tristan and {Cointepas}, Marion and {Dos Santos}, Leonardo A. and {Casewell}, Sarah L. and {Jenkins}, Jon M. and {Henning}, Thomas and {Nielsen}, Louise D. and {Psaridi}, Angelica and {Udry}, St{\'e}phane and {S{\'e}gransan}, Damien and {Eastman}, Jason D. and {Zhou}, George and {Abe}, Lyu and {Agabi}, Abelkrim and {Bakos}, Gaspar and {Charbonneau}, David and {Collins}, Kevin I. and {Colon}, Knicole D. and {Crouzet}, Nicolas and {Dransfield}, Georgina and {Evans}, Phil and {Goeke}, Robert F. and {Hart}, Rhodes and {Irwin}, Jonathan M. and {Jensen}, Eric L.~N. and {Jord{\'a}n}, Andr{\'e}s and {Kielkopf}, John F. and {Latham}, David W. and {Marie-Sainte}, Wenceslas and {M{\'e}karnia}, Djamel and {Nelson}, Peter and {Quinn}, Samuel N. and {Radford}, Don J. and {Rodriguez}, David R. and {Rowden}, Pamela and {Schmider}, Fran{\c{c}}ois-Xavier and {Schwarz}, Richard P. and {Smith}, Jeffrey C. and {Stockdale}, Chris and {Suarez}, Olga and {Tan}, Thiam-Guan and {Triaud}, Amaury H.~M.~J. and {Waalkes}, William and {Wingham}, Geof},
        title = "{Populating the brown dwarf and stellar boundary: Five stars with transiting companions near the hydrogen-burning mass limit}",
      journal = {\aap},
         year = 2021,
        month = aug,
       volume = {652},
          eid = {A127},
        pages = {A127},
          doi = {10.1051/0004-6361/202141145},
archivePrefix = {arXiv},
       eprint = {2107.03480},
 primaryClass = {astro-ph.SR},
       adsurl = {https://ui.adsabs.harvard.edu/abs/2021A&A...652A.127G}
}

@ARTICLE{2020AJ....160..133M,
       author = {{Mireles}, Ismael and {Shporer}, Avi and {Grieves}, Nolan and {Zhou}, George and {G{\"u}nther}, Maximilian N. and {Brahm}, Rafael and {Ziegler}, Carl and {Stassun}, Keivan G. and {Huang}, Chelsea X. and {Nielsen}, Louise and {dos Santos}, Leonardo A. and {Udry}, St{\'e}phane and {Bouchy}, Fran{\c{c}}ois and {Ireland}, Michael and {Wallace}, Alexander and {Sarkis}, Paula and {Henning}, Thomas and {Jord{\'a}n}, Andr{\'e}s and {Law}, Nicholas and {Mann}, Andrew W. and {Paredes}, Leonardo A. and {James}, Hodari-Sadiki and {Jao}, Wei-Chun and {Henry}, Todd J. and {Butler}, R. Paul and {Rodriguez}, Joseph E. and {Yu}, Liang and {Flowers}, Erin and {Ricker}, George R. and {Latham}, David W. and {Vanderspek}, Roland and {Seager}, Sara and {Winn}, Joshua N. and {Jenkins}, Jon M. and {Furesz}, Gabor and {Hesse}, Katharine and {Quintana}, Elisa V. and {Rose}, Mark E. and {Smith}, Jeffrey C. and {Tenenbaum}, Peter and {Vezie}, Michael and {Yahalomi}, Daniel A. and {Zhan}, Zhuchang},
        title = "{TOI 694b and TIC 220568520b: Two Low-mass Companions near the Hydrogen-burning Mass Limit Orbiting Sun-like Stars}",
      journal = {\aj},
         year = 2020,
        month = sep,
       volume = {160},
       number = {3},
          eid = {133},
        pages = {133},
          doi = {10.3847/1538-3881/aba526},
archivePrefix = {arXiv},
       eprint = {2006.14019},
 primaryClass = {astro-ph.SR},
       adsurl = {https://ui.adsabs.harvard.edu/abs/2020AJ....160..133M}
}

@ARTICLE{2011A&A...533A..83B,
       author = {{Bouchy}, F. and {Bonomo}, A.~S. and {Santerne}, A. and {Moutou}, C. and {Deleuil}, M. and {D{\'\i}az}, R.~F. and {Eggenberger}, A. and {Ehrenreich}, D. and {Gry}, C. and {Guillot}, T. and {Havel}, M. and {H{\'e}brard}, G. and {Udry}, S.},
        title = "{SOPHIE velocimetry of Kepler transit candidates. III. KOI-423b: an 18 M$_{Jup}$ transiting companion around an F7IV star}",
      journal = {\aap},
         year = 2011,
        month = sep,
       volume = {533},
          eid = {A83},
        pages = {A83},
          doi = {10.1051/0004-6361/201117095},
archivePrefix = {arXiv},
       eprint = {1106.3225},
 primaryClass = {astro-ph.EP},
       adsurl = {https://ui.adsabs.harvard.edu/abs/2011A&A...533A..83B}
}

@ARTICLE{2014A&A...572A.109D,
       author = {{D{\'\i}az}, R.~F. and {Montagnier}, G. and {Leconte}, J. and {Bonomo}, A.~S. and {Deleuil}, M. and {Almenara}, J.~M. and {Barros}, S.~C.~C. and {Bouchy}, F. and {Bruno}, G. and {Damiani}, C. and {H{\'e}brard}, G. and {Moutou}, C. and {Santerne}, A.},
        title = "{SOPHIE velocimetry of Kepler transit candidates. XIII. KOI-189 b and KOI-686 b: two very low-mass stars in long-period orbits}",
      journal = {\aap},
         year = 2014,
        month = dec,
       volume = {572},
          eid = {A109},
        pages = {A109},
          doi = {10.1051/0004-6361/201424406},
archivePrefix = {arXiv},
       eprint = {1410.5248},
 primaryClass = {astro-ph.SR},
       adsurl = {https://ui.adsabs.harvard.edu/abs/2014A&A...572A.109D}
}

@ARTICLE{2015A&A...575A..85B,
       author = {{Bonomo}, A.~S. and {Sozzetti}, A. and {Santerne}, A. and {Deleuil}, M. and {Almenara}, J. -M. and {Bruno}, G. and {D{\'\i}az}, R.~F. and {H{\'e}brard}, G. and {Moutou}, C.},
        title = "{Improved parameters of seven Kepler giant companions characterized with SOPHIE and HARPS-N}",
      journal = {\aap},
         year = 2015,
        month = mar,
       volume = {575},
          eid = {A85},
        pages = {A85},
          doi = {10.1051/0004-6361/201323042},
archivePrefix = {arXiv},
       eprint = {1501.02653},
 primaryClass = {astro-ph.EP},
       adsurl = {https://ui.adsabs.harvard.edu/abs/2015A&A...575A..85B}
}

@ARTICLE{2021AJ....161...97C,
       author = {{Carmichael}, Theron W. and {Quinn}, Samuel N. and {Zhou}, George and {Grieves}, Nolan and {Irwin}, Jonathan M. and {Stassun}, Keivan G. and {Vanderburg}, Andrew M. and {Winn}, Joshua N. and {Bouchy}, Francois and {Brasseur}, Clara E. and {Brice{\~n}o}, C{\'e}sar and {Caldwell}, Douglas A. and {Charbonneau}, David and {Collins}, Karen A. and {Colon}, Knicole D. and {Eastman}, Jason D. and {Fausnaugh}, Michael and {Fong}, William and {F{\H{u}}r{\'e}sz}, G{\'a}bor and {Huang}, Chelsea and {Jenkins}, Jon M. and {Kielkopf}, John F. and {Latham}, David W. and {Law}, Nicholas and {Lund}, Michael B. and {Mann}, Andrew W. and {Ricker}, George R. and {Rodriguez}, Joseph E. and {Schwarz}, Richard P. and {Shporer}, Avi and {Tenenbaum}, Peter and {Wood}, Mackenna L. and {Ziegler}, Carl},
        title = "{TOI-811b and TOI-852b: New Transiting Brown Dwarfs with Similar Masses and Very Different Radii and Ages from the TESS Mission}",
      journal = {\aj},
         year = 2021,
        month = feb,
       volume = {161},
       number = {2},
          eid = {97},
        pages = {97},
          doi = {10.3847/1538-3881/abd4e1},
archivePrefix = {arXiv},
       eprint = {2009.13515},
 primaryClass = {astro-ph.SR},
       adsurl = {https://ui.adsabs.harvard.edu/abs/2021AJ....161...97C}
}

@ARTICLE{2021AJ....162..144A,
       author = {{Artigau}, {\'E}tienne and {H{\'e}brard}, Guillaume and {Cadieux}, Charles and {Vandal}, Thomas and {Cook}, Neil J. and {Doyon}, Ren{\'e} and {Gagn{\'e}}, Jonathan and {Moutou}, Claire and {Martioli}, Eder and {Frasca}, Antonio and {Jahandar}, Farbod and {Lafreni{\`e}re}, David and {Malo}, Lison and {Donati}, Jean-Fran{\c{c}}ois and {Cort{\'e}s-Zuleta}, P{\'\i}a and {Boisse}, Isabelle and {Delfosse}, Xavier and {Carmona}, Andres and {Fouqu{\'e}}, Pascal and {Morin}, Julien and {Rowe}, Jason and {Marino}, Giuseppe and {Papini}, Riccardo and {Ciardi}, David R. and {Lund}, Michael B. and {Martins}, Jorge H.~C. and {Pelletier}, Stefan and {Arnold}, Luc and {Bouchy}, Fran{\c{c}}ois and {Forveille}, Thierry and {Santos}, Nuno C. and {Bonfils}, Xavier and {Figueira}, Pedro and {Fausnaugh}, Michael and {Ricker}, George and {Latham}, David W. and {Seager}, Sara and {Winn}, Joshua N. and {Jenkins}, Jon M. and {Ting}, Eric B. and {Torres}, Guillermo and {Gomes da Silva}, Jo{\~a}o},
        title = "{TOI-1278 B: SPIRou Unveils a Rare Brown Dwarf Companion in Close-in Orbit around an M Dwarf}",
      journal = {\aj},
         year = 2021,
        month = oct,
       volume = {162},
       number = {4},
          eid = {144},
        pages = {144},
          doi = {10.3847/1538-3881/ac096d},
archivePrefix = {arXiv},
       eprint = {2106.04536},
 primaryClass = {astro-ph.SR},
       adsurl = {https://ui.adsabs.harvard.edu/abs/2021AJ....162..144A}
}

@ARTICLE{2023MNRAS.523.6162L,
       author = {{Lin}, Zitao and {Gan}, Tianjun and {Wang}, Sharon X. and {Shporer}, Avi and {Rabus}, Markus and {Zhou}, George and {Psaridi}, Angelica and {Bouchy}, Fran{\c{c}}ois and {Bieryla}, Allyson and {Latham}, David W. and {Mao}, Shude and {Stassun}, Keivan G. and {Hellier}, Coel and {Howell}, Steve B. and {Ziegler}, Carl and {Caldwell}, Douglas A. and {Clark}, Catherine A. and {Collins}, Karen A. and {Curtis}, Jason L. and {Faherty}, Jacqueline K. and {Gnilka}, Crystal L. and {Grunblatt}, Samuel K. and {Jenkins}, Jon M. and {Johnson}, Marshall C. and {Law}, Nicholas and {Lendl}, Monika and {Littlefield}, Colin and {Lund}, Michael B. and {Lund}, Mikkel N. and {Mann}, Andrew W. and {McDermott}, Scott and {Mishra}, Lokesh and {Mounzer}, Dany and {Paegert}, Martin and {Pritchard}, Tyler and {Ricker}, George R. and {Seager}, Sara and {Srdoc}, Gregor and {Sun}, Qinghui and {Tang}, Jiaxin and {Udry}, St{\'e}phane and {Vanderspek}, Roland and {Watanabe}, David and {Winn}, Joshua N. and {Yu}, Jie},
        title = "{Three low-mass companions around aged stars discovered by TESS}",
      journal = {\mnras},
         year = 2023,
        month = aug,
       volume = {523},
       number = {4},
        pages = {6162-6185},
          doi = {10.1093/mnras/stad1745},
archivePrefix = {arXiv},
       eprint = {2210.13939},
 primaryClass = {astro-ph.SR},
       adsurl = {https://ui.adsabs.harvard.edu/abs/2023MNRAS.523.6162L}
}

@ARTICLE{2022A&A...664A..94P,
       author = {{Psaridi}, Angelica and {Bouchy}, Fran{\c{c}}ois and {Lendl}, Monika and {Grieves}, Nolan and {Stassun}, Keivan G. and {Carmichael}, Theron and {Gill}, Samuel and {Pe{\~n}a Rojas}, Pablo A. and {Gan}, Tianjun and {Shporer}, Avi and {Bieryla}, Allyson and {Brahm}, Rafael and {Christiansen}, Jessie L. and {Crossfield}, Ian J.~M. and {Galland}, Franck and {Hooton}, Matthew J. and {Jenkins}, Jon M. and {Jenkins}, James S. and {Latham}, David W. and {Lund}, Michael B. and {Rodriguez}, Joseph E. and {Ting}, Eric B. and {Udry}, St{\'e}phane and {Ulmer-Moll}, Sol{\`e}ne and {Wittenmyer}, Robert A. and {Zhang}, Yanzhe and {Zhou}, George and {Addison}, Brett and {Cointepas}, Marion and {Collins}, Karen A. and {Collins}, Kevin I. and {Deline}, Adrien and {Dressing}, Courtney D. and {Evans}, Phil and {Giacalone}, Steven and {Heitzmann}, Alexis and {Mireles}, Ismael and {Mounzer}, Dany and {Otegi}, Jon and {Radford}, Don J. and {Rudat}, Alexander and {Schlieder}, Joshua E. and {Schwarz}, Richard P. and {Srdoc}, Gregor and {Stockdale}, Chris and {Suarez}, Olga and {Wright}, Duncan J. and {Zhao}, Yinan},
        title = "{Three new brown dwarfs and a massive hot Jupiter revealed by TESS around early-type stars}",
      journal = {\aap},
         year = 2022,
        month = aug,
       volume = {664},
          eid = {A94},
        pages = {A94},
          doi = {10.1051/0004-6361/202243454},
archivePrefix = {arXiv},
       eprint = {2205.10854},
 primaryClass = {astro-ph.EP},
       adsurl = {https://ui.adsabs.harvard.edu/abs/2022A&A...664A..94P}
}

@ARTICLE{2022MNRAS.514.4944C,
       author = {{Carmichael}, Theron W. and {Irwin}, Jonathan M. and {Murgas}, Felipe and {Pall{\'e}}, Enric and {Stassun}, Keivan G. and {Bartnik}, Matthew and {Collins}, Karen A. and {de Leon}, Jerome and {Esparza-Borges}, Emma and {Fedewa}, Jeremy and {Fong}, William and {Fukui}, Akihiko and {Jenkins}, Jon M. and {Kagetani}, Taiki and {Latham}, David W. and {Lund}, Michael B. and {Mann}, Andrew W. and {Moldovan}, Dan and {Morgan}, Edward H. and {Narita}, Norio and {Painter}, Shane and {Parviainen}, Hannu and {Quintana}, Elisa V. and {Ricker}, George R. and {Schulte}, Jack and {Schwarz}, Richard P. and {Seager}, Sara and {Sokolovsky}, Kirill and {Twicken}, Joseph D. and {Winn}, Joshua N.},
        title = "{TOI-2119: a transiting brown dwarf orbiting an active M-dwarf from NASA's TESS mission}",
      journal = {\mnras},
         year = 2022,
        month = aug,
       volume = {514},
       number = {4},
        pages = {4944-4957},
          doi = {10.1093/mnras/stac1666},
archivePrefix = {arXiv},
       eprint = {2202.08842},
 primaryClass = {astro-ph.SR},
       adsurl = {https://ui.adsabs.harvard.edu/abs/2022MNRAS.514.4944C}
}

@ARTICLE{2023A&A...672L...7K,
       author = {{Khandelwal}, Akanksha and {Sharma}, Rishikesh and {Chakraborty}, Abhijit and {Chaturvedi}, Priyanka and {Ulmer-Moll}, Sol{\`e}ne and {Ciardi}, David R. and {Boyle}, Andrew W. and {Baliwal}, Sanjay and {Bieryla}, Allyson and {Latham}, David W. and {Prasad}, Neelam J.~S.~S.~V. and {Nayak}, Ashirbad and {Lendl}, Monika and {Mordasini}, Christoph},
        title = "{Discovery of a massive giant planet with extreme density around the sub-giant star TOI-4603}",
      journal = {\aap},
         year = 2023,
        month = apr,
       volume = {672},
          eid = {L7},
        pages = {L7},
          doi = {10.1051/0004-6361/202245608},
archivePrefix = {arXiv},
       eprint = {2303.11841},
 primaryClass = {astro-ph.EP},
       adsurl = {https://ui.adsabs.harvard.edu/abs/2023A&A...672L...7K}
}

@ARTICLE{2023AJ....165..218L,
       author = {{Lambert}, Mika and {Bender}, Chad F. and {Kanodia}, Shubham and {Ca{\~n}as}, Caleb I. and {Monson}, Andrew and {Stef{\'a}nsson}, Gudmundur and {Cochran}, William D. and {Everett}, Mark E. and {Gupta}, Arvind F. and {Hearty}, Fred and {Kobulnicky}, Henry A. and {Libby-Roberts}, Jessica E. and {Lin}, Andrea S.~J. and {Mahadevan}, Suvrath and {Ninan}, Joe P. and {Parker}, Brock A. and {Robertson}, Paul and {Schwab}, Christian and {Terrien}, Ryan C.},
        title = "{TOI-5375 B: A Very Low Mass Star at the Hydrogen-burning Limit Orbiting an Early M-type Star}",
      journal = {\aj},
         year = 2023,
        month = may,
       volume = {165},
       number = {5},
          eid = {218},
        pages = {218},
          doi = {10.3847/1538-3881/acc651},
archivePrefix = {arXiv},
       eprint = {2303.16193},
 primaryClass = {astro-ph.SR},
       adsurl = {https://ui.adsabs.harvard.edu/abs/2023AJ....165..218L}
}

@ARTICLE{2011ApJ...726L..19A,
       author = {{Anderson}, D.~R. and {Collier Cameron}, A. and {Hellier}, C. and {Lendl}, M. and {Maxted}, P.~F.~L. and {Pollacco}, D. and {Queloz}, D. and {Smalley}, B. and {Smith}, A.~M.~S. and {Todd}, I. and {Triaud}, A.~H.~M.~J. and {West}, R.~G. and {Barros}, S.~C.~C. and {Enoch}, B. and {Gillon}, M. and {Lister}, T.~A. and {Pepe}, F. and {S{\'e}gransan}, D. and {Street}, R.~A. and {Udry}, S.},
        title = "{WASP-30b: A 61 M $_{Jup}$ Brown Dwarf Transiting a V = 12, F8 Star}",
      journal = {\apjl},
         year = 2011,
        month = jan,
       volume = {726},
       number = {2},
          eid = {L19},
        pages = {L19},
          doi = {10.1088/2041-8205/726/2/L19},
archivePrefix = {arXiv},
       eprint = {1010.3006},
 primaryClass = {astro-ph.SR},
       adsurl = {https://ui.adsabs.harvard.edu/abs/2011ApJ...726L..19A}
}

@ARTICLE{2018MNRAS.481.5091H,
       author = {{Hod{\v{z}}i{\'c}}, Vedad and {Triaud}, Amaury H.~M.~J. and {Anderson}, David R. and {Bouchy}, Fran{\c{c}}ois and {Collier Cameron}, Andrew and {Delrez}, Laetitia and {Gillon}, Micha{\"e}l and {Hellier}, Coel and {Jehin}, Emmanu{\"e}l and {Lendl}, Monika and {Maxted}, Pierre F.~L. and {Pepe}, Francesco and {Pollacco}, Don and {Queloz}, Didier and {S{\'e}gransan}, Damien and {Smalley}, Barry and {Udry}, St{\'e}phane and {West}, Richard},
        title = "{WASP-128b: a transiting brown dwarf in the dynamical-tide regime}",
      journal = {\mnras},
         year = 2018,
        month = dec,
       volume = {481},
       number = {4},
        pages = {5091-5097},
          doi = {10.1093/mnras/sty2512},
archivePrefix = {arXiv},
       eprint = {1807.07557},
 primaryClass = {astro-ph.EP},
       adsurl = {https://ui.adsabs.harvard.edu/abs/2018MNRAS.481.5091H}
}

@ARTICLE{2008CeMDA.101..171F,
       author = {{Ferraz-Mello}, Sylvio and {Rodr{\'\i}guez}, Adri{\'a}n and {Hussmann}, Hauke},
        title = "{Tidal friction in close-in satellites and exoplanets: The Darwin theory re-visited}",
      journal = {Celestial Mechanics and Dynamical Astronomy},
         year = 2008,
        month = may,
       volume = {101},
       number = {1-2},
        pages = {171-201},
          doi = {10.1007/s10569-008-9133-x},
archivePrefix = {arXiv},
       eprint = {0712.1156},
 primaryClass = {astro-ph},
       adsurl = {https://ui.adsabs.harvard.edu/abs/2008CeMDA.101..171F}
}

@ARTICLE{2024AJ....168..145F,
       author = {{Ferreira dos Santos}, Thiago and {Rice}, Malena and {Wang}, Xian-Yu and {Wang}, Songhu},
        title = "{SOLES XII. The Aligned Orbit of TOI-2533 b, a Transiting Brown Dwarf Orbiting an F8-type Star}",
      journal = {\aj},
         year = 2024,
        month = oct,
       volume = {168},
       number = {4},
          eid = {145},
        pages = {145},
          doi = {10.3847/1538-3881/ad6b7f},
archivePrefix = {arXiv},
       eprint = {2408.00725},
 primaryClass = {astro-ph.EP},
       adsurl = {https://ui.adsabs.harvard.edu/abs/2024AJ....168..145F}
}

@ARTICLE{2009ApJ...698L..42G,
       author = {{Greenberg}, Richard},
        title = "{Frequency Dependence of Tidal q}",
      journal = {\apjl},
         year = 2009,
        month = jun,
       volume = {698},
       number = {1},
        pages = {L42-L45},
          doi = {10.1088/0004-637X/698/1/L42},
       adsurl = {https://ui.adsabs.harvard.edu/abs/2009ApJ...698L..42G}
}

@INCOLLECTION{2016cole.book..169F,
       author = {{Ferraz-Mello}, S. and {CoRot Team}},
        title = "{III.9-2 Tidal evolution of CoRoT massive planets and brown dwarfs and of their host stars}",
    booktitle = {The CoRoT Legacy Book: The Adventure of the Ultra High Precision Photometry from Space},
         year = 2016,
        pages = {169},
          doi = {10.1051/978-2-7598-1876-1.c239},
       adsurl = {https://ui.adsabs.harvard.edu/abs/2016cole.book..169F}
}

@ARTICLE{2025arXiv250709461G,
       author = {{Gan}, Tianjun and {Cadieux}, Charles and {Ida}, Shigeru and {Wang}, Sharon X. and {Mao}, Shude and {Lin}, Zitao and {Stassun}, Keivan G. and {Burgasser}, Adam J. and {Howell}, Steve B. and {Clark}, Catherine A. and {Strakhov}, Ivan A. and {Benni}, Paul and {Ricker}, George R. and {Vanderspek}, Roland and {Latham}, David W. and {Seager}, Sara and {Winn}, Joshua N. and {Jenkins}, Jon M. and {Arnold}, Luc and {Artigau}, {\'E}tienne and {Charbonneau}, David and {Collins}, Karen A. and {Cook}, Neil J. and {de Beurs}, Zo{\"e} L. and {Deveny}, Sarah J. and {Doty}, John P. and {Doyon}, Ren{\'e} and {Littlefield}, Colin and {Pritchard}, Tyler and {Ross}, Gabrielle and {Shporer}, Avi and {Theissen}, Christopher R. and {Tofflemire}, Benjamin M. and {Vanderburg}, Andrew and {Watanabe}, David},
        title = "{A New Brown Dwarf Orbiting an M star and An Investigation on the Eccentricity Distribution of Transiting Long-Period Brown Dwarfs}",
      journal = {arXiv e-prints},
         year = 2025,
        month = jul,
          eid = {arXiv:2507.09461},
        pages = {arXiv:2507.09461},
          doi = {10.48550/arXiv.2507.09461},
archivePrefix = {arXiv},
       eprint = {2507.09461},
 primaryClass = {astro-ph.EP},
       adsurl = {https://ui.adsabs.harvard.edu/abs/2025arXiv250709461G}
}

@ARTICLE{1910AN....183..345V,
       author = {{von Zeipel}, H.},
        title = "{Sur l'application des s{\'e}ries de M. Lindstedt {\`a} l'{\'e}tude du mouvement des com{\`e}tes p{\'e}riodiques}",
      journal = {Astronomische Nachrichten},
         year = 1910,
        month = mar,
       volume = {183},
       number = {22},
        pages = {345},
          doi = {10.1002/asna.19091832202},
       adsurl = {https://ui.adsabs.harvard.edu/abs/1910AN....183..345V}
}

@ARTICLE{2021AJ....161...84M,
       author = {{Miyazaki}, Shota and {Johnson}, Samson A. and {Sumi}, Takahiro and {Penny}, Matthew T. and {Koshimoto}, Naoki and {Yamawaki}, Tsubasa},
        title = "{Revealing Short-period Exoplanets and Brown Dwarfs in the Galactic Bulge Using the Microlensing Xallarap Effect with the Nancy Grace Roman Space Telescope}",
      journal = {\aj},
         year = 2021,
        month = feb,
       volume = {161},
       number = {2},
          eid = {84},
        pages = {84},
          doi = {10.3847/1538-3881/abcec2},
archivePrefix = {arXiv},
       eprint = {2010.10315},
 primaryClass = {astro-ph.EP},
       adsurl = {https://ui.adsabs.harvard.edu/abs/2021AJ....161...84M}
}

@ARTICLE{2023arXiv230612363H,
       author = {{Holwerda}, Benne and {Pirzkal}, Nor and {Burgasser}, Adam and {Hsu}, Chih-Chun},
        title = "{Detection and characterization of M-L-T-Y dwarfs belonging to the Milky Way Disks and Stellar Halo with the Roman Space Telescope}",
      journal = {arXiv e-prints},
         year = 2023,
        month = jun,
          eid = {arXiv:2306.12363},
        pages = {arXiv:2306.12363},
          doi = {10.48550/arXiv.2306.12363},
archivePrefix = {arXiv},
       eprint = {2306.12363},
 primaryClass = {astro-ph.GA},
       adsurl = {https://ui.adsabs.harvard.edu/abs/2023arXiv230612363H}
}
\bibliographystyle{aasjournalv7}

\appendix 

\section{Other Eccentricity Distributions}

\subsection{Jovian Planets}\label{app:jovianplanets}

To contextualise and compare our findings from $\S$\ref{sec:BDs}, we applied the same HBM model introduced in $\S$\ref{sec:genframework} to the set of Jovian planets ($M > 0.3~M_{\rm Jup}$) reported by the {\it California Legacy Survey} \citep{2021ApJS..255....8R}. We filtered confirmed planets with $e > 0$ and masses below the deuterium-burning limit of $13~M_{\rm Jup}$, resulting in a sample of 167 planets. Applying a KS test to search for a mass boundary that maximises the contrast in eccentricity behaviour, we identified a threshold at $M = 5~M_{\rm Jup}$, for which the test returns $D = 0.43$ with a $p-$value of 0.004, which indicates a statistically significant difference in the eccentricity distributions across this division. Specifically, the higher-mass ($M > 5~M_{\rm Jup}$) planets are best fit by a Beta distribution with peak eccentricity $e\sim0.24$ (see Figure \ref{fig:BDs_Beta}), in rough agreement with the findings of \citet{blunt2026evidence}. Adopting this threshold yields a massive subgroup totalling $N = 15$ and a low-mass one with $N = 152$. Therefore, in principle, distinct physical processes may shape the orbital architectures of low-mass and massive Jovians. 

We then inferred the shape of the underlying eccentricity distributions for each subgroup using Beta distribution hyperparameters ($\alpha,\beta)$ within our HBM framework. We found the average eccentricity for low-mass Jovians to be $\langle e \rangle_{\rm SP, GPs} = 0.20\pm0.02$ with optimal Beta parameters $\mathcal{B}_{\rm LM,J} = (0.844^{+0.087}_{-0.081}, 3.313^{+0.421}_{-0.388})$, indicating a slight tendency towards circular orbits ($\alpha < 1, \beta > 1$). For massive Jovians, however, the average eccentricity was found to be $\langle e \rangle_{\rm M,J} = 0.36\pm0.14$, with parameters $\mathcal{B}_{\rm M,~J} = (1.545^{+0.545}_{-0.432}, 2.716^{+1.013}_{-0.804})$, consistent with a broader eccentricity distribution ($\alpha,\beta > 1$) (see Figure \ref{fig:BDs_Beta}).

\subsection{Wide \emph{Gaia} Stellar Binaries}\label{app:binaries}

We also applied our HBM approach to the sample of \emph{Gaia}-wide binaries reported in \cite{2022MNRAS.512.3383H}, aiming to provide another comparison point for the BD eccentricity distributions derived in $\S$\ref{sec:BDs}. Following Section \ref{app:jovianplanets}, the sample was divided into binaries with semi-major axis $a < 100$ au ($N = 62$ binaries) and $a > 100$ au ($N = 9559$), for which the KS test diagnostic $D = 0.38$ and $p = 10^{-8}$ suggest distinct physical processes acting upon each population. For the $a < 100$ au binaries, the results indicate a bimodal U-shaped eccentricity distribution ($\mathcal{B}_{\rm \small a < 100 au} = (0.704^{+0.123}_{-0.110}, 0.573^{+0.095}_{-0.085})$, with a strong but asymmetric preference toward near-circular orbits ($\langle e\rangle_{\rm \small a \leq 100 au} = 0.55\pm0.11$, $\alpha < 1$, $\beta < 1$, but $\alpha < \beta$). In contrast, $a > 100$ au binaries exhibit a heavily skewed distribution toward high eccentricity ($\mathcal{B}_{\rm \small a > 100 au} = 3.047^{+0.049}_{-0.047}, 0.675^{+0.008}_{-0.008}$; i.e., $\alpha > 1$ and $\beta < 1$), with $\langle e \rangle_{\rm \small a \geq 100 au} = 0.82\pm0.01$. This corresponds to a Beta distribution with $\alpha > 1$ and $\beta \lesssim 1$, consistent with our widest analysed binaries, which likely dominate this very large sample (see Figure \ref{fig:BDs_Beta}). The high typical eccentricities that we obtain for the $a>100$ au population agree with the results of \citet{2022MNRAS.512.3383H}, given that the sample is dominated by wide-separation ($s>1000$ au) binaries with superthermal {($f(e) \propto e^\eta,~\eta > 1$)} eccentricity distributions.

\section{Keplerian Models for the Transiting BDs and Very Low-Mass Companions Near the Hydrogen-Burning Limit}\label{app:keplerian_models}

Phase-folded Keplerian orbital models fitted to the radial velocity data for the sample are presented below. All data are public and were taken directly from each system's discovery paper (see $\S$\ref{sec:re-fit}). 

\begin{deluxetable*}{l|ccccccc}
\tablecaption{Orbital and Physical Radial Velocity-Derived Parameters of Transiting Brown Dwarfs and {Very Low-Mass Stars}.}
\label{tab:bdparams}
\tabletypesize{\footnotesize}
\tablehead{
\colhead{{\bf ID}} & \colhead{{\bf $P$}} & \colhead{{\bf $e$}} & \colhead{{\bf $\omega$}} & \colhead{{\bf $m\sin{i}$}} & \colhead{{\bf $a$}} & \colhead{{\bf $M_\star$}} & \colhead{{\bf $T_{\rm eff}$}} \\
\colhead{} & \colhead{(d)} & \colhead{} & \colhead{(deg)} & \colhead{($M_{\rm Jup}$)} & \colhead{(km s$^{-1}$)} & \colhead{($M_\odot$)} & \colhead{(K)}
}
\startdata
TOI-4603 b & 7.254 $\pm$ $3.540\times10^{-3}$ & 0.315 $\pm$ 0.060 & 18.70 $\pm$ 7.66 & 12.83 $\pm$ 0.67 & 0.96 $\pm$ 0.05 & 1.77 $\pm$ 0.06 & 6264 $\pm$ 95 \\
HATS-70 b & 1.880 $\pm$ $5.448\times10^{-3}$ & 0.689 $\pm$ 0.330 & 80.97 $\pm$ 92.77 & 17.31 $\pm$ 6.36 & 2.82 $\pm$ 3.18 & 1.78 $\pm$ 0.000 & 7930 $\pm$ 820 \\
TOI-2119 b & 7.201 $\pm$ 0.000 & 0.885 $\pm$ 0.059 & -13.73 $\pm$ 16.18 & 19.05 $\pm$ 8.02 & 6.15 $\pm$ 3.83 & 0.53 $\pm$ 0.02 & 3621 $\pm$ 48 \\
TOI-1278 B & 14.475 $\pm$ $1.995\times10^{-3}$ & 0.009 $\pm$ $5.261\times10^{-3}$ & -89.41 $\pm$ 42.01 & 19.06 $\pm$ 0.11 & 2.32 $\pm$ 0.01 & 0.55 $\pm$ 0.000 & 3799 $\pm$ 42 \\
Kepler-39 b & 21.646 $\pm$ 0.215 & 0.136 $\pm$ 0.027 & 123.85 $\pm$ 15.30 & 20.19 $\pm$ 0.97 & 1.24 $\pm$ 0.04 & 1.29 $\pm$ 0.07 & 6350 $\pm$ 100 \\
GPX-1 b & 1.745 $\pm$ $5.873\times10^{-4}$ & 0.083 $\pm$ 0.075 & -47.49 $\pm$ 98.21 & 20.69 $\pm$ 1.92 & 2.47 $\pm$ 0.23 & 1.68 $\pm$ 0.000 & 7000 $\pm$ 200 \\
CoRoT-3 b & 4.257 $\pm$ $3.961\times10^{-4}$ & 0.010 $\pm$ 0.012 & 43.51 $\pm$ 113.60 & 22.43 $\pm$ 1.05 & 2.18 $\pm$ 0.03 & 1.44 $\pm$ 0.09 & 6710 $\pm$ 140 \\
KELT-1 b & 1.218 $\pm$ $2.214\times10^{-4}$ & 0.016 $\pm$ 0.014 & -61.08 $\pm$ 147.64 & 27.27 $\pm$ 0.38 & 4.22 $\pm$ 0.06 & 1.34 $\pm$ 0.000 & 6516 $\pm$ 49 \\
NLTT 41135 b & 2.889 $\pm$ 0.000 & 0.006 $\pm$ $4.595\times10^{-3}$ & -107.51 $\pm$ 79.61 & 34.20 $\pm$ 2.74 & 13.25 $\pm$ 0.07 & 0.19 $\pm$ 0.02 & 3230 $\pm$ 130 \\
CWW 89 Ab & 5.293 $\pm$ $9.395\times10^{-4}$ & 0.192 $\pm$ $3.348\times10^{-3}$ & -14.35 $\pm$ 0.87 & 37.03 $\pm$ 0.11 & 4.27 $\pm$ 0.01 & 1.01 $\pm$ 0.000 & 5850 $\pm$ 85 \\
WASP-128 b & 2.209 $\pm$ 0.000 & 0.014 $\pm$ 0.012 & 94.44 $\pm$ 145.93 & 37.72 $\pm$ 1.01 & 5.22 $\pm$ 0.07 & 1.16 $\pm$ 0.04 & 5950 $\pm$ 50 \\
KOI-205 b & 11.720 $\pm$ 0.000 & 0.018 $\pm$ 0.012 & -29.91 $\pm$ 47.76 & 41.12 $\pm$ 1.19 & 3.77 $\pm$ 0.05 & 0.93 $\pm$ 0.03 & 5237 $\pm$ 60 \\
TOI-1406 b & 10.573 $\pm$ $6.243\times10^{-3}$ & 0.027 $\pm$ 0.013 & 126.72 $\pm$ 36.25 & 46.40 $\pm$ 2.21 & 3.76 $\pm$ 0.06 & 1.18 $\pm$ 0.08 & 6290 $\pm$ 100 \\
TOI-503 b & 3.677 $\pm$ 0.000 & 0.006 $\pm$ $3.625\times10^{-3}$ & -129.49 $\pm$ 162.84 & 53.11 $\pm$ 1.19 & 4.64 $\pm$ 0.02 & 1.80 $\pm$ 0.06 & 7650 $\pm$ 159 \\
TOI-852 b & 4.946 $\pm$ $1.020\times10^{-3}$ & 0.005 $\pm$ $7.127\times10^{-3}$ & 9.44 $\pm$ 98.17 & 53.89 $\pm$ 1.91 & 5.19 $\pm$ 0.04 & 1.32 $\pm$ 0.05 & 5768 $\pm$ 84 \\
EPIC 212036875 b & 5.170 $\pm$ $6.500\times10^{-5}$ & 0.133 $\pm$ $1.895\times10^{-3}$ & 163.24 $\pm$ 1.04 & 54.31 $\pm$ 0.11 & 5.29 $\pm$ 0.01 & 1.29 $\pm$ 0.000 & 6238 $\pm$ 60 \\
TOI-263 b & 0.557 $\pm$ 0.000 & 0.150 $\pm$ 0.016 & -55.70 $\pm$ 8.09 & 54.49 $\pm$ 3.90 & 21.73 $\pm$ 0.78 & 0.44 $\pm$ 0.04 & 3471 $\pm$ 33 \\
AD 3116 b & 1.983 $\pm$ 0.000 & 0.133 $\pm$ 0.025 & -5.35 $\pm$ 13.38 & 56.29 $\pm$ 3.49 & 18.92 $\pm$ 0.66 & 0.28 $\pm$ 0.02 & 3165 $\pm$ 106 \\
TOI-811 b & 25.178 $\pm$ 0.034 & 0.347 $\pm$ 0.095 & 136.50 $\pm$ 11.51 & 57.37 $\pm$ 4.44 & 3.45 $\pm$ 0.30 & 1.32 $\pm$ 0.07 & 6107 $\pm$ 77 \\
CoRoT-33 b & 5.819 $\pm$ $1.089\times10^{-4}$ & 0.071 $\pm$ $4.606\times10^{-3}$ & 176.50 $\pm$ 179.06 & 58.87 $\pm$ 1.81 & 7.06 $\pm$ 0.03 & 0.86 $\pm$ 0.04 & 5225 $\pm$ 80 \\
WASP-30 b & 4.157 $\pm$ $6.389\times10^{-4}$ & 0.003 $\pm$ $2.428\times10^{-3}$ & -77.30 $\pm$ 95.23 & 60.04 $\pm$ 0.89 & 6.63 $\pm$ 0.02 & 1.17 $\pm$ 0.03 & 6201 $\pm$ 97 \\
LHS 6343c & 12.600 $\pm$ 0.000 & 0.088 $\pm$ 0.018 & 14.81 $\pm$ 15.28 & 61.50 $\pm$ 1.55 & 9.42 $\pm$ 0.15 & 0.37 $\pm$ 0.01 & 3130 $\pm$ 20 \\
TOI-569 b & 6.554 $\pm$ $5.772\times10^{-4}$ & 0.003 $\pm$ $1.724\times10^{-3}$ & 21.33 $\pm$ 36.86 & 63.58 $\pm$ 1.04 & 5.88 $\pm$ 0.01 & 1.21 $\pm$ 0.03 & 5705 $\pm$ 76 \\
NGTS-7A & 0.670 $\pm$ 0.000 & 0.000 $\pm$ 0.000 & 0.000 $\pm$ 0.000 & 63.74 $\pm$ 7.00 & 22.14 $\pm$ 1.11 & 0.48 $\pm$ 0.07 & 3359 $\pm$ 97 \\
TOI-629 b & 8.718 $\pm$ $9.097\times10^{-4}$ & 0.284 $\pm$ 0.018 & 9.22 $\pm$ 3.91 & 64.78 $\pm$ 3.09 & 3.92 $\pm$ 0.09 & 2.16 $\pm$ 0.13 & 9100 $\pm$ 200 \\
CoRoT-15 b & 3.060 $\pm$ $6.314\times10^{-4}$ & 0.017 $\pm$ 0.016 & 23.31 $\pm$ 67.15 & 65.19 $\pm$ 4.08 & 7.35 $\pm$ 0.14 & 1.32 $\pm$ 0.12 & 6350 $\pm$ 200 \\
HIP33609 & 39.470 $\pm$ $6.105\times10^{-3}$ & 0.555 $\pm$ 0.275 & 148.53 $\pm$ 80.95 & 65.43 $\pm$ 16.17 & 2.70 $\pm$ 9.39 & 2.38 $\pm$ 0.10 & 10400 $\pm$ 700 \\
TOI-1982 b & 17.182 $\pm$ 0.041 & 0.286 $\pm$ 0.345 & -84.80 $\pm$ 14.34 & 65.78 $\pm$ 4.56 & 4.20 $\pm$ 7.89 & 1.41 $\pm$ 0.08 & 6325 $\pm$ 110 \\
KOI-415 b & 166.699 $\pm$ 0.068 & 0.743 $\pm$ 0.028 & 28.94 $\pm$ 7.16 & 65.96 $\pm$ 7.01 & 3.64 $\pm$ 0.33 & 0.94 $\pm$ 0.06 & 5810 $\pm$ 80 \\
EPIC 201702477 b & 40.643 $\pm$ 0.047 & 0.229 $\pm$ $5.135\times10^{-3}$ & -162.86 $\pm$ 1.27 & 67.01 $\pm$ 1.66 & 4.26 $\pm$ 0.02 & 0.87 $\pm$ 0.03 & 5517 $\pm$ 70 \\
LP 261-75 b & 1.882 $\pm$ $3.720\times10^{-4}$ & 0.015 $\pm$ 0.014 & -9.89 $\pm$ 123.57 & 68.68 $\pm$ 3.45 & 21.87 $\pm$ 0.55 & 0.30 $\pm$ 0.02 & 3100 $\pm$ 50 \\
TOI-2336 & 7.709 $\pm$ $6.526\times10^{-3}$ & 0.014 $\pm$ 0.440 & -62.89 $\pm$ 126.14 & 69.83 $\pm$ 12.47 & 5.54 $\pm$ 9.12 & 1.40 $\pm$ 0.07 & 6433 $\pm$ 84 \\
NGTS-28A b & 1.254 $\pm$ $5.945\times10^{-3}$ & 0.095 $\pm$ 0.098 & -98.21 $\pm$ 116.09 & 70.31 $\pm$ 6.69 & 18.17 $\pm$ 1.39 & 0.56 $\pm$ 0.02 & 3626 $\pm$ 47 \\
NGTS-19 b & 17.844 $\pm$ $5.287\times10^{-3}$ & 0.377 $\pm$ 0.012 & -31.94 $\pm$ 2.82 & 70.83 $\pm$ 2.45 & 6.49 $\pm$ 0.09 & 0.81 $\pm$ 0.04 & 4716 $\pm$ 39 \\
TOI-2490 b & 60.333 $\pm$ 0.000 & 0.752 $\pm$ 0.118 & -143.20 $\pm$ 66.41 & 70.87 $\pm$ 22.28 & 5.35 $\pm$ 13.84 & 1.00 $\pm$ 0.03 & 5558 $\pm$ 80 \\
TOI-148 b & 4.867 $\pm$ 0.000 & 0.108 $\pm$ 0.064 & 85.86 $\pm$ 164.10 & 70.99 $\pm$ 4.50 & 8.05 $\pm$ 0.40 & 1.03 $\pm$ 0.06 & 5836 $\pm$ 286 \\
TOI-2533 & 6.679 $\pm$ $5.983\times10^{-3}$ & 0.250 $\pm$ 0.013 & 102.36 $\pm$ 2.81 & 72.58 $\pm$ 3.69 & 7.59 $\pm$ 0.10 & 1.02 $\pm$ 0.06 & 6228 $\pm$ 114 \\
\enddata
\tablecomments{Columns indicate orbital period $P$ in days, orbital eccentricity $e$, argument of periastron $\omega$ in degrees, the BD's minimum mass $m\sin{i}$ in Jupiter masses, stellar mass $M_\star$ in Solar masses, and the host star's effective temperature $T_{\rm eff}$ in Kelvin. Values in the two final columns were extracted from the original BD discovery studies.}
\end{deluxetable*}

\begin{deluxetable*}{l|ccccccc}
\label{tab:bdparams}
\tabletypesize{\footnotesize}
\tablehead{
\colhead{{\bf ID}} & \colhead{{\bf $P$}} & \colhead{{\bf $e$}} & \colhead{{\bf $\omega$}} & \colhead{{\bf $m\sin{i}$}} & \colhead{{\bf $a$}} & \colhead{{\bf $M_\star$}} & \colhead{{\bf $T_{\rm eff}$}} \\
\colhead{} & \colhead{(d)} & \colhead{} & \colhead{(deg.)} & \colhead{($M_{\rm Jup}$)} & \colhead{(km s$^{-1}$)} & \colhead{($M_\odot$)} & \colhead{(K)}
}
\startdata
KOI-189 b & 30.368 $\pm$ 0.018 & 0.273 $\pm$ $3.548\times10^{-3}$ & -119.09 $\pm$ 0.90 & 78.09 $\pm$ 3.35 & 5.95 $\pm$ 0.03 & 0.76 $\pm$ 0.05 & 4952 $\pm$ 40 \\
TOI-2521 & 5.550 $\pm$ 0.018 & 0.013 $\pm$ 0.012 & 110.75 $\pm$ 124.83 & 78.12 $\pm$ 3.72 & 8.78 $\pm$ 0.10 & 0.95 $\pm$ 0.06 & 5625 $\pm$ 74 \\
CoRoT-34 b & 2.119 $\pm$ $6.026\times10^{-3}$ & 0.265 $\pm$ 0.091 & 86.89 $\pm$ 38.67 & 79.40 $\pm$ 9.03 & 9.07 $\pm$ 1.04 & 1.66 $\pm$ 0.10 & 7820 $\pm$ 160 \\
TOI-2543 b & 7.543 $\pm$ $4.464\times10^{-3}$ & 0.017 $\pm$ $6.695\times10^{-3}$ & -15.90 $\pm$ 29.38 & 80.77 $\pm$ 3.35 & 6.78 $\pm$ 0.04 & 1.29 $\pm$ 0.08 & 6060 $\pm$ 82 \\
TOI-587 b & 8.040 $\pm$ 0.000 & 0.078 $\pm$ 0.074 & -4.45 $\pm$ 135.64 & 81.19 $\pm$ 7.69 & 4.64 $\pm$ 0.39 & 2.32 $\pm$ 0.14 & 10400 $\pm$ 300 \\
TOI-5375 & 1.720 $\pm$ $1.010\times10^{-3}$ & 0.021 $\pm$ 0.012 & -134.84 $\pm$ 47.32 & 83.36 $\pm$ 8.39 & 17.54 $\pm$ 0.20 & 0.64 $\pm$ 0.10 & 3885 $\pm$ 25 \\
ELM-J0555-57A b & 7.758 $\pm$ $6.585\times10^{-4}$ & 0.110 $\pm$ 0.024 & 134.03 $\pm$ 13.32 & 83.92 $\pm$ 4.73 & 7.44 $\pm$ 0.13 & 1.18 $\pm$ 0.08 & 6368 $\pm$ 124 \\
OGLE-TR-123 & 1.804 $\pm$ 0.000 & 0.000 $\pm$ 0.000 & 0.000 $\pm$ 0.000 & 83.98 $\pm$ 26.41 & 11.54 $\pm$ 3.09 & 1.29 $\pm$ 0.26 & 6700 $\pm$ 300 \\
TOI-746 b & 10.980 $\pm$ $3.670\times10^{-3}$ & 0.198 $\pm$ $4.525\times10^{-3}$ & 116.26 $\pm$ 1.36 & 84.49 $\pm$ 3.36 & 7.56 $\pm$ 0.02 & 0.98 $\pm$ 0.06 & 5593 $\pm$ 215 \\
TOI-1608 b & 2.470 $\pm$ $1.187\times10^{-3}$ & 0.027 $\pm$ 0.020 & -17.31 $\pm$ 52.69 & 88.57 $\pm$ 3.54 & 10.65 $\pm$ 0.16 & 1.31 $\pm$ 0.07 & 6028 $\pm$ 82 \\
TOI-694 b & 48.049 $\pm$ 0.012 & 0.519 $\pm$ 0.228 & 31.79 $\pm$ 54.81 & 89.01 $\pm$ 18.00 & 5.64 $\pm$ 11.98 & 0.97 $\pm$ 0.05 & 5496 $\pm$ 87 \\
TOI-681 b & 15.778 $\pm$ 0.017 & 0.099 $\pm$ 0.057 & -76.01 $\pm$ 30.59 & 90.33 $\pm$ 4.29 & 5.31 $\pm$ 0.21 & 1.54 $\pm$ 0.06 & 7440 $\pm$ 150 \\
KOI-607 b & 5.894 $\pm$ $1.523\times10^{-3}$ & 0.385 $\pm$ 0.011 & 114.87 $\pm$ 1.36 & 94.24 $\pm$ 3.27 & 10.90 $\pm$ 0.14 & 0.99 $\pm$ 0.05 & 5418 $\pm$ 87 \\
J1219-39 b & 6.760 $\pm$ $6.350\times10^{-6}$ & 0.056 $\pm$ $3.962\times10^{-4}$ & 20.98 $\pm$ 0.40 & 95.58 $\pm$ 2.22 & 10.83 $\pm$ 0.01 & 0.83 $\pm$ 0.03 & 5412 $\pm$ 81 \\
\enddata
\tablecomments{(Continued).}
\end{deluxetable*}

\begin{figure*}[h]
    \centering
    \includegraphics[width=0.32\linewidth]{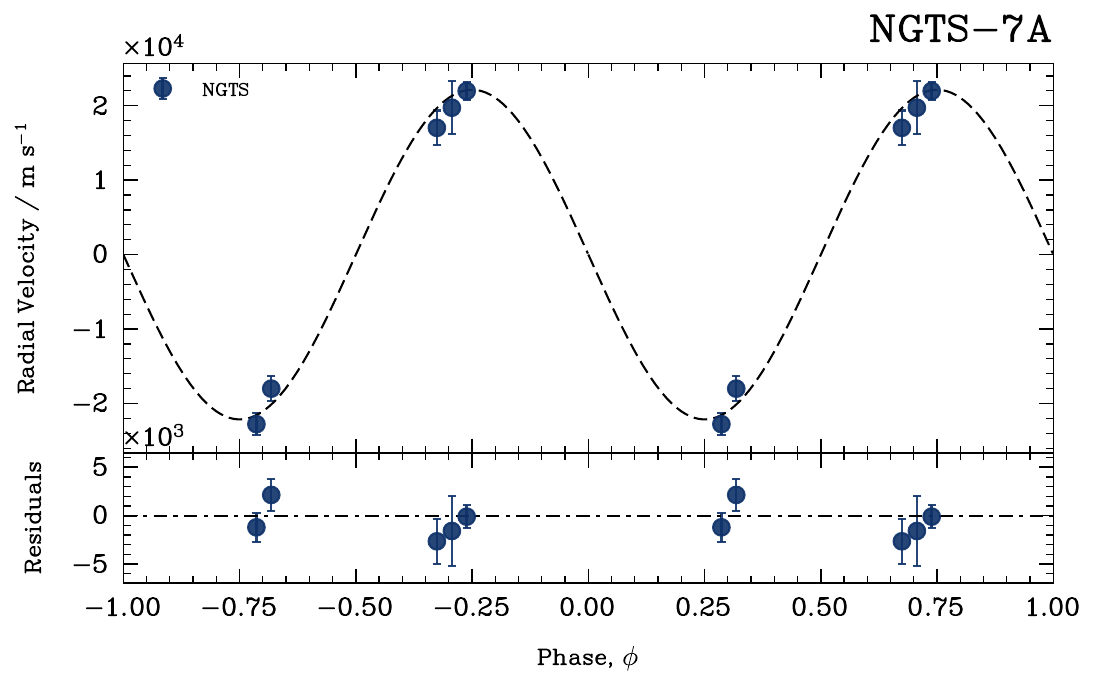}
    \includegraphics[width=0.32\linewidth]{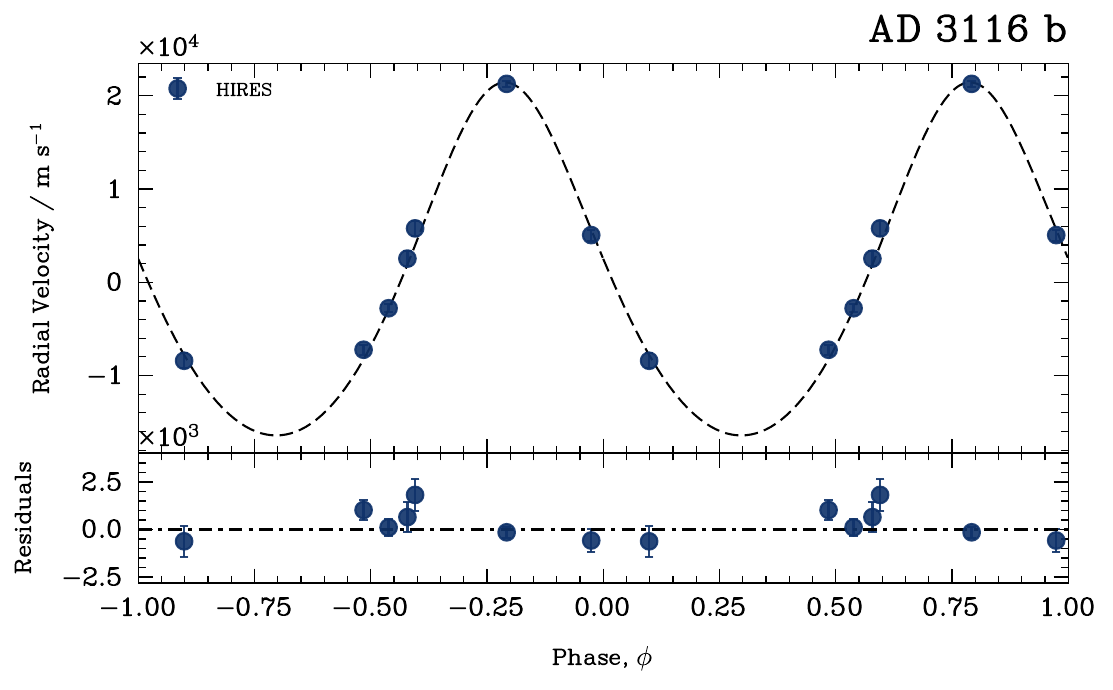}
    \includegraphics[width=0.32\linewidth]{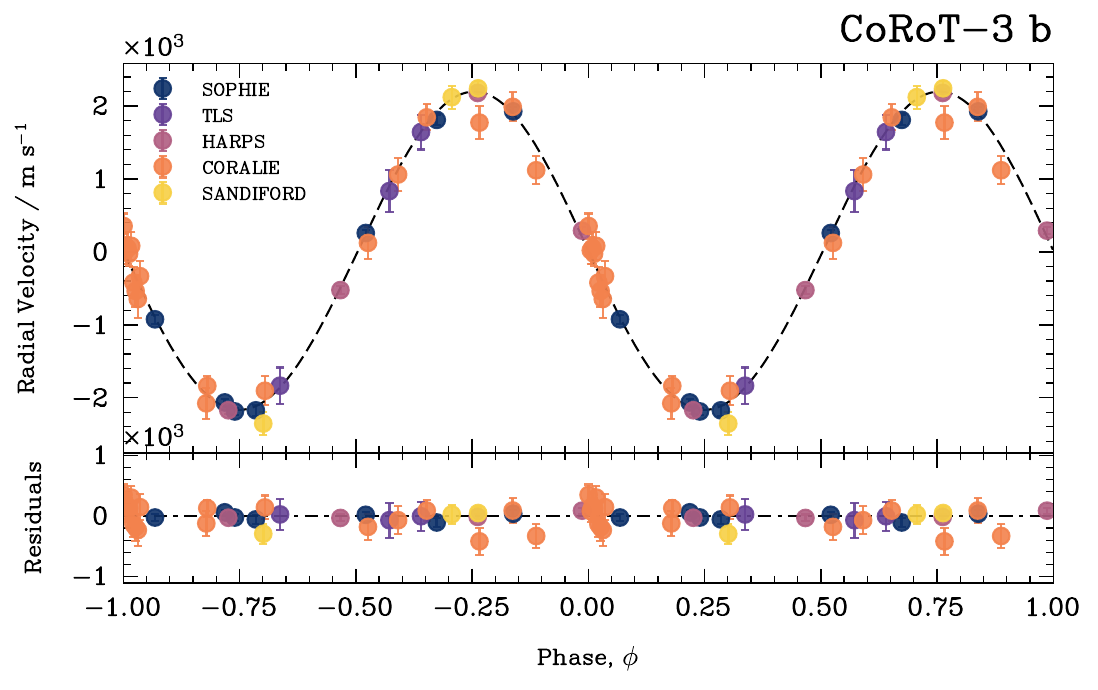} \\
    \includegraphics[width=0.32\linewidth]{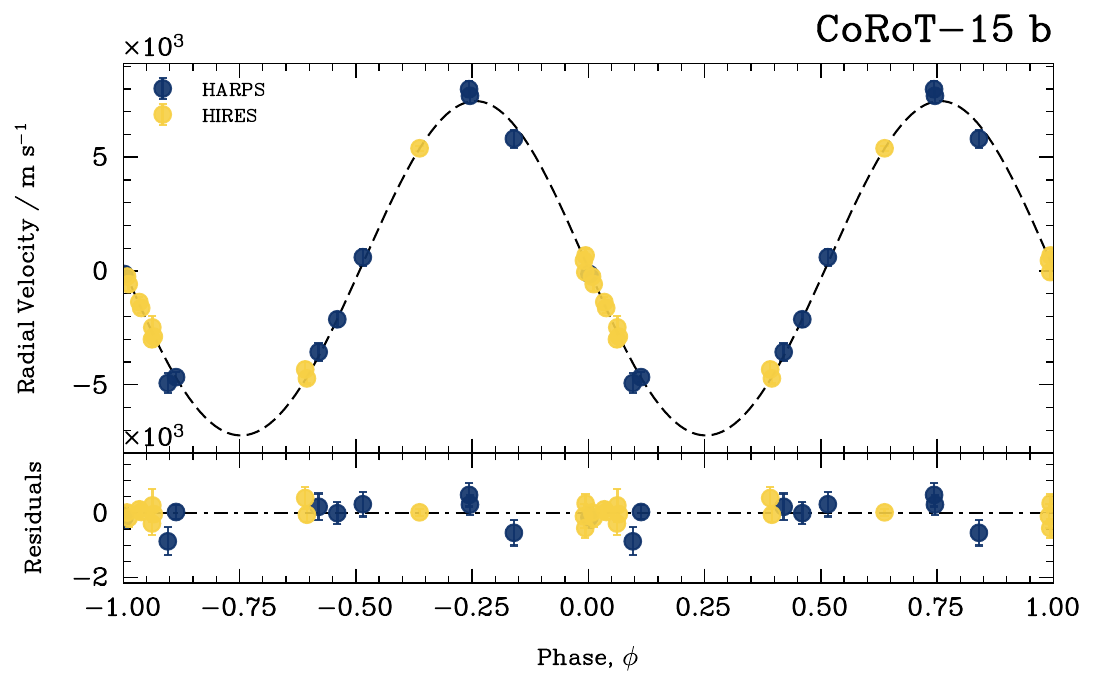}
    \includegraphics[width=0.32\linewidth]{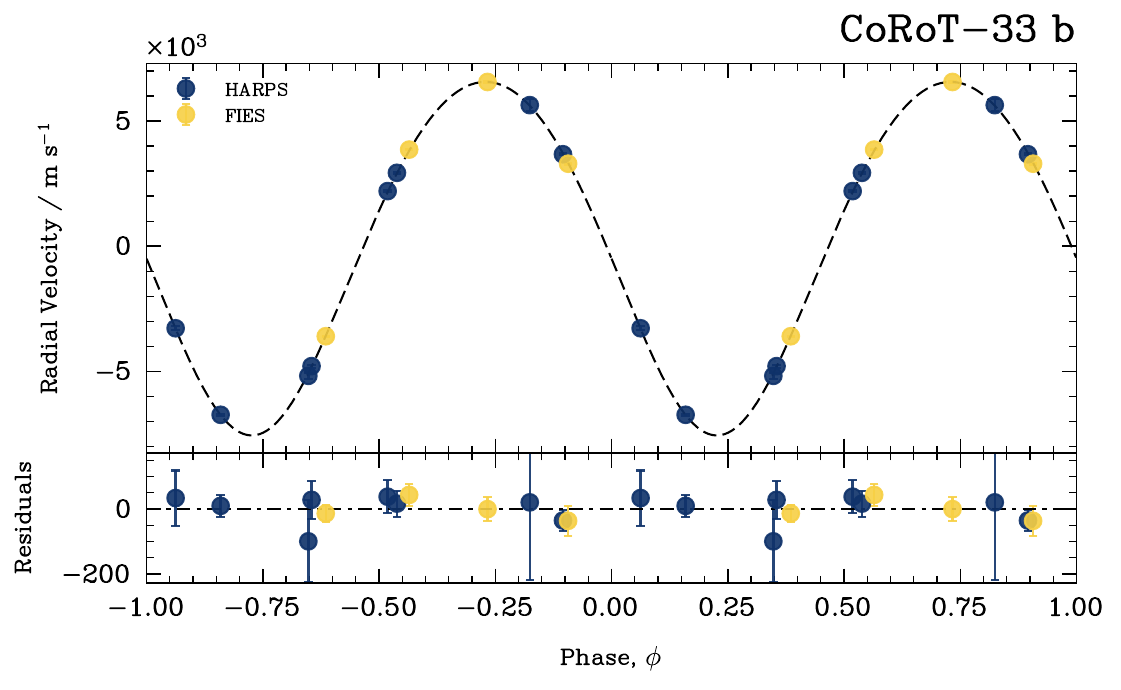}
    \includegraphics[width=0.32\linewidth]{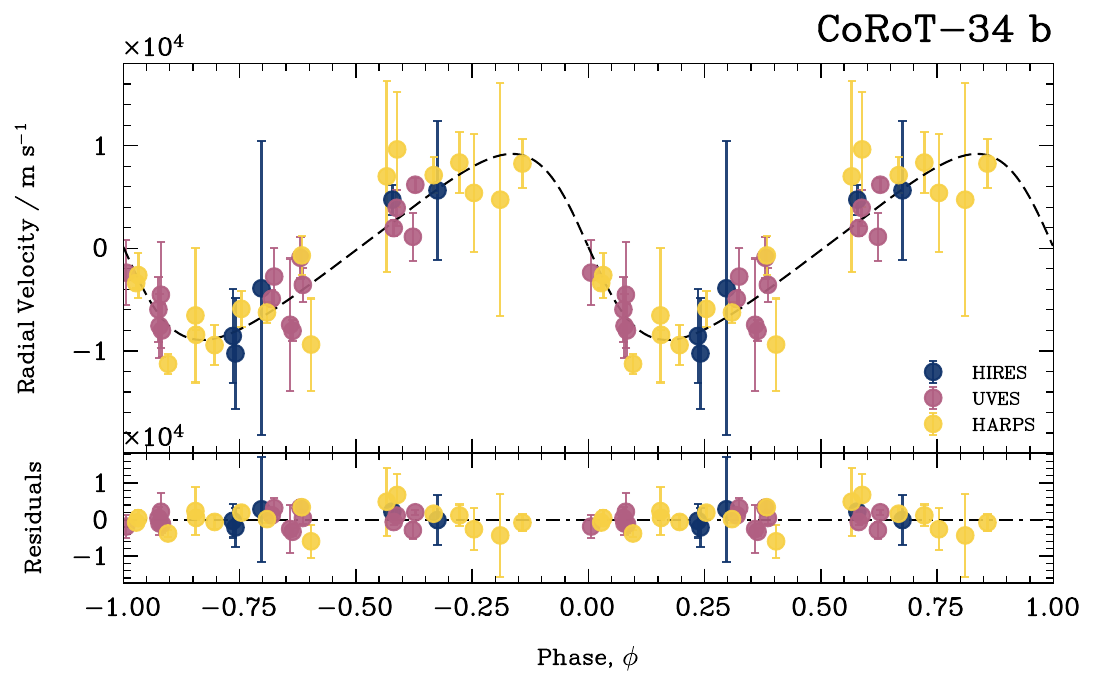} \\
    \includegraphics[width=0.32\linewidth]{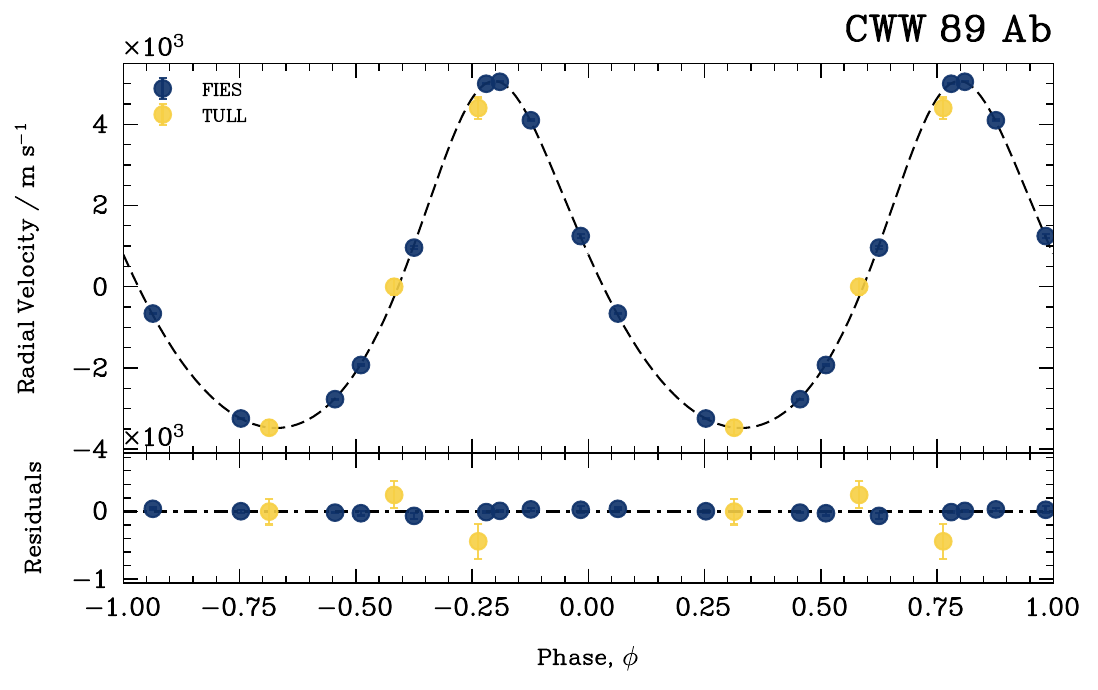}
    \includegraphics[width=0.32\linewidth]{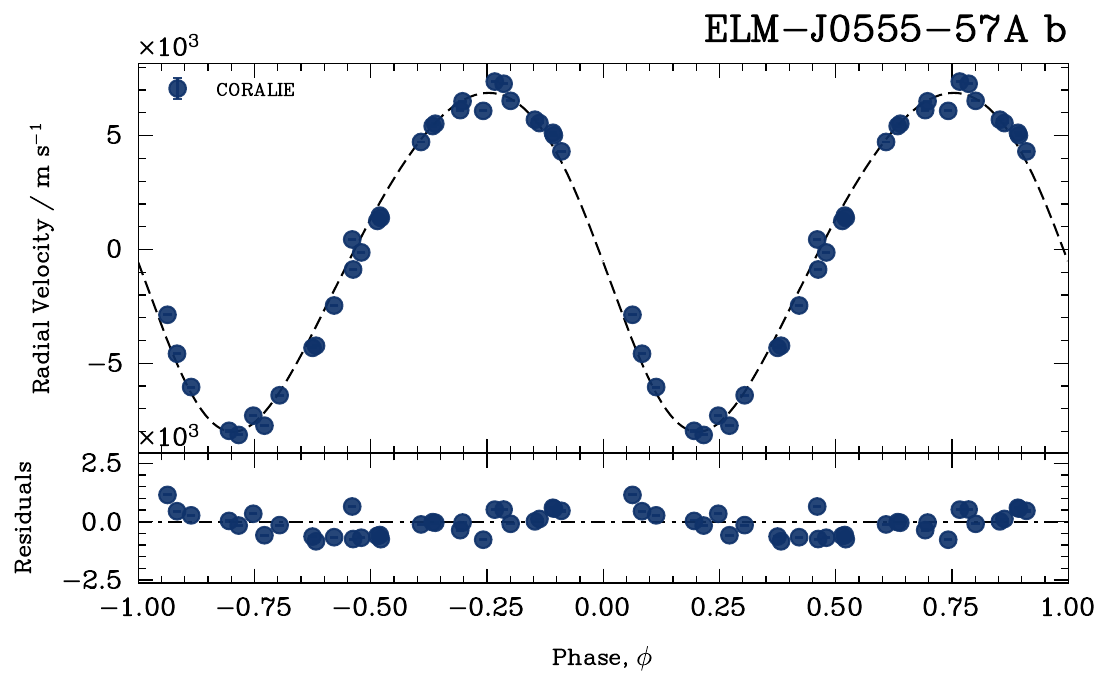}
    \includegraphics[width=0.32\linewidth]{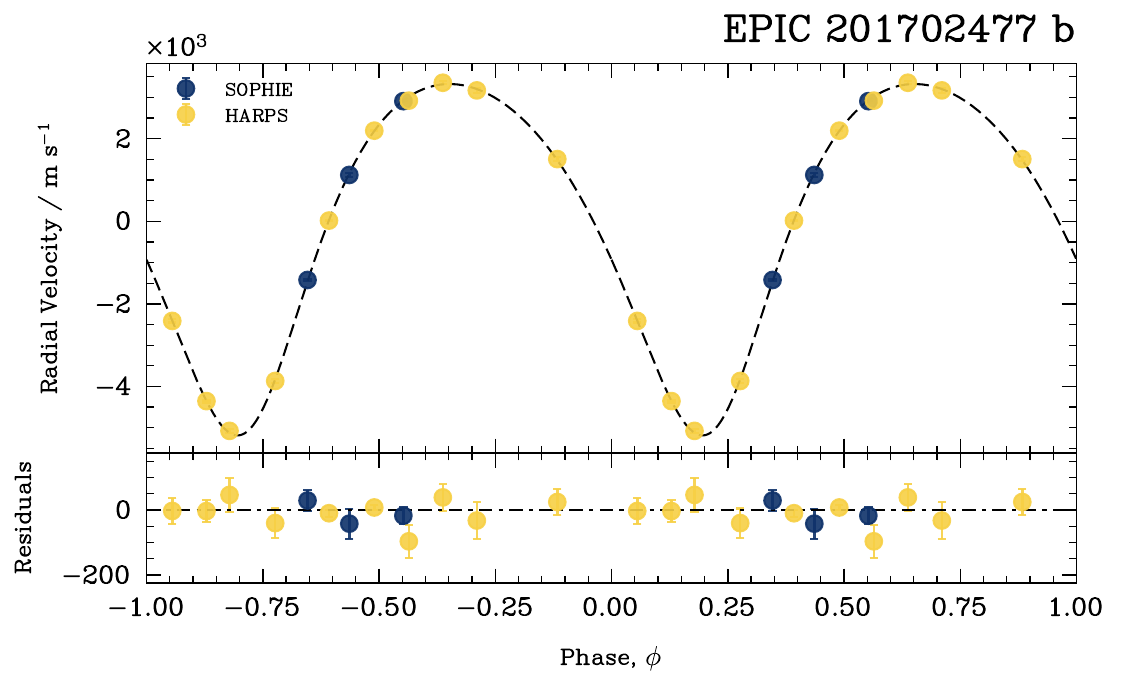} \\
    \includegraphics[width=0.32\linewidth]{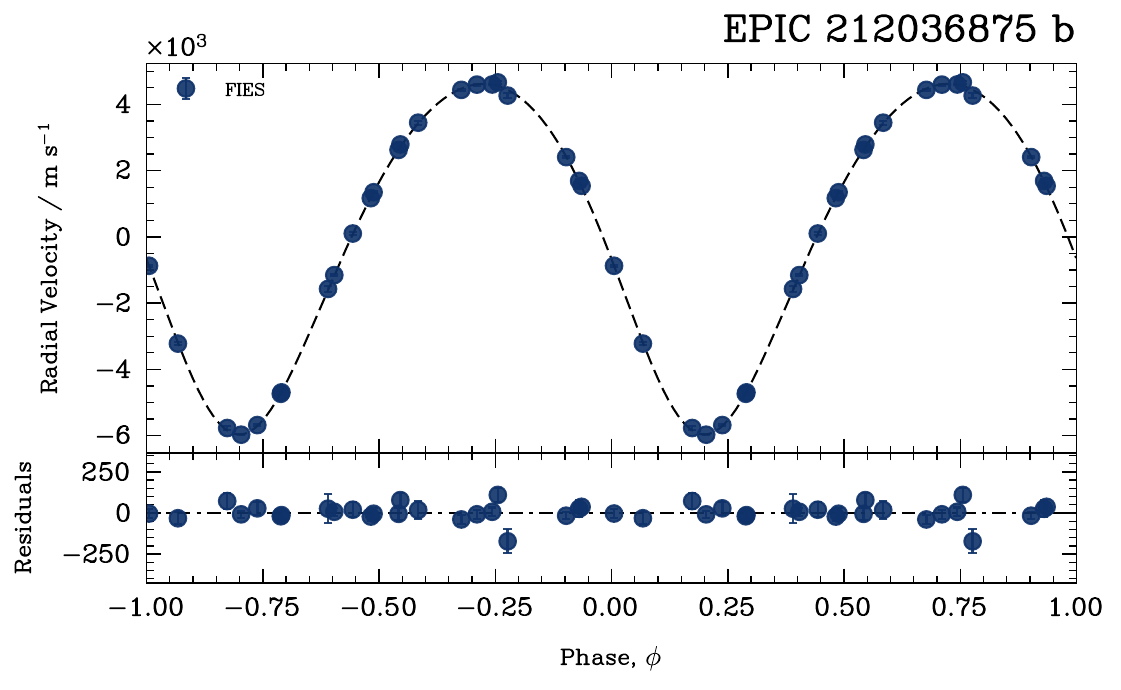}
    \includegraphics[width=0.32\linewidth]{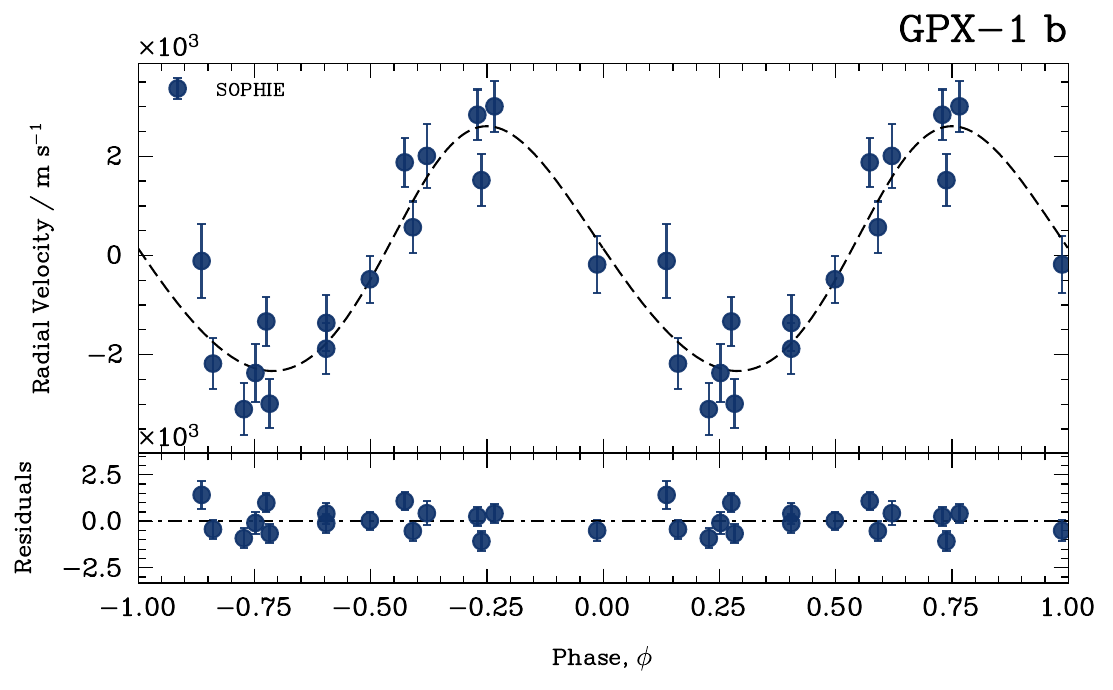}
    \includegraphics[width=0.32\linewidth]{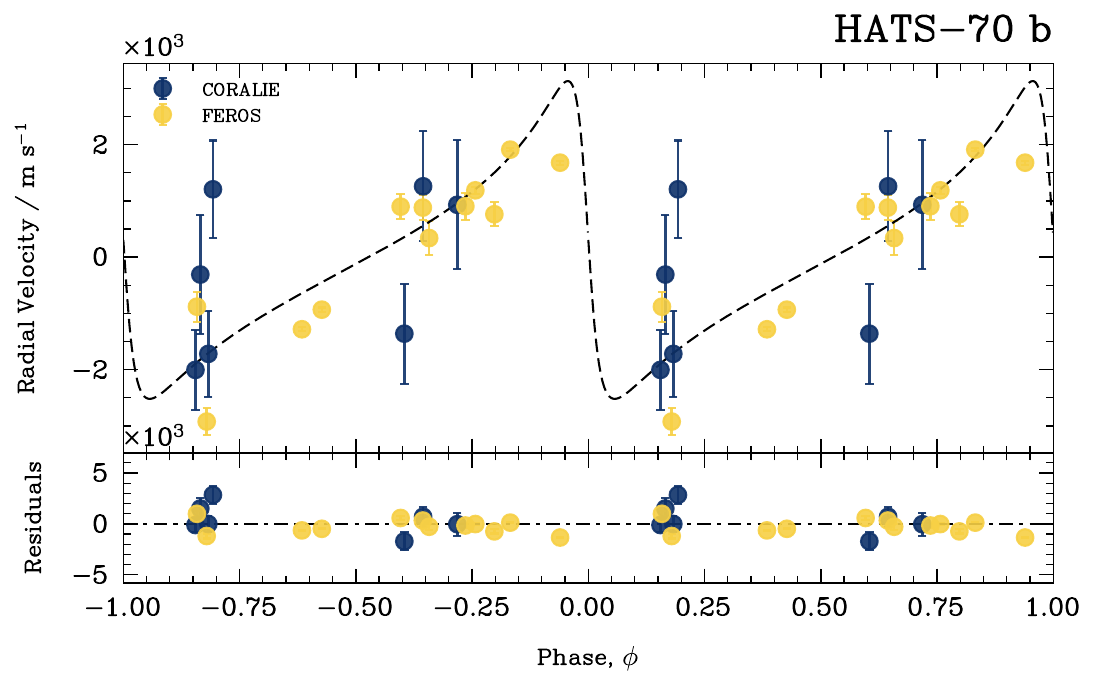} \\
    \includegraphics[width=0.32\linewidth]{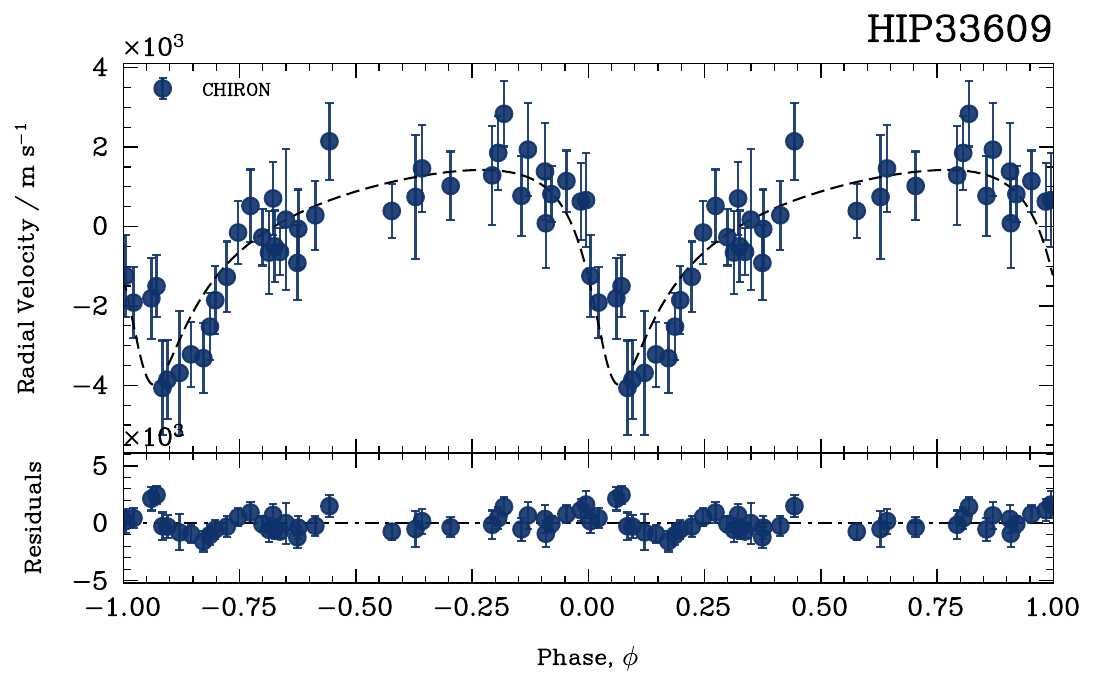}
    \includegraphics[width=0.32\linewidth]{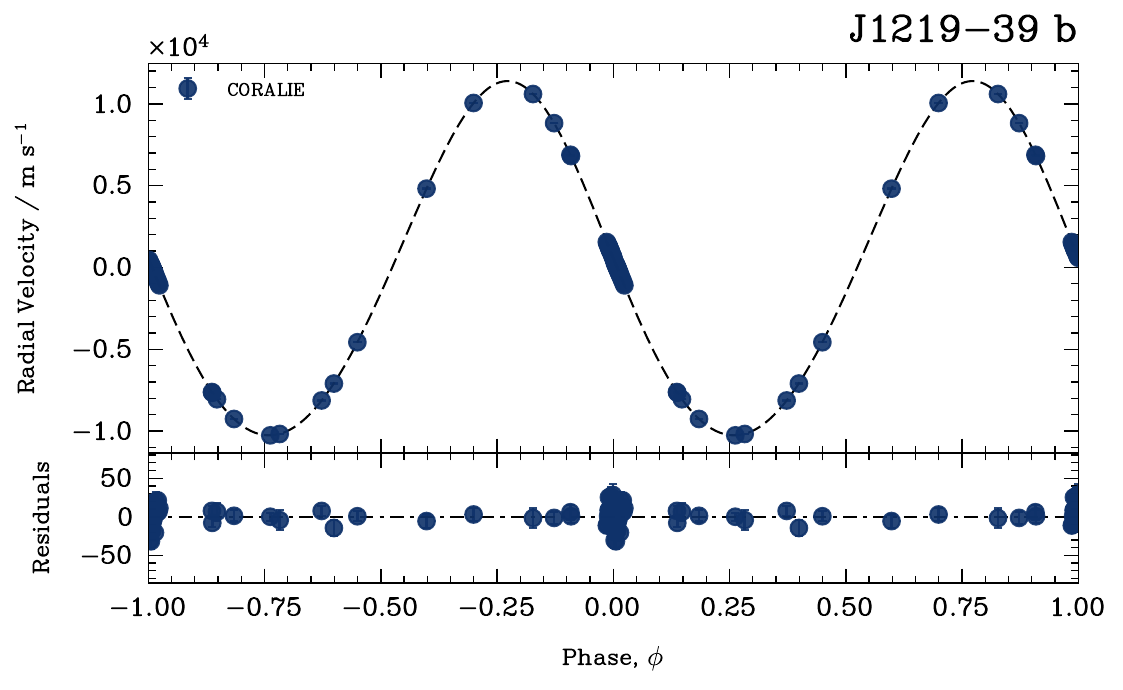}
    \includegraphics[width=0.32\linewidth]{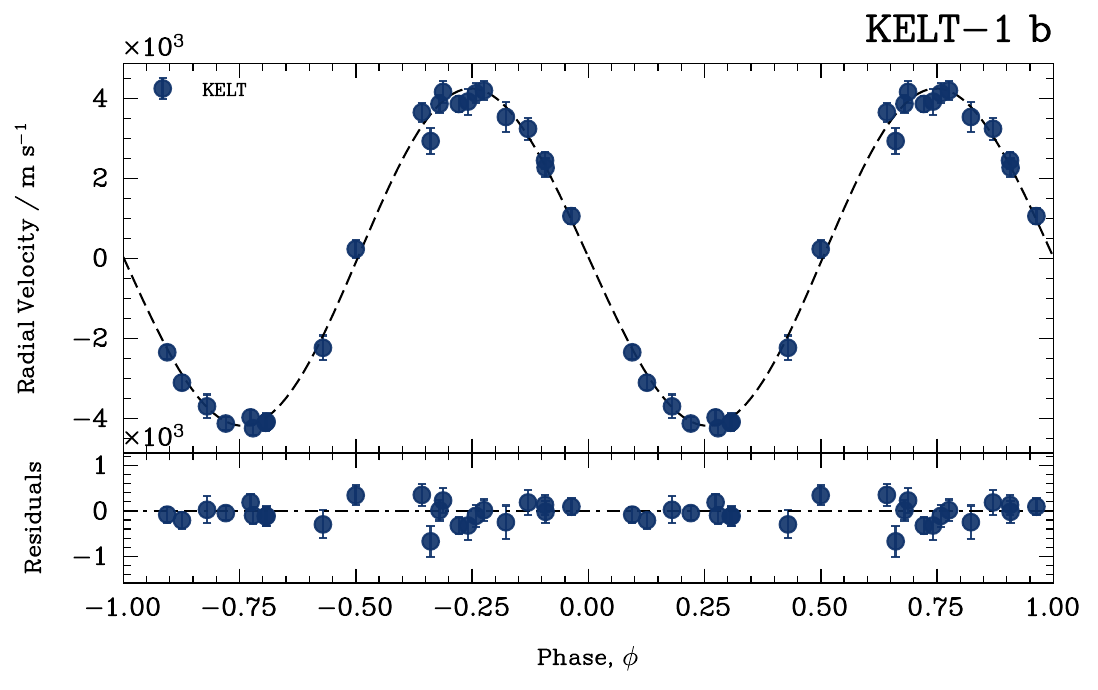} \\
    \caption{Phase-folded radial velocity diagrams for the transiting systems: NGTS-7 A $b$ \citep{2019MNRAS.489.5146J}, AD 3116 $b$ \citep{2017ApJ...849...11G}, CoRoT-3 $b$ \citep{2008A&A...491..889D}, CoRoT-15 $b$ \citep{2011A&A...525A..68B}, CoRoT-33 $b$ \citep{2015A&A...584A..13C}, CoRoT-34 $b$ \citep{2022MNRAS.516..636S}, CWW 89 A $b$ \citep{2018AJ....156..168B}, ELM-J0555-57 A $b$ \citep{2017A&A...604L...6V}, EPIC 201702477 $b$ \citep{2017AJ....153...15B}, EPIC 212036875 $b$ \citep{2019AJ....158...38C}, GPX-1 $b$ \citep{2021MNRAS.505.4956B}, HATS-70 $b$ \citep{2019AJ....157...31Z}, HIP33609 $b$ \citep{2023AJ....165..268V}, J1219-39 $b$ \citep{2013A&A...549A..18T}, and KELT-1 $b$ \citep{2012ApJ...761..123S}.}
    \label{fig:phased1}
\end{figure*}

\begin{figure*}[h]
    \centering
    \includegraphics[width=0.32\linewidth]{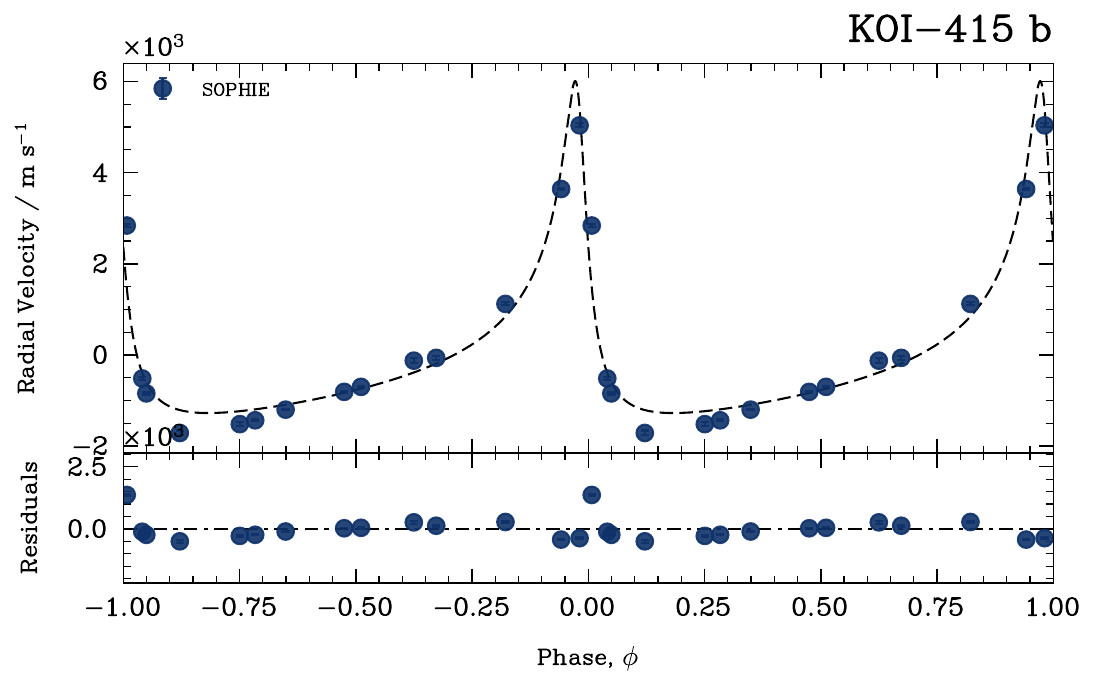}
    \includegraphics[width=0.32\linewidth]{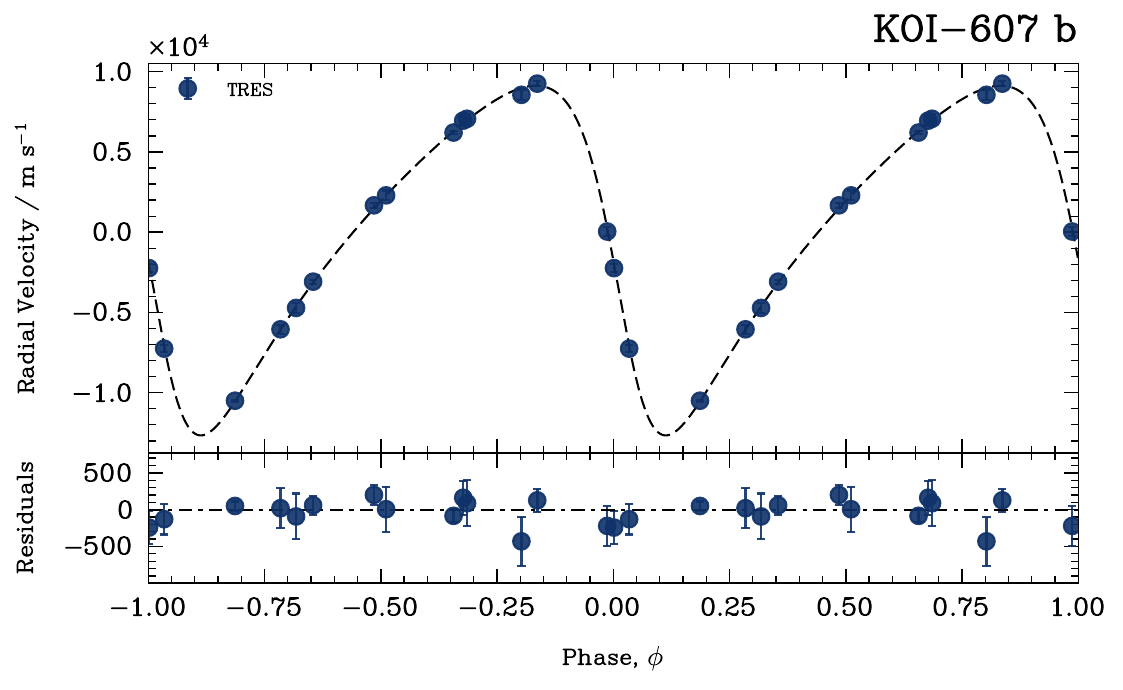}
    \includegraphics[width=0.32\linewidth]{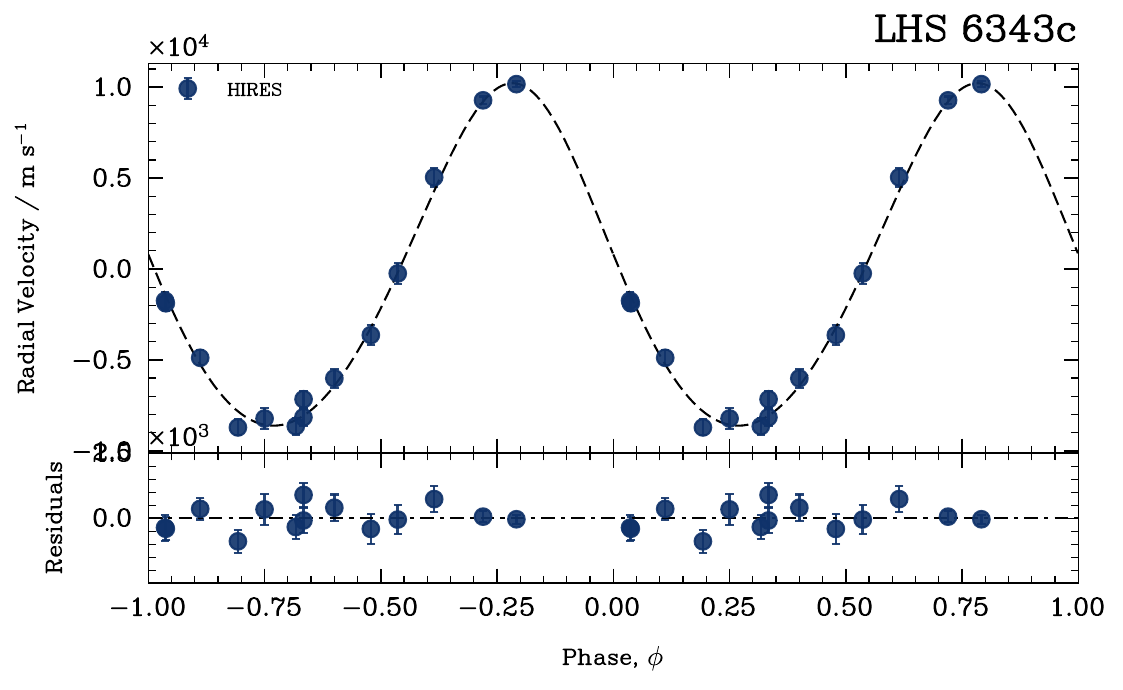} \\
    \includegraphics[width=0.32\linewidth]{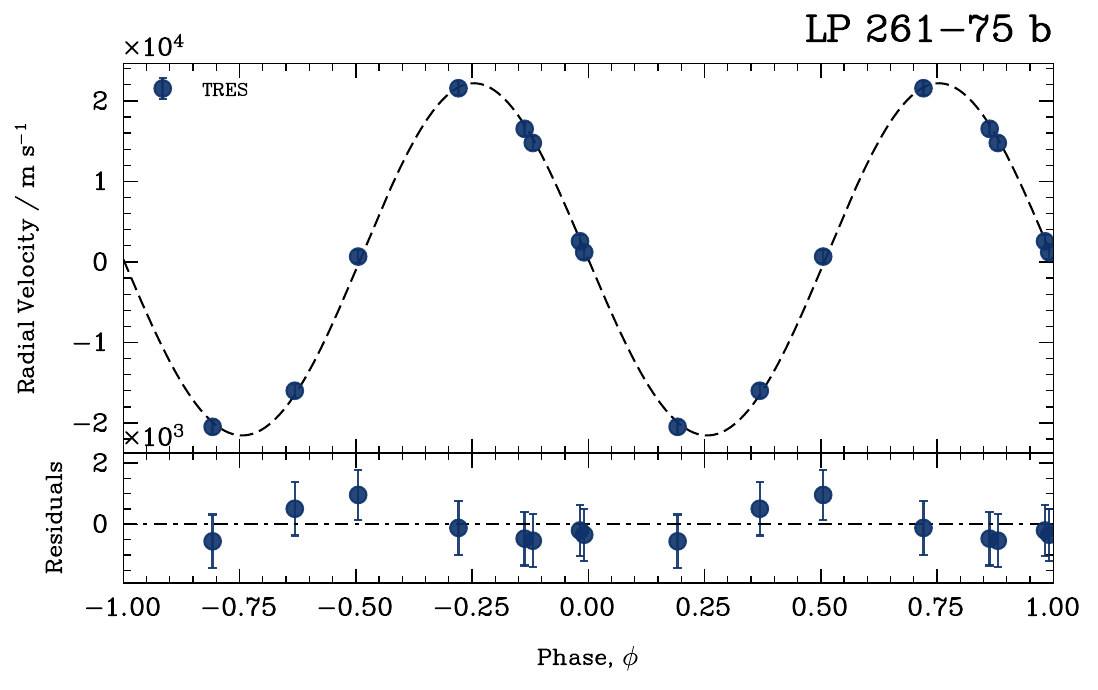}
    \includegraphics[width=0.32\linewidth]{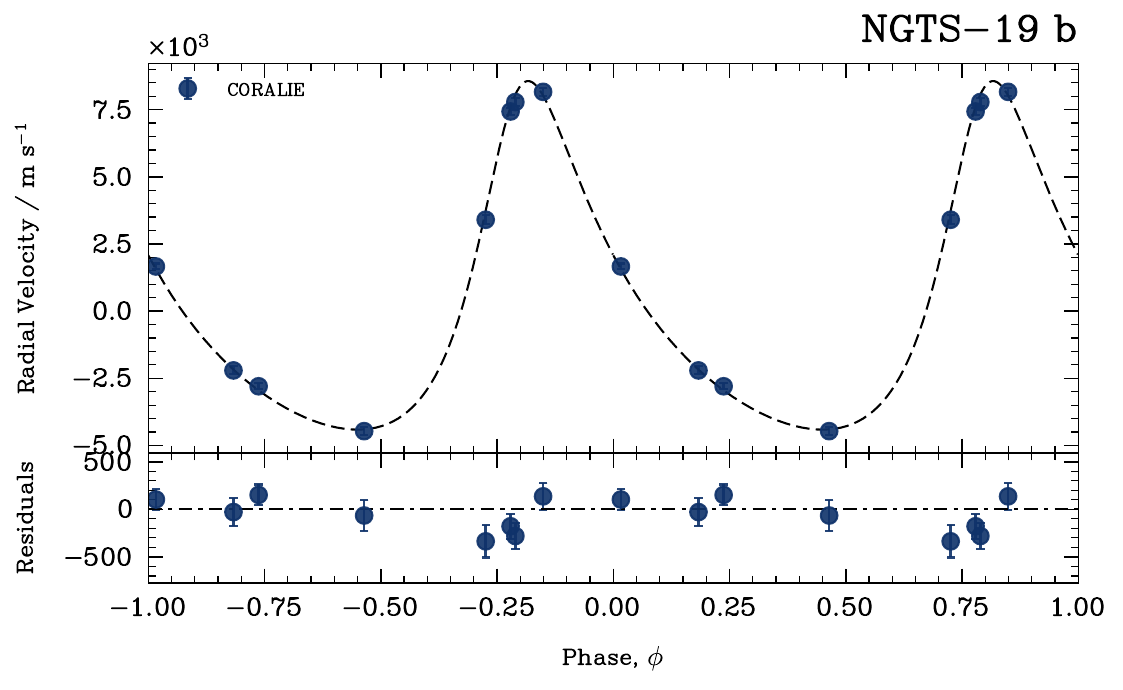}
    \includegraphics[width=0.32\linewidth]{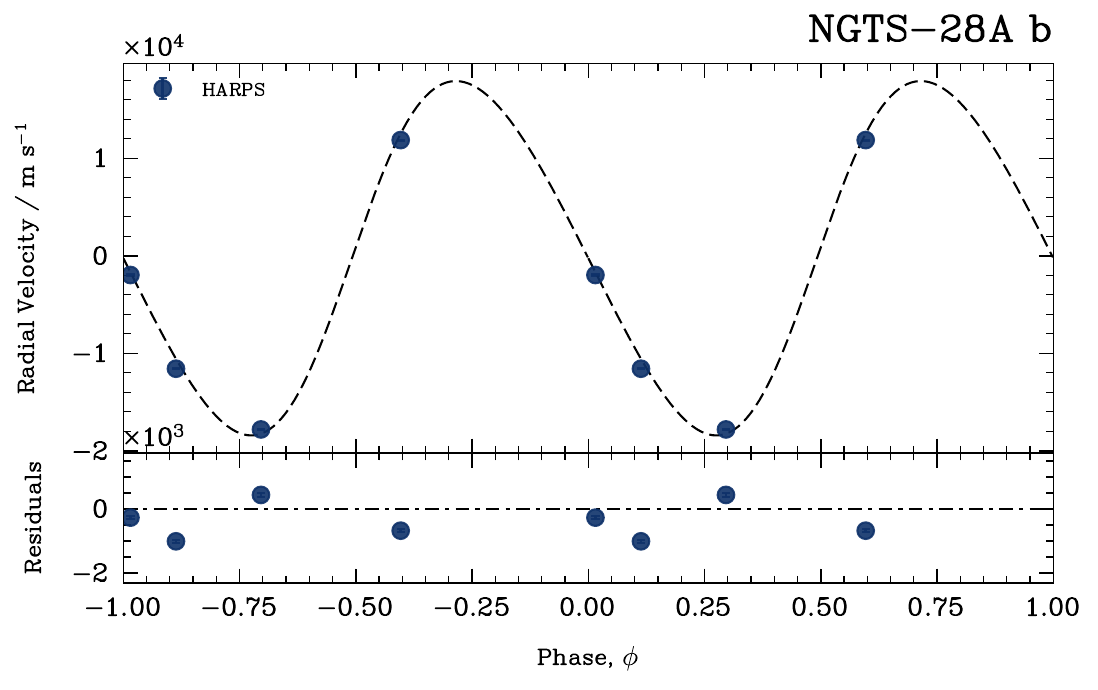} \\
    \includegraphics[width=0.32\linewidth]{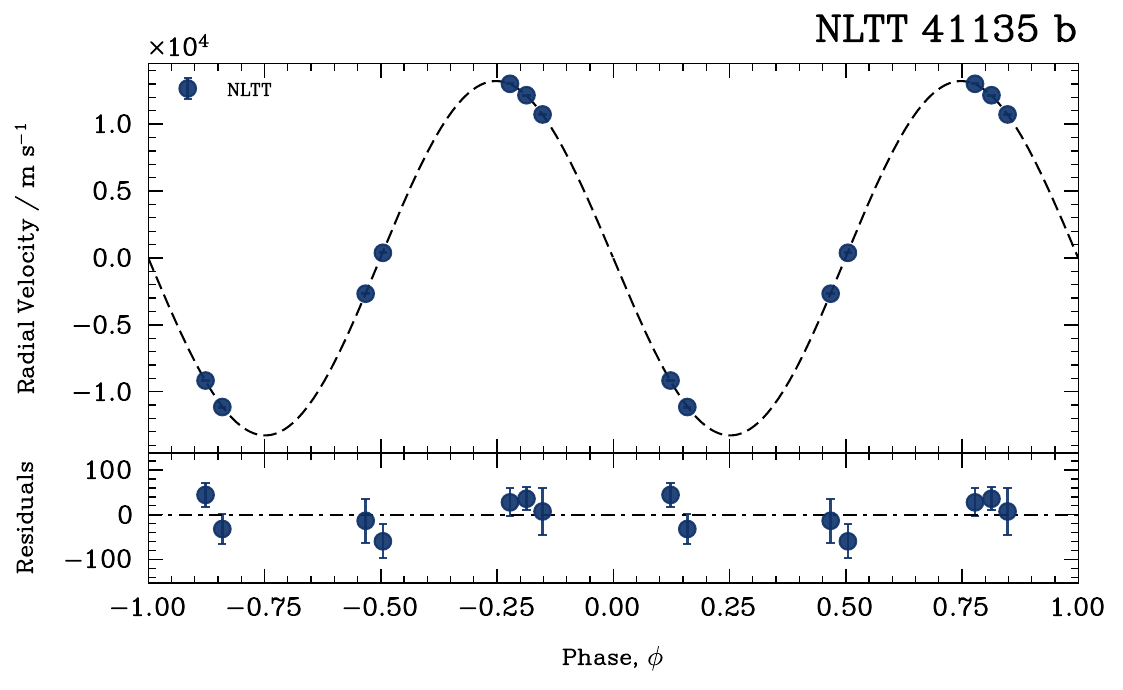}
    \includegraphics[width=0.32\linewidth]{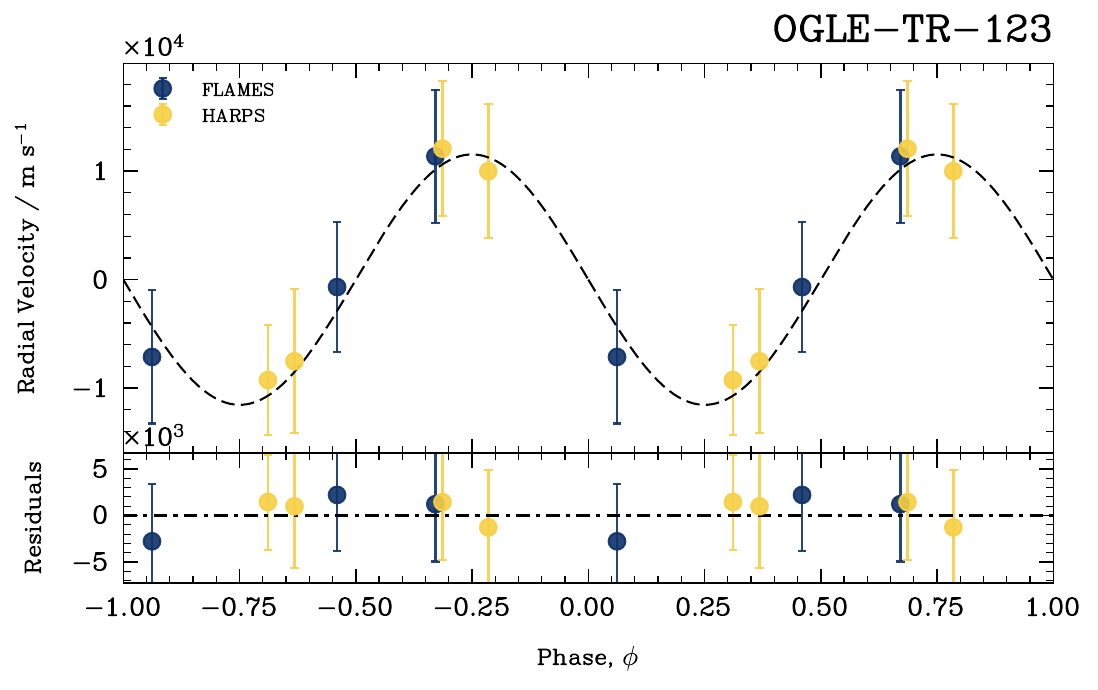}
    \includegraphics[width=0.32\linewidth]{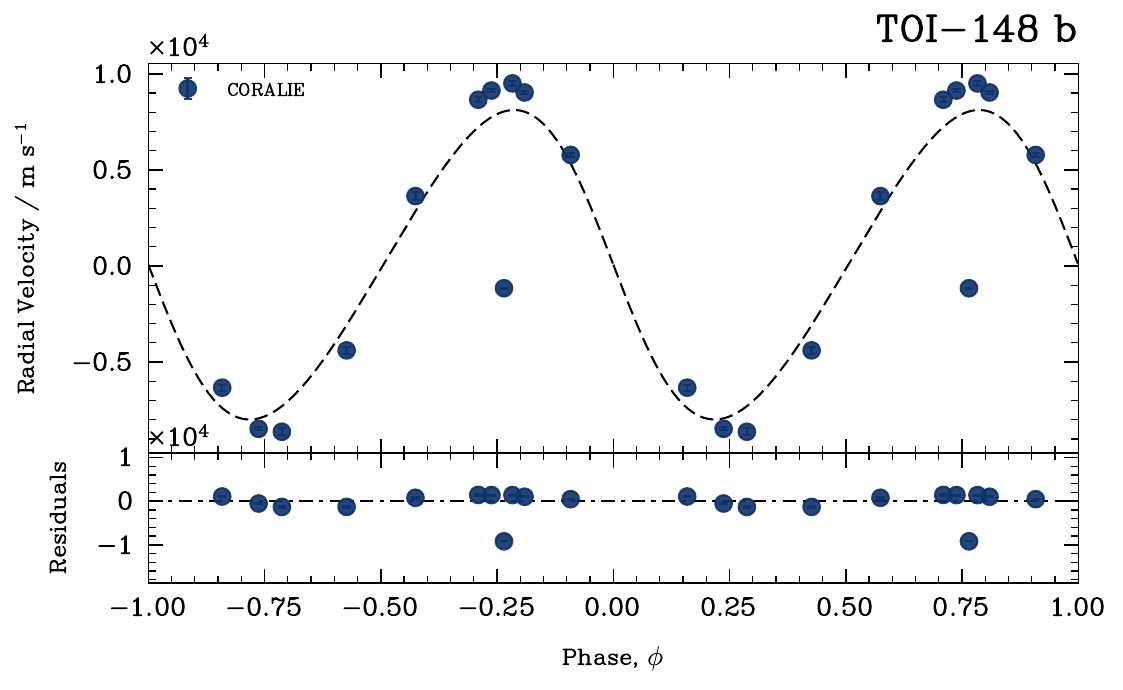} \\
    \includegraphics[width=0.32\linewidth]{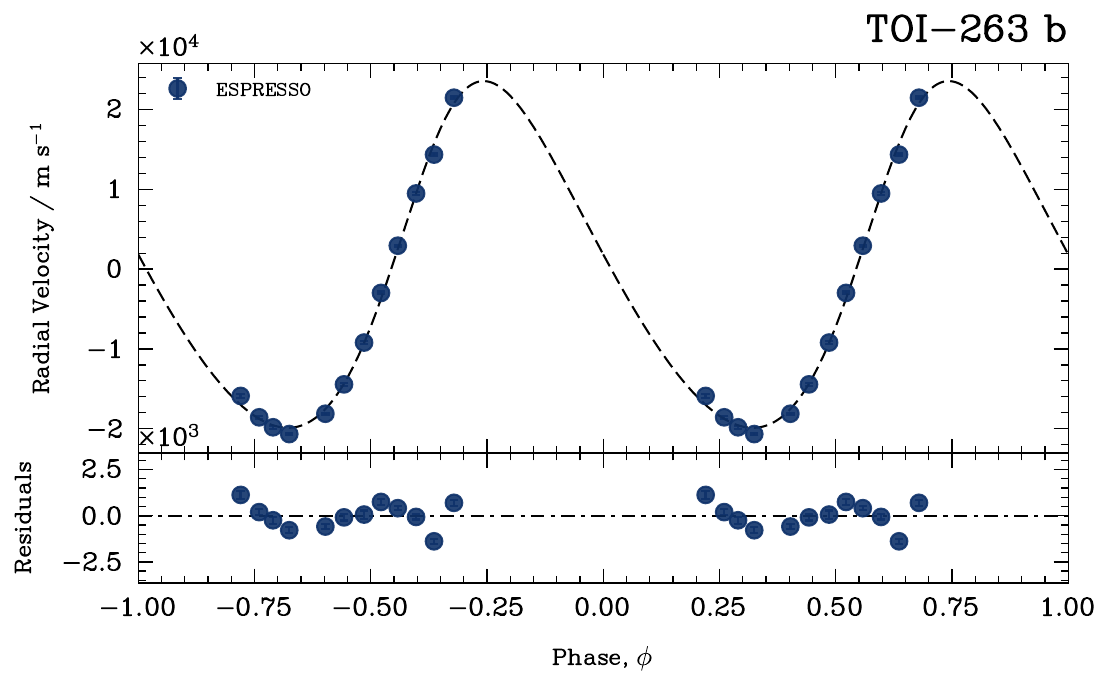}
    \includegraphics[width=0.32\linewidth]{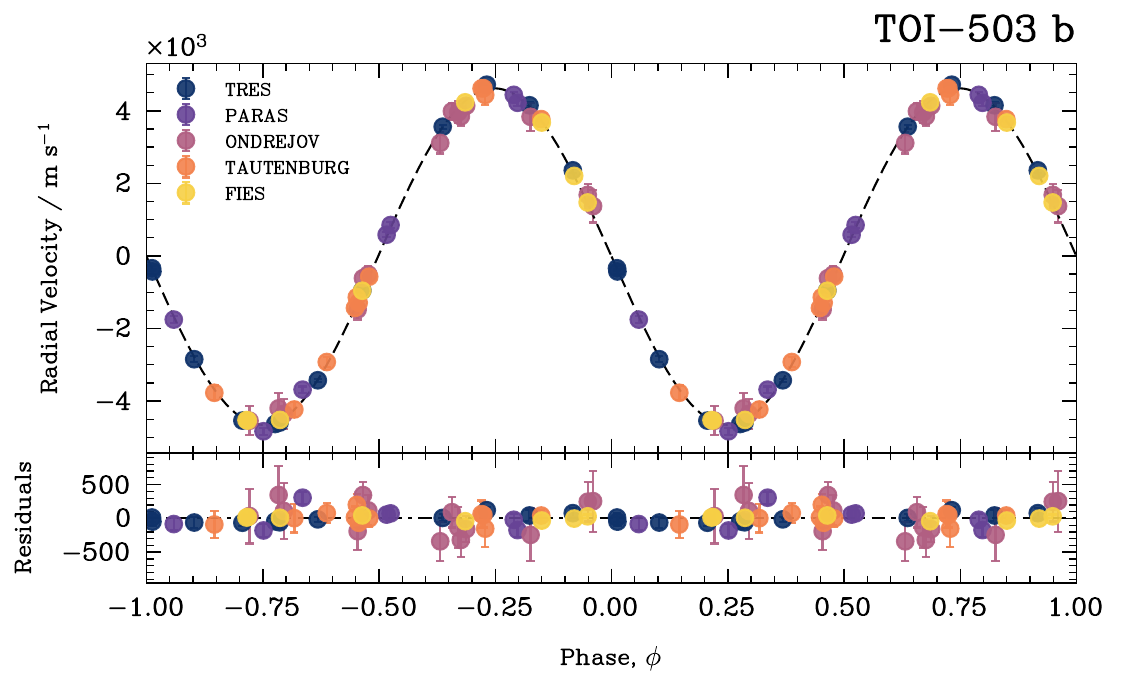}
    \includegraphics[width=0.32\linewidth]{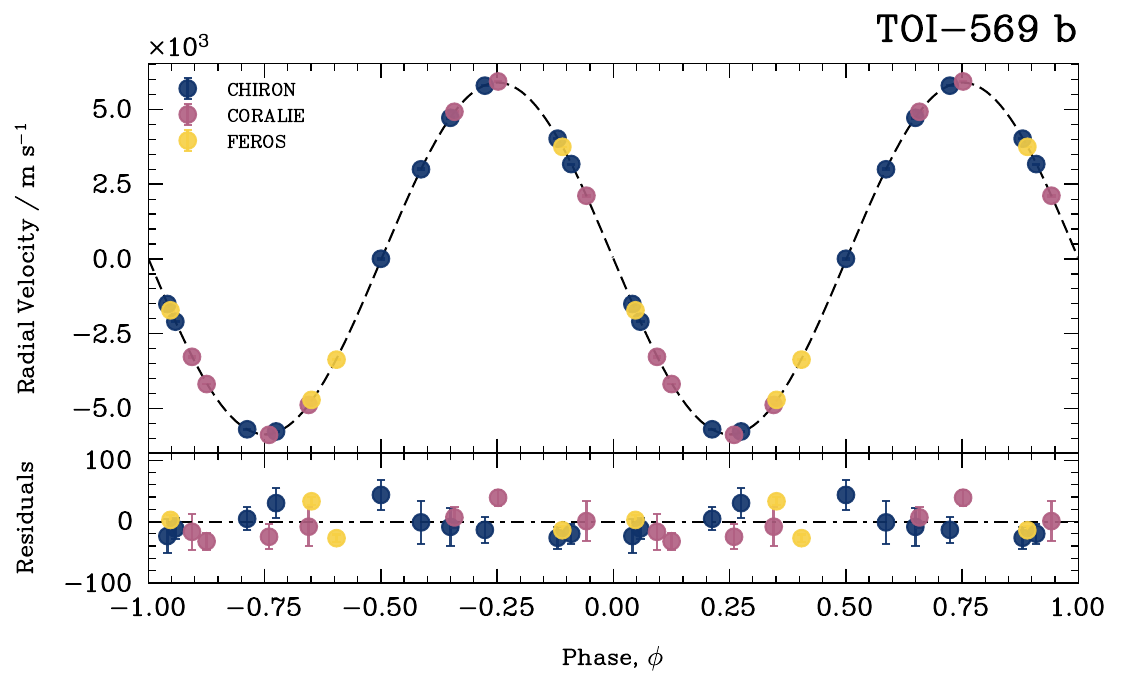} \\
    \includegraphics[width=0.32\linewidth]{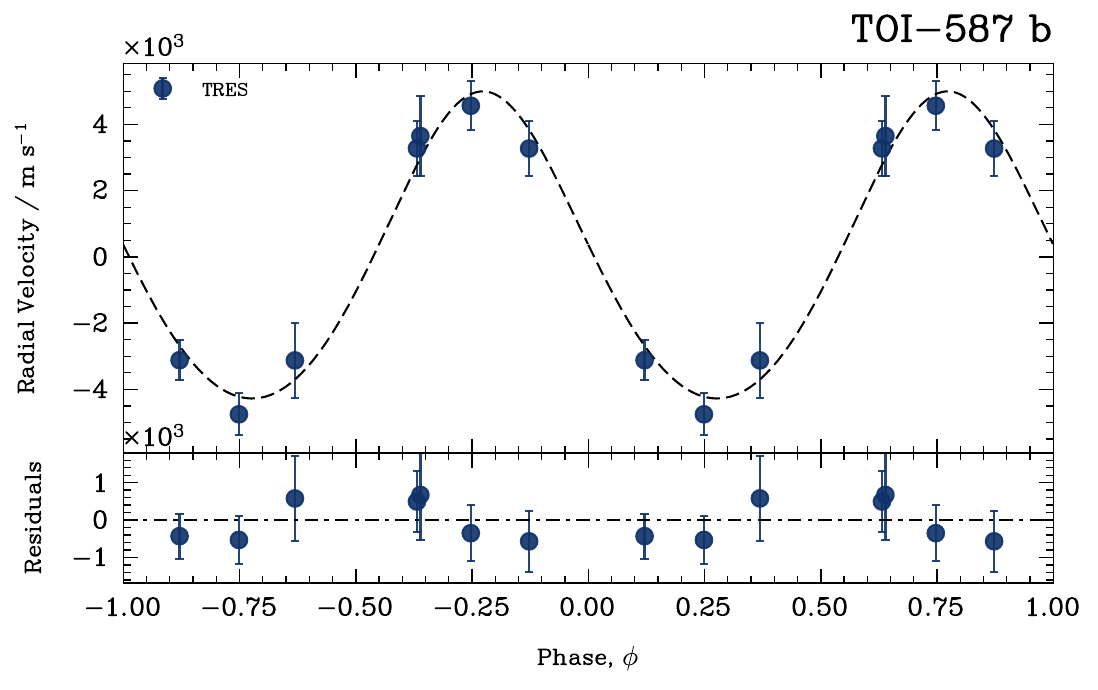}
    \includegraphics[width=0.32\linewidth]{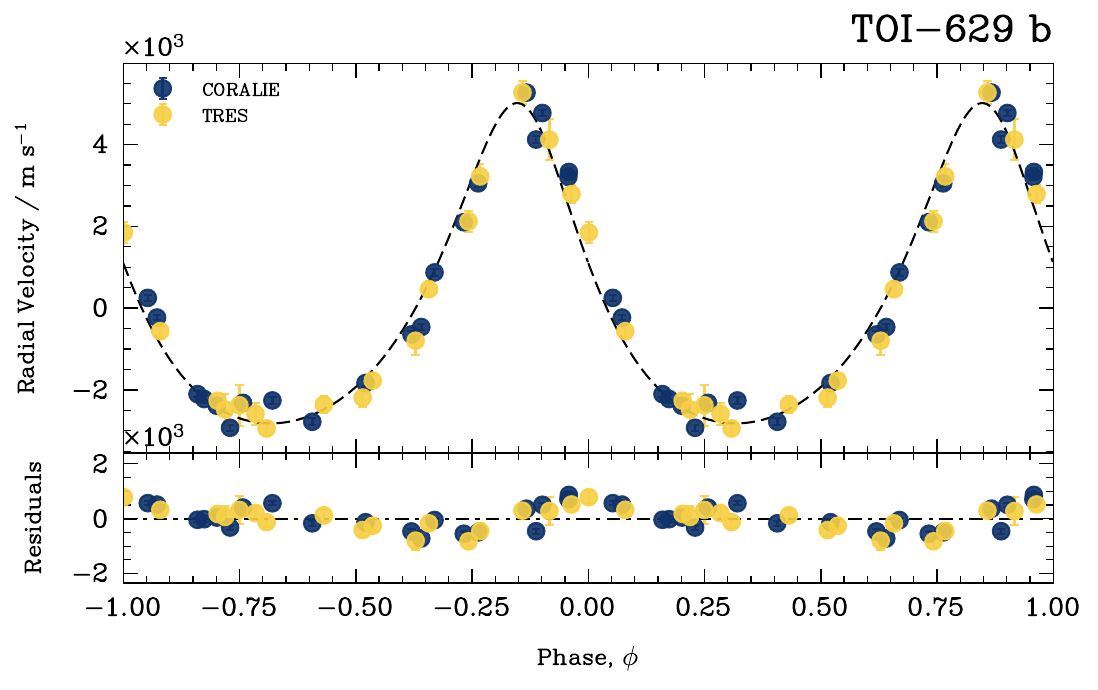}
    \includegraphics[width=0.32\linewidth]{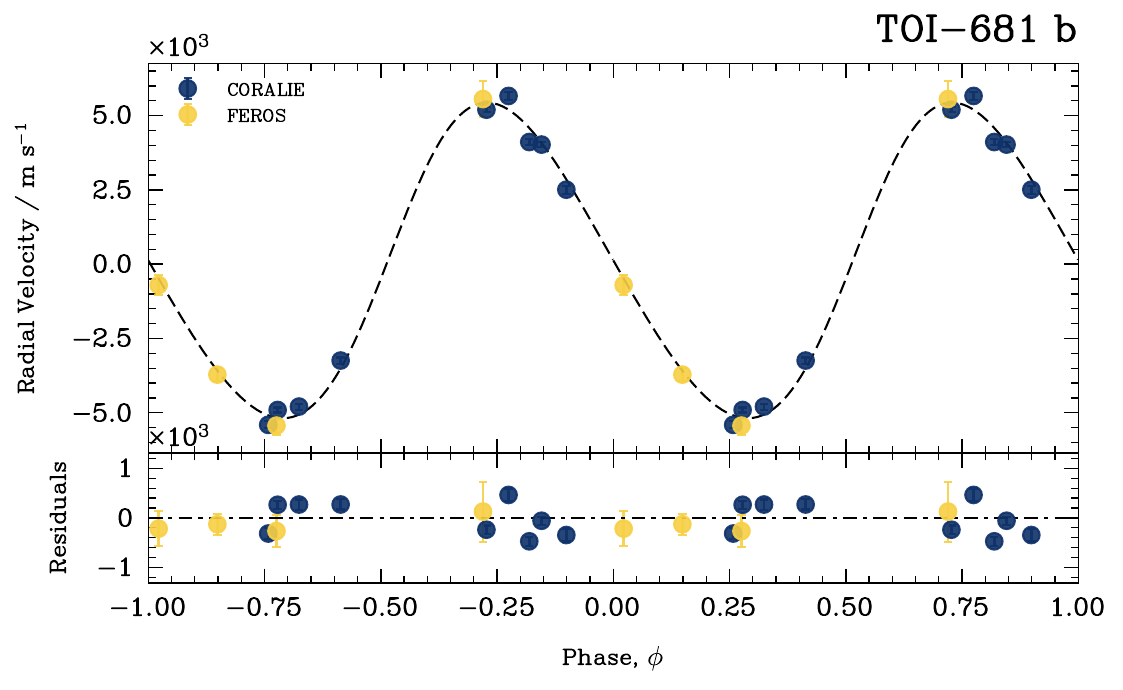} \\
    \includegraphics[width=0.32\linewidth]{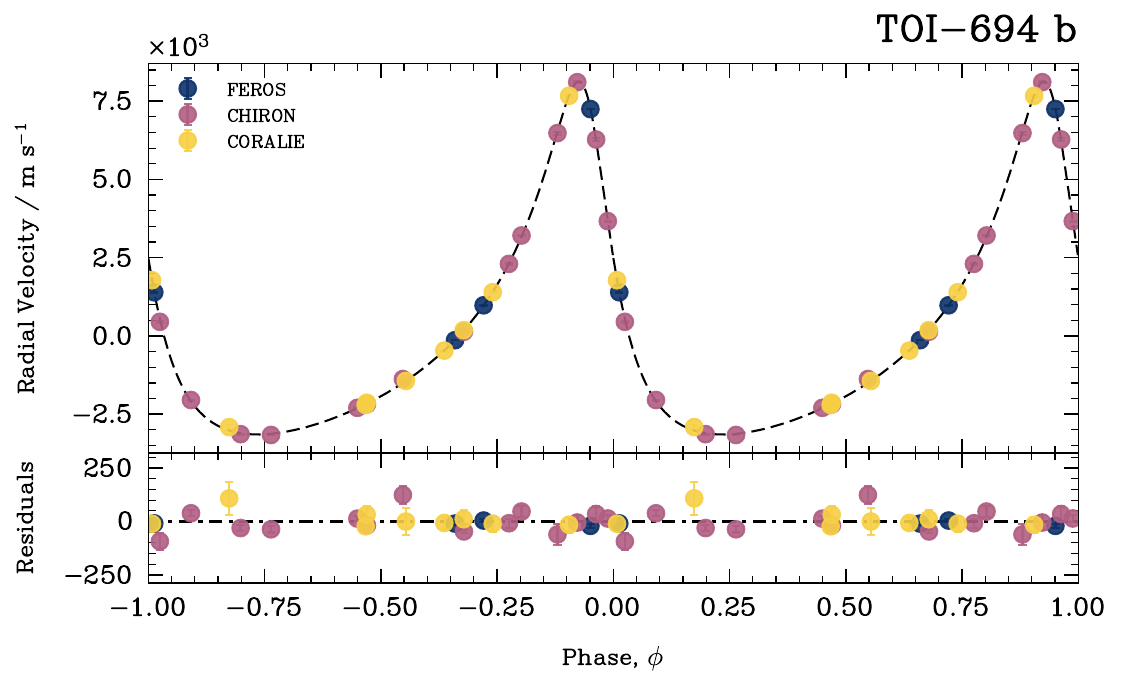}
    \includegraphics[width=0.32\linewidth]{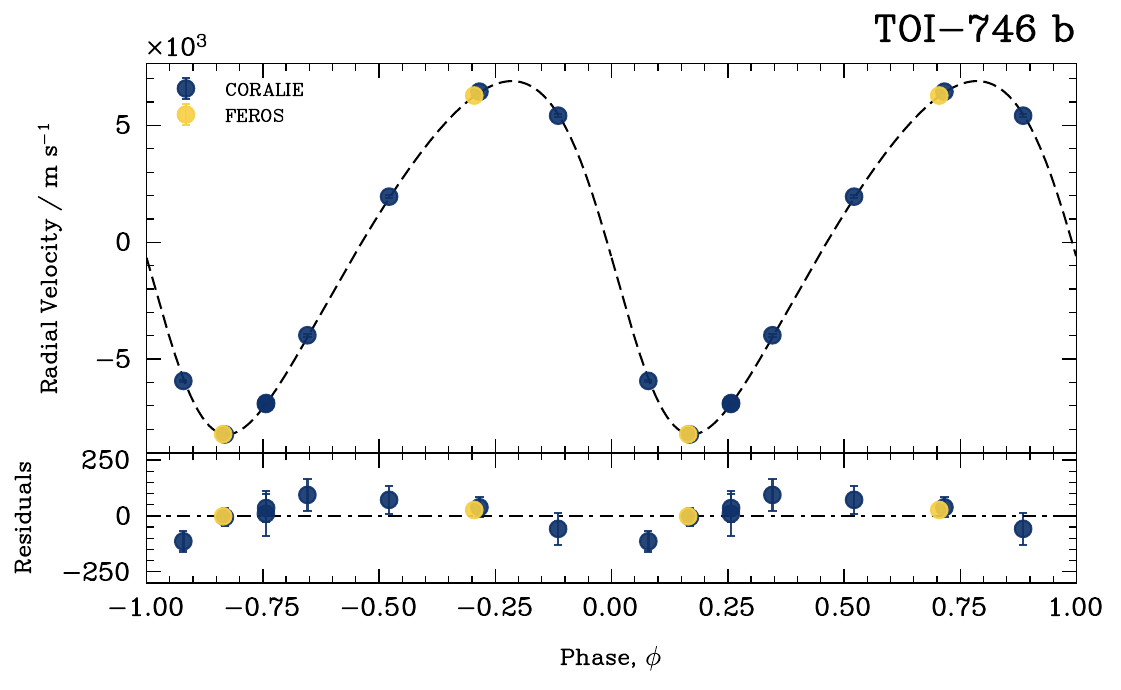}
    \includegraphics[width=0.32\linewidth]{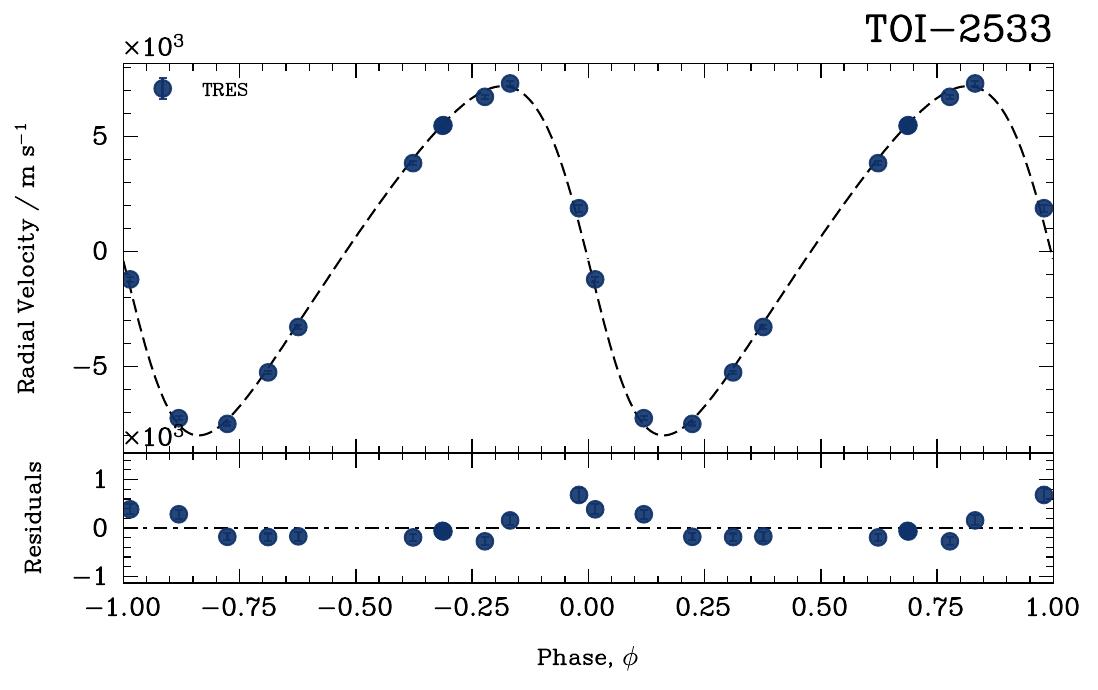}
    \caption{(Continued) KOI-415 $b$ \citep{2013A&A...558L...6M}, KOI-607 $b$ \citep{2019AJ....158...38C}, LHS 6343 C \citep{2011ApJ...730...79J}, LP 261-75 $b$ \citep{2018AJ....156..140I}, NGTS-19 $b$ \citep{2021MNRAS.505.2741A}, NGTS-28 A $b$ \citep{2024MNRAS.530..318H}, NLTT 41135 $b$ \citep{2010ApJ...718.1353I}, OGLE-TR-123 $b$ \citep{2006A&A...447.1035P}, TOI-263 $b$ \citep{2021A&A...650A..55P}, TOI-503 $b$ \citep{2020AJ....159..151S}, TOI-569 $b$ \citep{2020AJ....160...53C},  TOI-148 $b$, TOI-587 $b$, TOI-681 $b$, TOI-746 $b$ \citep{2021A&A...652A.127G}, TOI-2533 $b$ \citep{2023AJ....166..225S, 2024AJ....168..145F}, TOI-629 $b$ \citep{2022A&A...664A..94P}, and TOI-694 $b$ \citep{2020AJ....160..133M}.}
    \label{fig:phased2}
\end{figure*}

\begin{figure*}[h]
    \centering
    \includegraphics[width=0.32\linewidth]{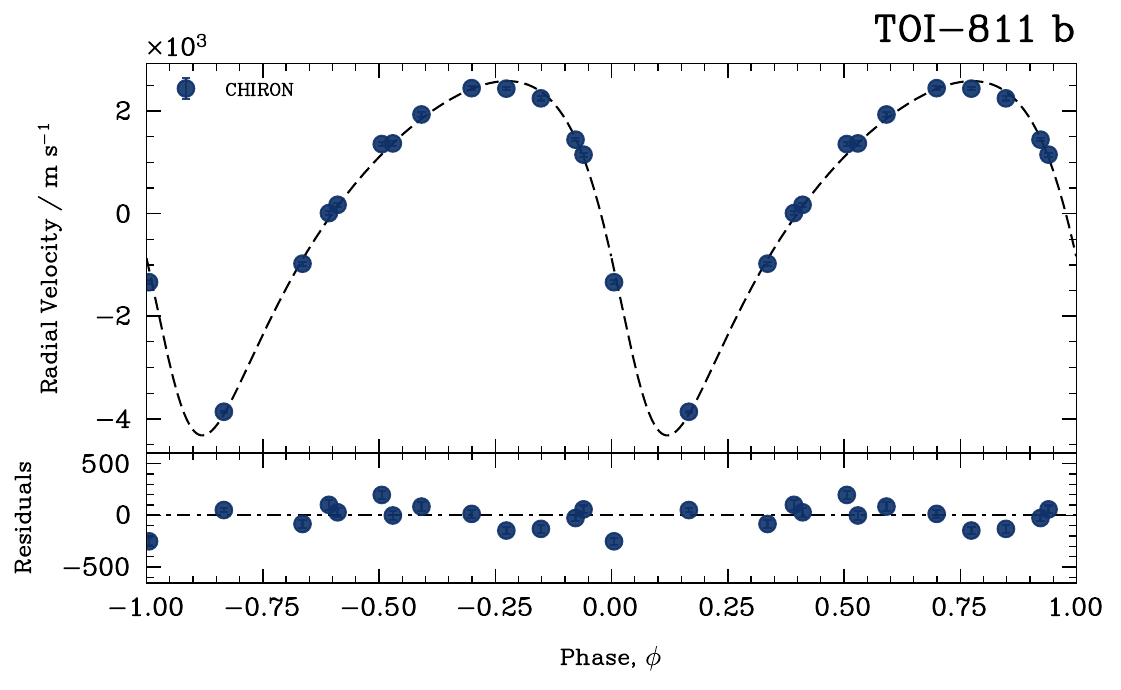}
    \includegraphics[width=0.32\linewidth]{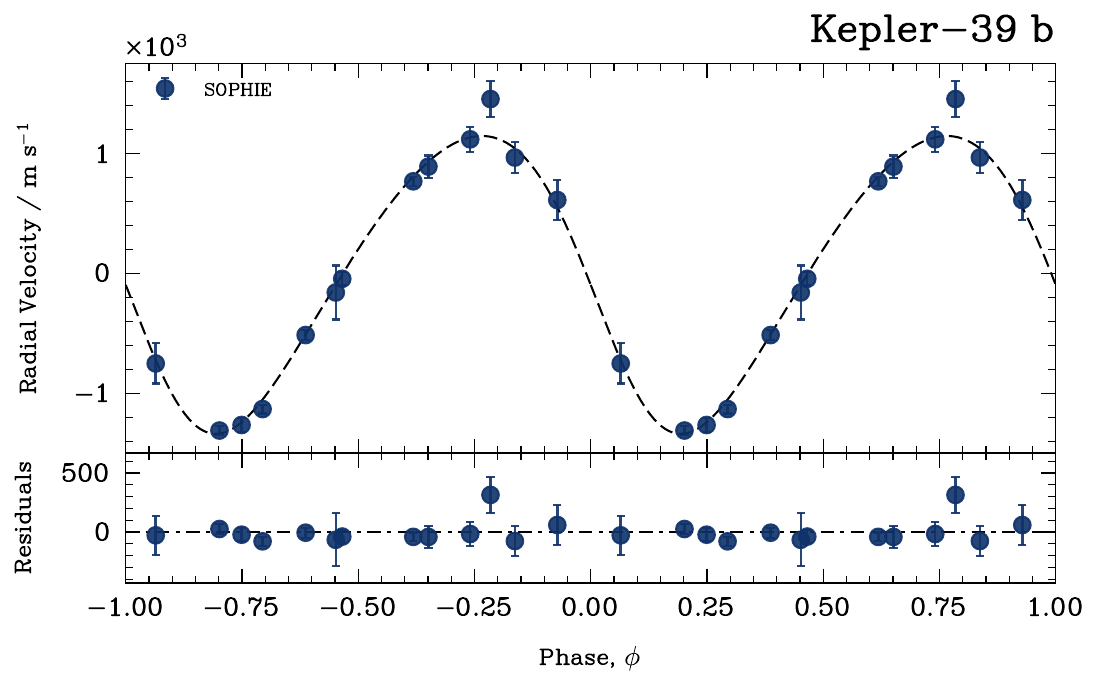}
    \includegraphics[width=0.32\linewidth]{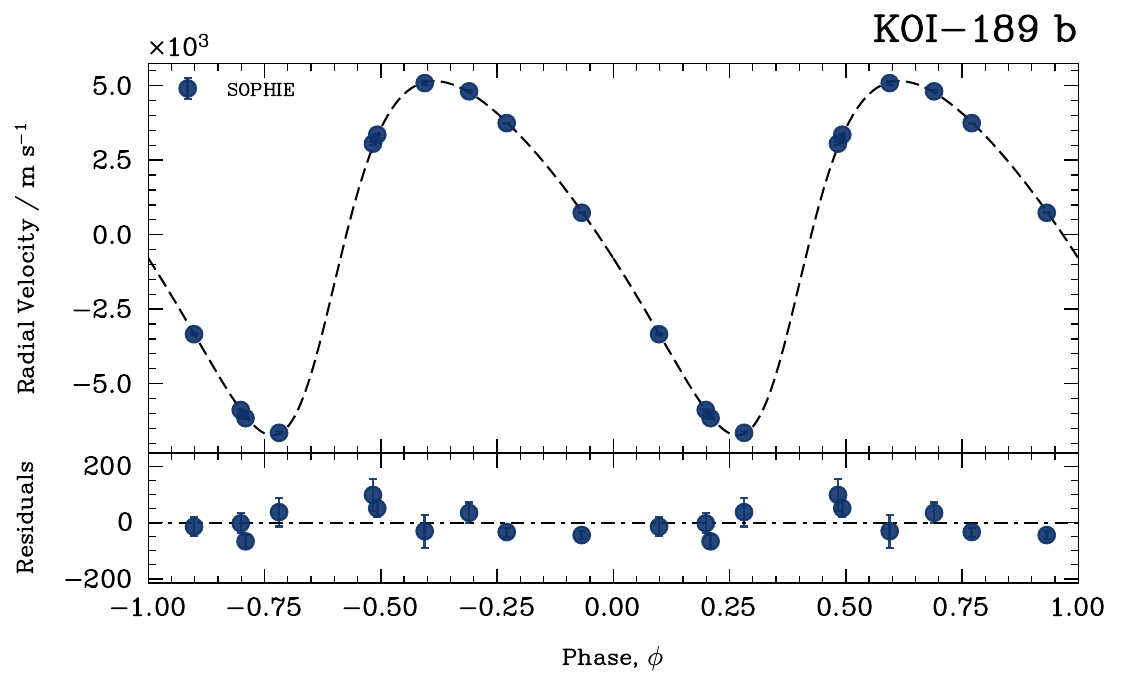} \\
    \includegraphics[width=0.32\linewidth]{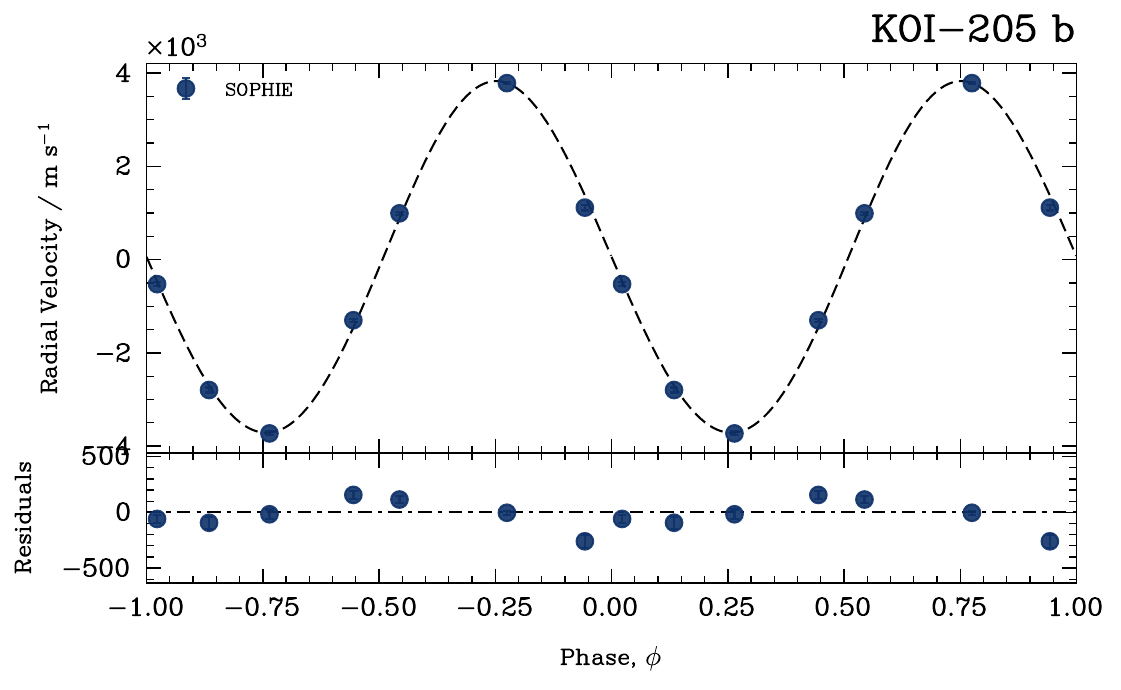} 
    \includegraphics[width=0.32\linewidth]{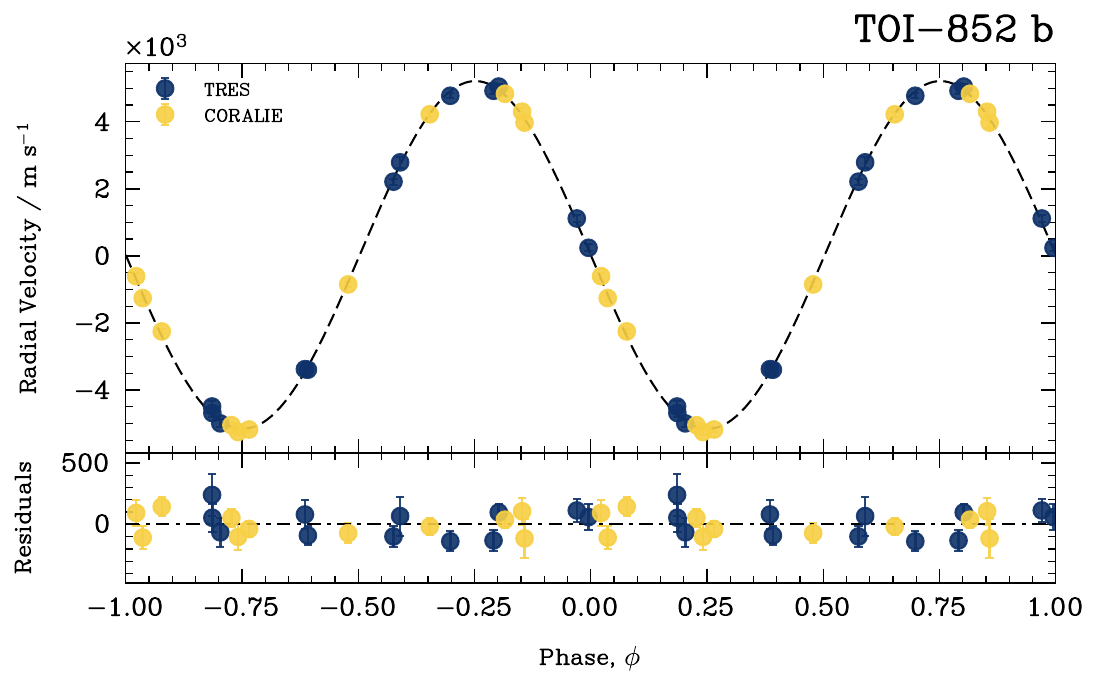} 
    \includegraphics[width=0.32\linewidth]{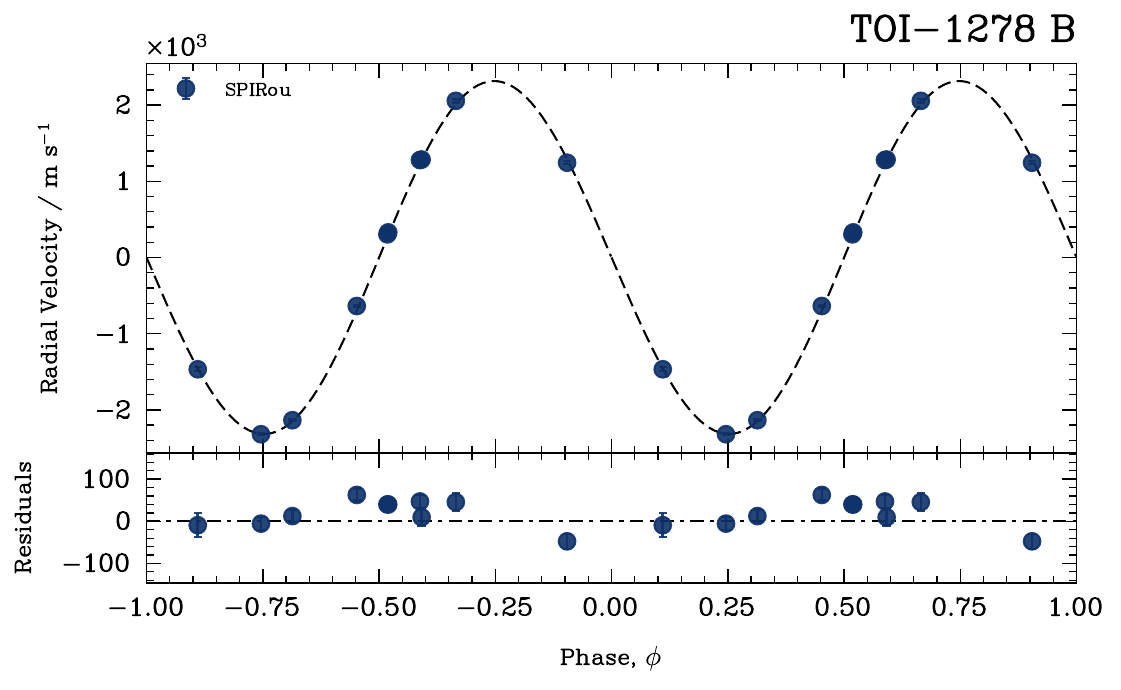} \\
    \includegraphics[width=0.32\linewidth]{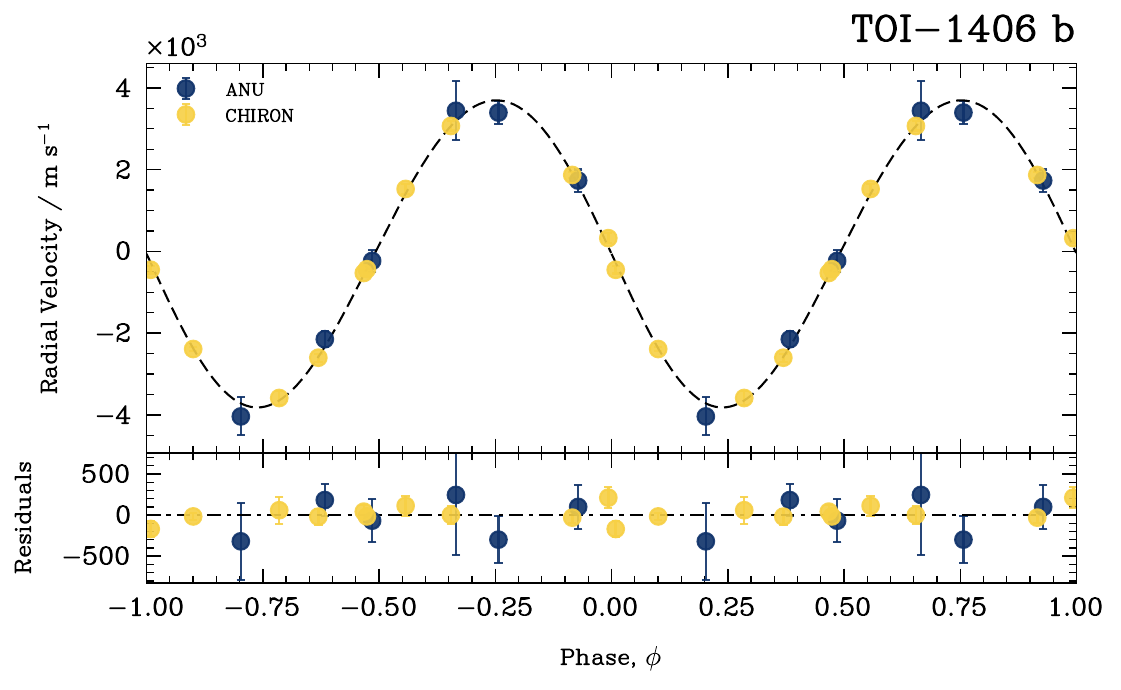}
    \includegraphics[width=0.32\linewidth]{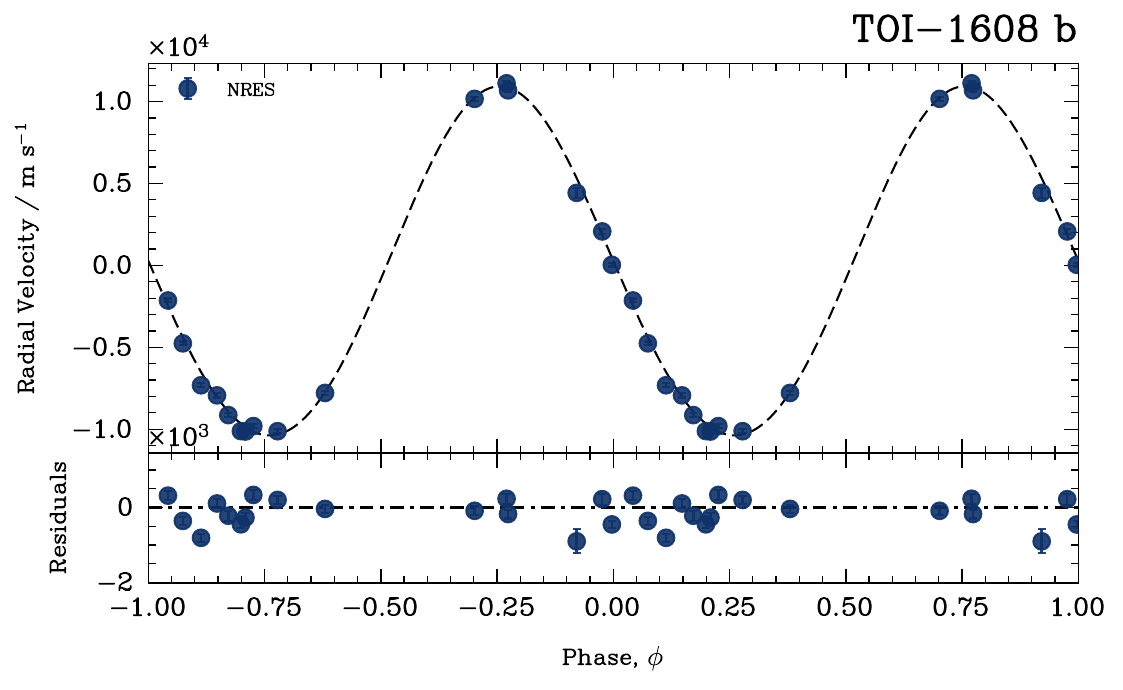}
    \includegraphics[width=0.32\linewidth]{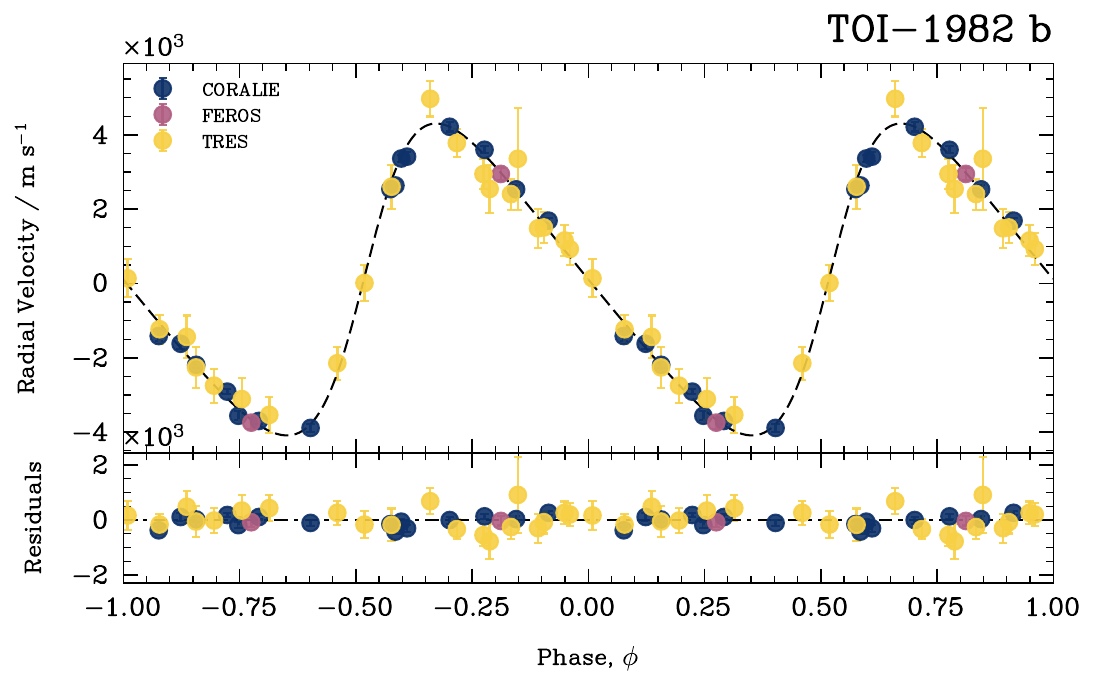} \\
    \includegraphics[width=0.32\linewidth]{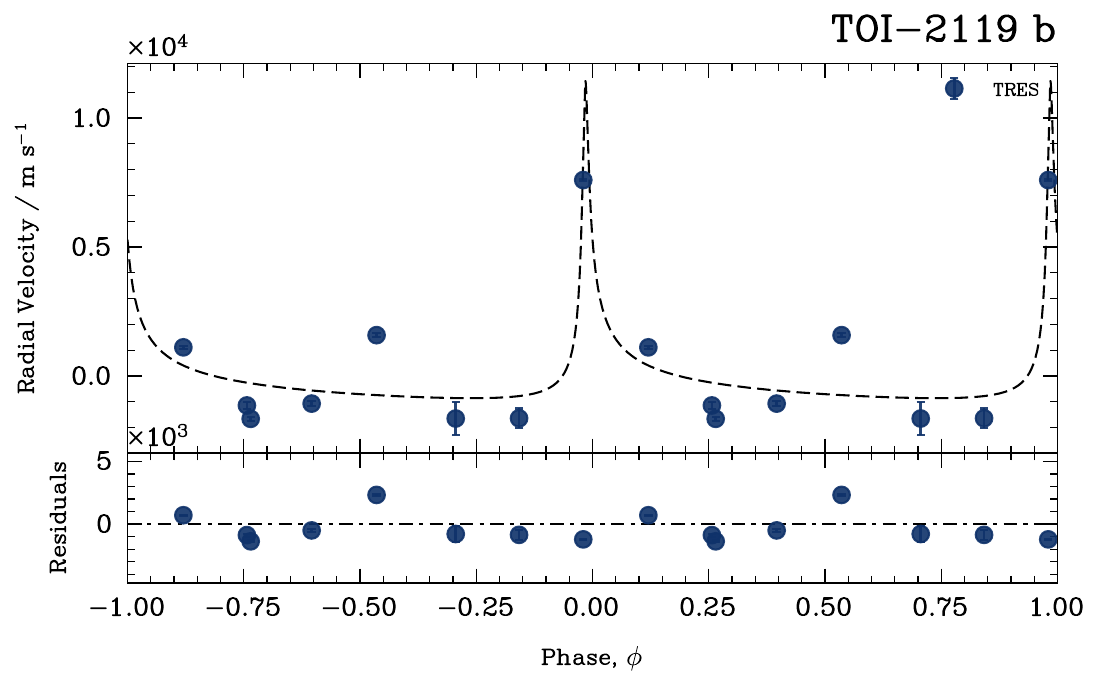} 
    \includegraphics[width=0.32\linewidth]{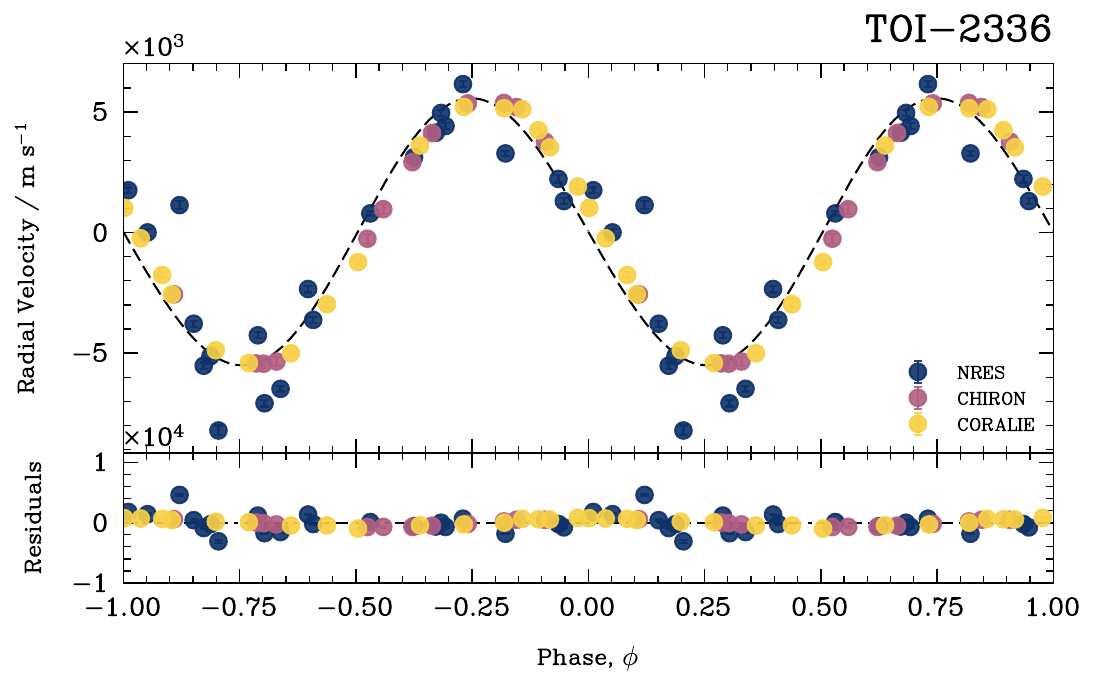}
    \includegraphics[width=0.32\linewidth]{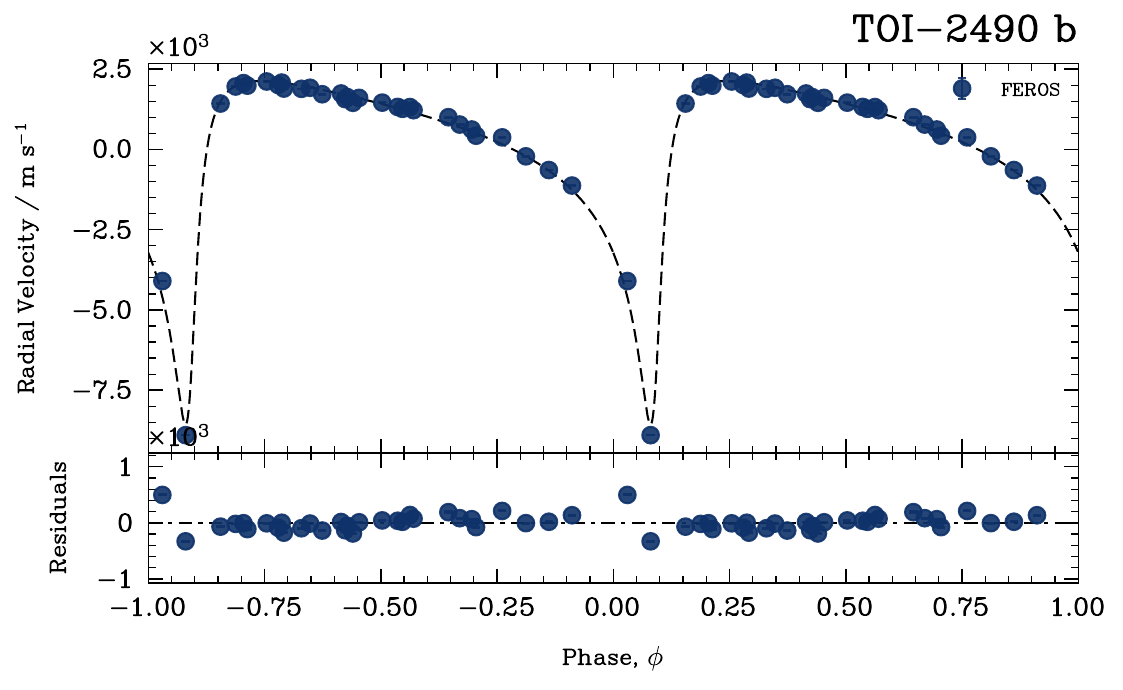} \\
    \includegraphics[width=0.32\linewidth]{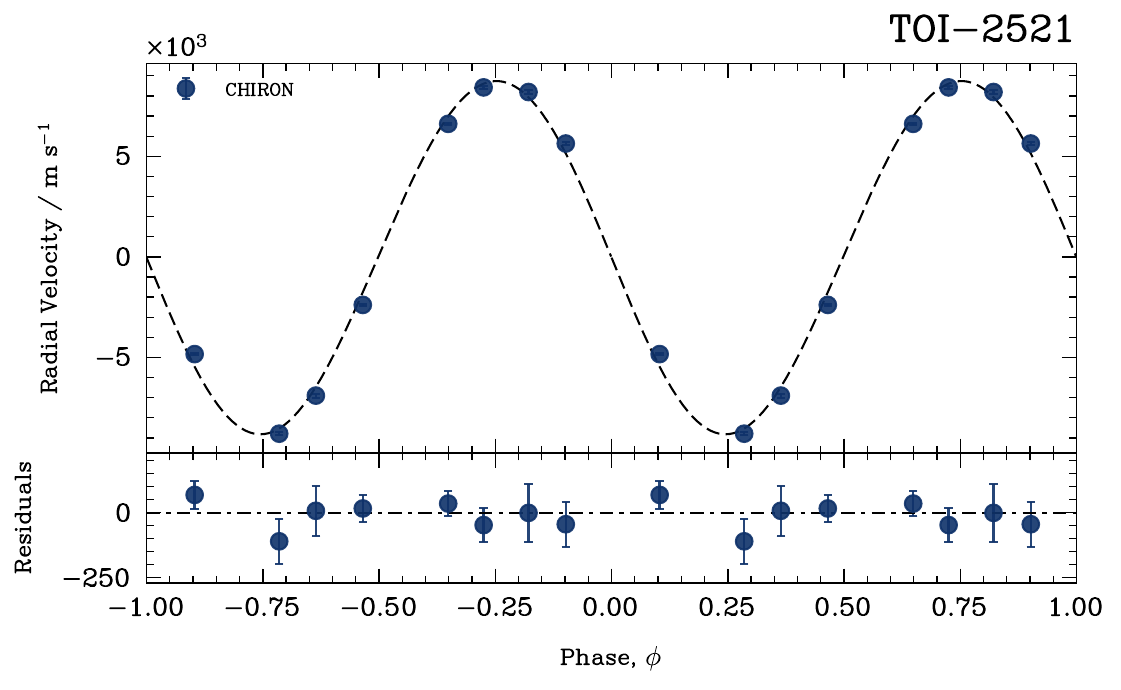}
    \includegraphics[width=0.32\linewidth]{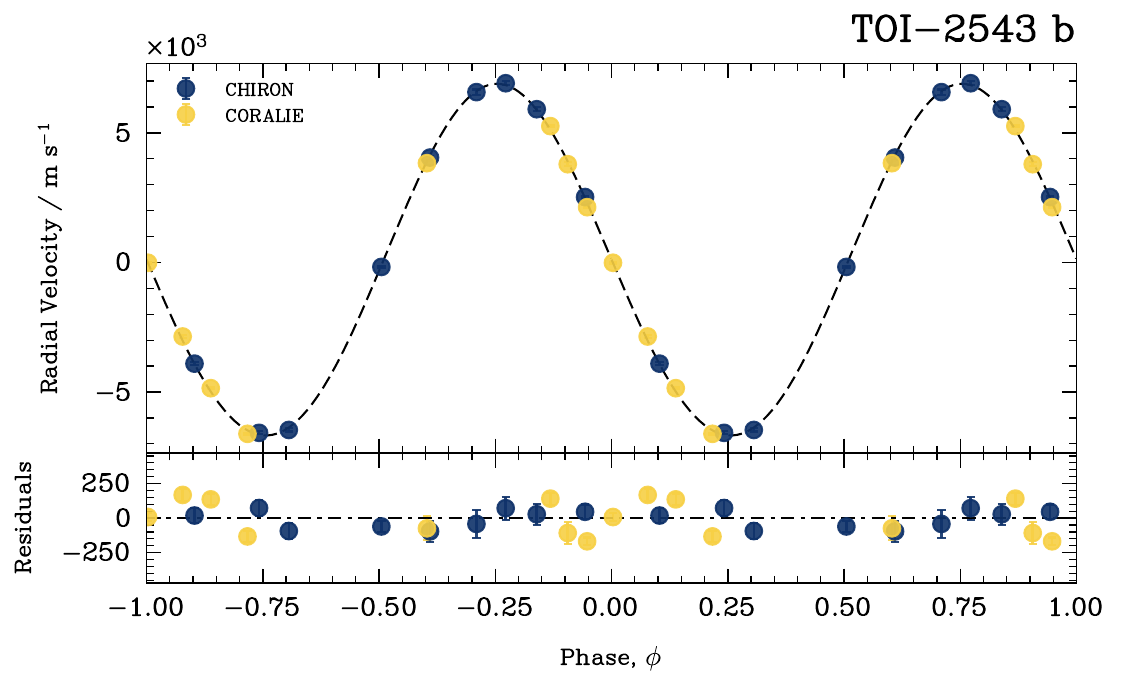}
    \includegraphics[width=0.32\linewidth]{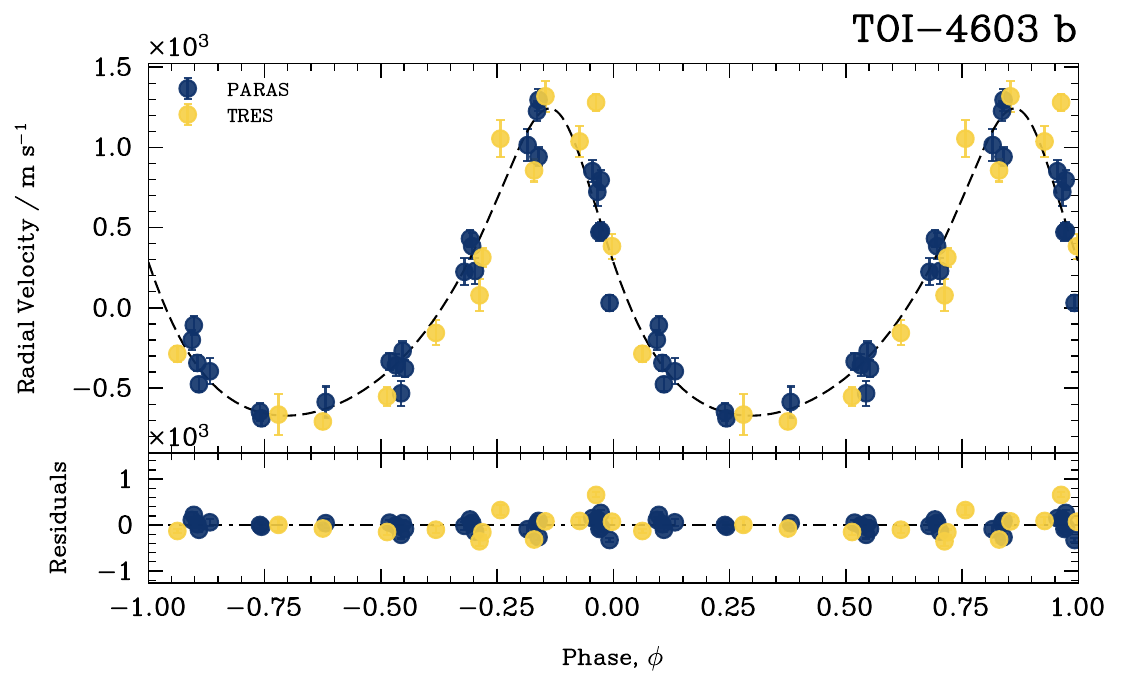} \\
    \includegraphics[width=0.32\linewidth]{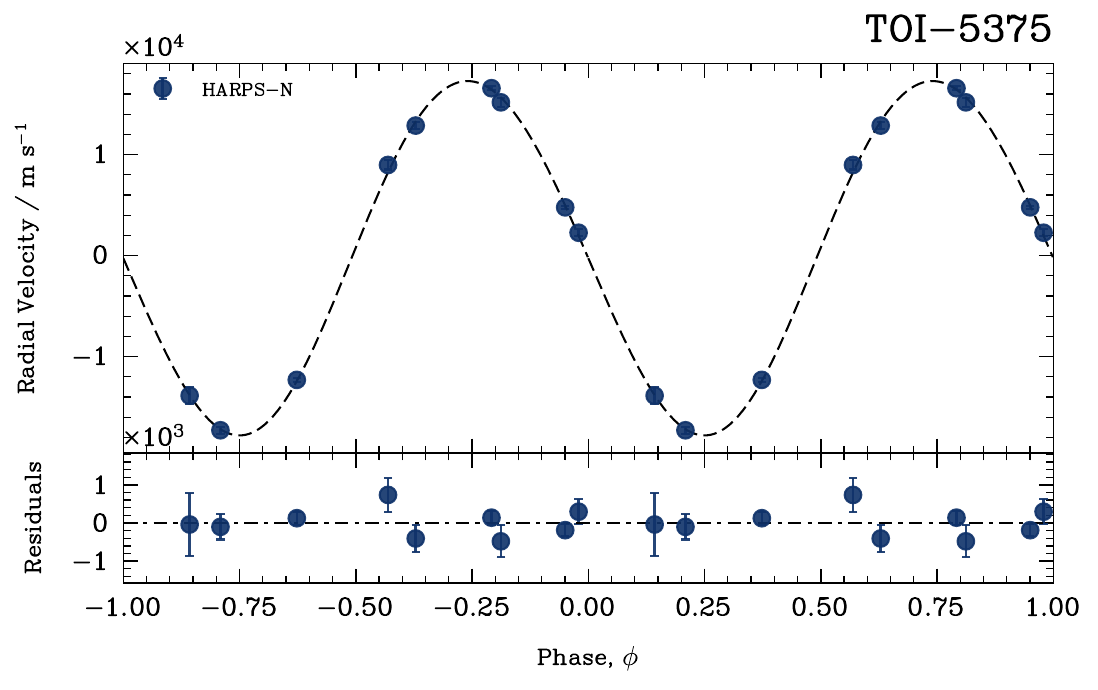}
    \includegraphics[width=0.32\linewidth]{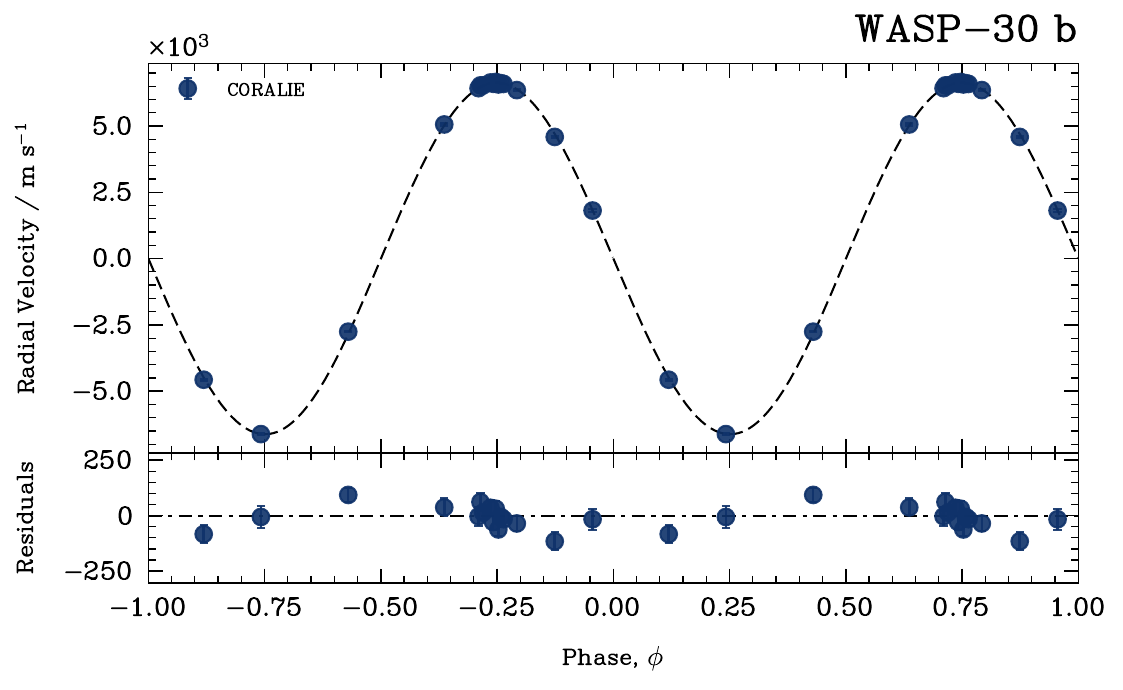}
    \includegraphics[width=0.32\linewidth]{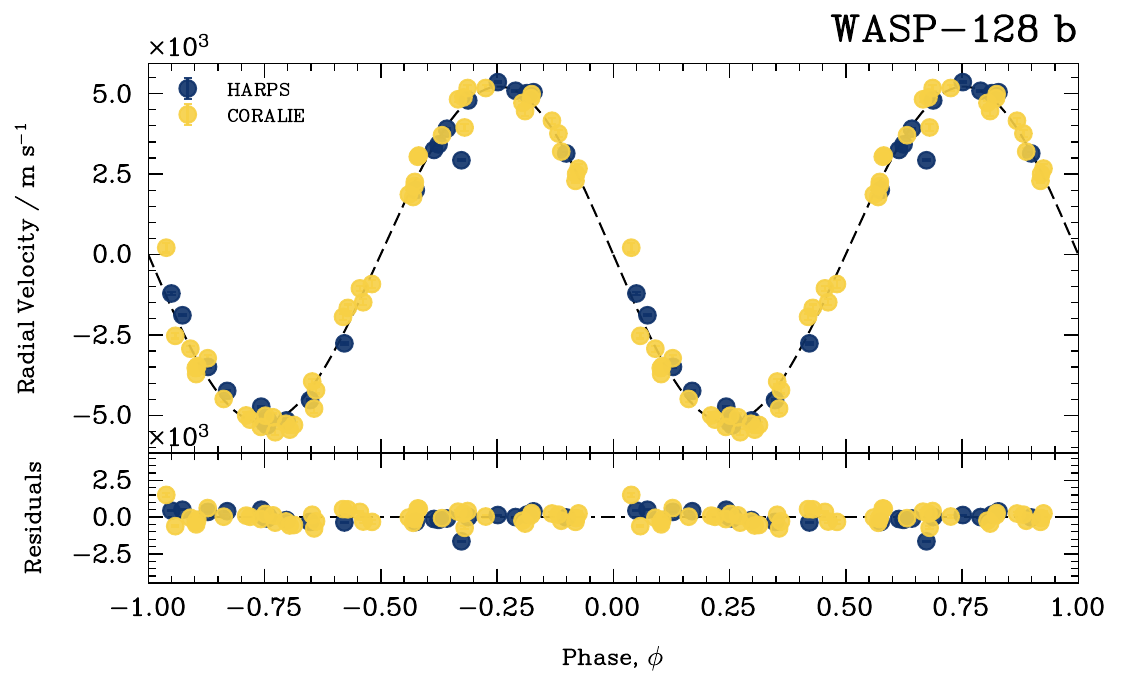}
    \caption{(Continued) TOI-811 $b$ \citep{2021AJ....161...97C}, Kepler-39 $b$ \citep{2011A&A...533A..83B}, KOI-189 $b$ \citep{2014A&A...572A.109D}, KOI-205 $b$ \citep{2015A&A...575A..85B}, TOI-852 $b$ \citep{2021AJ....161...97C}, TOI-1278 $b$ \citep{2021AJ....162..144A}, TOI-1406 $b$ \citep{2020AJ....160...53C}, TOI-2521 $b$, TOI-2336 $b$, TOI-1608 $b$ \citep{2023MNRAS.523.6162L}, TOI-2543 $b$, TOI-1982 $b$ \citep{2022A&A...664A..94P}, TOI-2119 $b$ \citep{2022MNRAS.514.4944C}, TOI-2490 $b$ \citep{2024MNRAS.533.2823H}, TOI-4603 $b$ \citep{2023A&A...672L...7K}, TOI-5375 $b$ \citep{2023AJ....165..218L}, WASP-30 $b$ \citep{2011ApJ...726L..19A}, and WASP-128 $b$ \citep{2018MNRAS.481.5091H}.}
    \label{fig:phased3}
\end{figure*}

\begin{deluxetable*}{lcccc}
\tablecaption{Priors used in the uniform Keplerian re-fitting of transiting systems considered in this work.}
\tablehead{\colhead{ID} & \colhead{$T_0$} & \colhead{$P$} & \colhead{$K$} & \colhead{RV zero-point offset} \\
\colhead{} & \colhead{(BJD$-$2450000)} & \colhead{(d)} & \colhead{(km s$^{-1}$)} & \colhead{(km s$^{-1}$)}}
\startdata
AD 3116 b & $\mathcal{U}[7508.75, 7548.75]$ & $\mathcal{U}[1.983, 1.983]$ & $\mathcal{U}[0.00, 29.68]$ & $\mathcal{U}[25.78, 56.46]$ \\
CWW 89 A b & $\mathcal{U}[7519.75, 7559.75]$ & $\mathcal{U}[5.280, 5.300]$ & $\mathcal{U}[0.00, 8.95]$ & $\mathcal{U}[41.90, 51.19]$; $\mathcal{U}[41.24, 50.11]$ \\
CoRoT-15 b & $\mathcal{U}[5203.96, 5243.96]$ & $\mathcal{U}[3.059, 3.061]$ & $\mathcal{U}[0.00, 15.00]$ & $\mathcal{U}[-3.18, 10.73]$; $\mathcal{U}[-4.09, 7.03]$ \\
CoRoT-3 b & $\mathcal{U}[4371.30, 4411.30]$ & $\mathcal{U}[4.255, 4.257]$ & $\mathcal{U}[0.00, 10.00]$ & $\mathcal{U}[-58.89, -53.77]$; $\mathcal{U}[-57.63, -53.15]$; $\mathcal{U}[-58.83, -53.48]$; $\mathcal{U}[-59.20, -54.13]$; $\mathcal{U}[-58.29, -52.69]$ \\
CoRoT-33 b & $\mathcal{U}[6497.64, 6537.64]$ & $\mathcal{U}[5.800, 5.820]$ & $\mathcal{U}[0.00, 13.33]$ & $\mathcal{U}[14.29, 27.66]$; $\mathcal{U}[17.47, 28.62]$ \\
CoRoT-34 b & $\mathcal{U}[6383.83, 6423.83]$ & $\mathcal{U}[2.100, 2.130]$ & $\mathcal{U}[0.00, 20.00]$ & $\mathcal{U}[26.95, 43.84]$; $\mathcal{U}[30.61, 45.85]$; $\mathcal{U}[25.27, 47.17]$ \\
ELM-J0555-57 A b & $\mathcal{U}[6720.59, 6760.59]$ & $\mathcal{U}[7.560, 7.960]$ & $\mathcal{U}[0.00, 30.00]$ & $\mathcal{U}[10.88, 27.40]$ \\
EPIC 201702477 b & $\mathcal{U}[7390.77, 7430.77]$ & $\mathcal{U}[40.600, 40.800]$ & $\mathcal{U}[0.00, 8.43]$ & $\mathcal{U}[32.74, 38.07]$; $\mathcal{U}[29.17, 38.60]$ \\
EPIC 212036875 b & $\mathcal{U}[8233.44, 8273.44]$ & $\mathcal{U}[5.160, 5.180]$ & $\mathcal{U}[0.00, 10.64]$ & $\mathcal{U}[-27.77, -16.13]$ \\
GPX-1 b & $\mathcal{U}[8065.07, 8105.07]$ & $\mathcal{U}[1.744, 1.746]$ & $\mathcal{U}[0.00, 6.11]$ & $\mathcal{U}[-16.93, -9.82]$ \\
HATS-70 b & $\mathcal{U}[6910.00, 6911.00]$ & $\mathcal{U}[1.860, 1.900]$ & $\mathcal{U}[0.00, 20.00]$ & $\mathcal{U}[33.75, 38.01]$; $\mathcal{U}[32.50, 38.33]$ \\
HIP33609 b & $\mathcal{U}[8915.98, 8915.98]$ & $\mathcal{U}[39.460, 39.480]$ & $\mathcal{U}[0.00, 50.00]$ & $\mathcal{U}[26.49, 34.40]$ \\
J1219-39 b & $\mathcal{U}[5624.69, 5664.69]$ & $\mathcal{U}[6.750, 6.770]$ & $\mathcal{U}[0.00, 20.86]$ & $\mathcal{U}[23.04, 44.89]$ \\
KELT-1 b & $\mathcal{U}[5911.63, 5913.63]$ & $\mathcal{U}[1.216, 1.218]$ & $\mathcal{U}[0.00, 8.44]$ & $\mathcal{U}[-18.91, -9.47]$ \\
KOI-189 b & $\mathcal{U}[5760.96, 5800.96]$ & $\mathcal{U}[30.200, 30.500]$ & $\mathcal{U}[0.00, 11.74]$ & $\mathcal{U}[-79.74, -67.00]$ \\
KOI-205 b & $\mathcal{U}[6051.52, 6091.52]$ & $\mathcal{U}[11.720, 11.730]$ & $\mathcal{U}[0.00, 7.00]$ & $\mathcal{U}[10.71, 19.22]$ \\
KOI-415 b & $\mathcal{U}[6203.28, 6243.28]$ & $\mathcal{U}[166.600, 166.800]$ & $\mathcal{U}[0.00, 12.00]$ & $\mathcal{U}[-3.68, 4.07]$ \\
KOI-607 b & $\mathcal{U}[6803.83, 6843.83]$ & $\mathcal{U}[5.840, 5.940]$ & $\mathcal{U}[0.00, 20.00]$ & $\mathcal{U}[-8.76, 11.99]$ \\
Kepler-39 b & $\mathcal{U}[5415.45, 5455.45]$ & $\mathcal{U}[21.000, 23.000]$ & $\mathcal{U}[0.00, 2.76]$ & $\mathcal{U}[-1.82, 1.94]$ \\
LHS 6343 c & $\mathcal{U}[5370.00, 5390.00]$ & $\mathcal{U}[12.600, 12.800]$ & $\mathcal{U}[0.00, 10.00]$ & $\mathcal{U}[-56.89, -37.01]$ \\
LP 261-75 b & $\mathcal{U}[8093.51, 8133.51]$ & $\mathcal{U}[1.860, 1.890]$ & $\mathcal{U}[0.00, 42.05]$ & $\mathcal{U}[-26.22, 16.82]$ \\
NGTS-19 b & $\mathcal{U}[8898.80, 8938.80]$ & $\mathcal{U}[17.830, 17.850]$ & $\mathcal{U}[0.00, 12.63]$ & $\mathcal{U}[-38.90, -25.27]$ \\
NGTS-28A b & $\mathcal{U}[9390.04, 9430.04]$ & $\mathcal{U}[1.245, 1.265]$ & $\mathcal{U}[0.00, 20.00]$ & $\mathcal{U}[-20.72, 9.96]$ \\
NGTS-7 A b & $\mathcal{U}[8372.00, 8374.00]$ & $\mathcal{U}[0.670, 0.670]$ & $\mathcal{U}[0.00, 100.00]$ & $\mathcal{U}[-25.87, 19.89]$ \\
NLTT 41135 b & $\mathcal{U}[5031.38, 5071.38]$ & $\mathcal{U}[2.889, 2.892]$ & $\mathcal{U}[0.00, 24.16]$ & $\mathcal{U}[-2.56, 22.60]$ \\
OGLE-TR-123 b & $\mathcal{U}[3000.00, 3500.00]$ & $\mathcal{U}[1.804, 1.810]$ & $\mathcal{U}[0.00, 30.00]$ & $\mathcal{U}[-6.71, 12.78]$; $\mathcal{U}[-9.37, 12.95]$ \\
TOI-1278 b & $\mathcal{U}[8988.50, 9028.50]$ & $\mathcal{U}[14.460, 14.480]$ & $\mathcal{U}[0.00, 4.38]$ & $\mathcal{U}[-32.18, -26.80]$ \\
TOI-1406 b & $\mathcal{U}[8523.55, 8563.55]$ & $\mathcal{U}[10.560, 10.580]$ & $\mathcal{U}[0.00, 7.48]$ & $\mathcal{U}[-20.07, -11.59]$; $\mathcal{U}[-19.59, -11.93]$ \\
TOI-148 b & $\mathcal{U}[8485.53, 8505.53]$ & $\mathcal{U}[4.867, 4.880]$ & $\mathcal{U}[0.00, 28.13]$ & $\mathcal{U}[-38.63, -19.50]$ \\
TOI-1608 b & $\mathcal{U}[9554.22, 9594.22]$ & $\mathcal{U}[2.450, 2.490]$ & $\mathcal{U}[0.00, 30.00]$ & $\mathcal{U}[71.62, 93.88]$ \\
TOI-1982 b & $\mathcal{U}[9325.78, 9345.78]$ & $\mathcal{U}[17.000, 17.300]$ & $\mathcal{U}[0.00, 45.00]$ & $\mathcal{U}[-42.13, -33.03]$; $\mathcal{U}[-41.00, -33.32]$; $\mathcal{U}[-4.05, 5.45]$ \\
TOI-2119 b & $\mathcal{U}[9131.74, 9133.10]$ & $\mathcal{U}[7.201, 7.210]$ & $\mathcal{U}[0.00, 20.00]$ & $\mathcal{U}[-0.50, 9.75]$ \\
TOI-2336 b & $\mathcal{U}[9358.00, 9360.00]$ & $\mathcal{U}[7.600, 7.800]$ & $\mathcal{U}[0.00, 50.00]$ & $\mathcal{U}[25.10, 40.46]$; $\mathcal{U}[24.33, 36.14]$; $\mathcal{U}[26.53, 38.16]$ \\
\enddata
\tablecomments{
For all systems, the parameters $\sqrt{e}\cos\omega$ and $\sqrt{e}\sin\omega$ are sampled uniformly over the interval $[-1, +1]$. Columns indicate: Time of inferior conjunction $T_0$, orbital period $P$, RV semi-amplitude $K$, and radial velocity zero-point offset for instruments: AD 3116 b: HIRES;  CWW 89 A b: FIES, TULL;  CoRoT-15 b: HARPS, HIRES;  CoRoT-3 b: SOPHIE, TLS, HARPS, CORALIE, SANDIFORD;  CoRoT-33 b: HARPS, FIES;  CoRoT-34 b: HIRES, UVES, HARPS;  ELM-J0555-57 A b: CORALIE;  EPIC 201702477 b: SOPHIE, HARPS;  EPIC 212036875 b: FIES;  GPX-1 b: SOPHIE;  HATS-70 b: CORALIE, FEROS;  HIP33609 b: CHIRON;  J1219-39 b: CORALIE;  KELT-1 b: KELT;  KOI-189 b: SOPHIE;  KOI-205 b: SOPHIE;  KOI-415 b: SOPHIE;  KOI-607 b: TRES;  Kepler-39 b: SOPHIE;  LHS 6343 c: HIRES;  LP 261-75 b: TRES;  NGTS-19 b: CORALIE;  NGTS-28A b: HARPS;  NGTS-7 A b: NGTS;  NLTT 41135 b: NLTT;  OGLE-TR-123 b: FLAMES, HARPS;  TOI-1278 b: SPIRou;  TOI-1406 b: ANU, CHIRON;  TOI-148 b: CORALIE;  TOI-1608 b: NRES;  TOI-1982 b: CORALIE, FEROS, TRES;  TOI-2119 b: TRES;  TOI-2336 b: NRES, CHIRON, CORALIE.}
\end{deluxetable*}

\begin{deluxetable*}{lcccc}
\tablehead{\colhead{ID} & \colhead{$T_0$} & \colhead{$P$} & \colhead{$K$} & \colhead{RV zero-point offset} \\
\colhead{} & \colhead{(BJD$-$2450000)} & \colhead{(d)} & \colhead{(km s$^{-1}$)} & \colhead{(km s$^{-1}$)}}
\startdata
TOI-2490 b & $\mathcal{U}[2180.70, 2180.80]$ & $\mathcal{U}[60.333, 61.000]$ & $\mathcal{U}[0.00, 100.00]$ & $\mathcal{U}[14.11, 26.13]$ \\
TOI-2521 b & $\mathcal{U}[9227.25, 9227.25]$ & $\mathcal{U}[5.500, 5.600]$ & $\mathcal{U}[0.00, 30.00]$ & $\mathcal{U}[-40.62, -22.40]$ \\
TOI-2533 b & $\mathcal{U}[9510.00, 9530.00]$ & $\mathcal{U}[6.670, 6.690]$ & $\mathcal{U}[0.00, 40.00]$ & $\mathcal{U}[-13.49, 2.31]$ \\
TOI-2543 b & $\mathcal{U}[2964.00, 2966.00]$ & $\mathcal{U}[7.530, 7.550]$ & $\mathcal{U}[0.00, 13.96]$ & $\mathcal{U}[24.96, 39.45]$; $\mathcal{U}[27.04, 39.92]$ \\
TOI-263 b & $\mathcal{U}[8790.93, 8811.75]$ & $\mathcal{U}[0.557, 0.560]$ & $\mathcal{U}[0.00, 30.00]$ & $\mathcal{U}[-22.30, 20.86]$ \\
TOI-4603 b & $\mathcal{U}[2527.20, 2527.40]$ & $\mathcal{U}[7.230, 7.270]$ & $\mathcal{U}[0.00, 2.21]$ & $\mathcal{U}[-0.83, 2.16]$; $\mathcal{U}[-1.05, 1.97]$ \\
TOI-503 b & $\mathcal{U}[8575.74, 8577.74]$ & $\mathcal{U}[3.677, 3.690]$ & $\mathcal{U}[0.00, 20.00]$ & $\mathcal{U}[-0.56, 9.79]$; $\mathcal{U}[-5.38, 4.89]$; $\mathcal{U}[-1.52, 8.16]$; $\mathcal{U}[-7.50, 2.35]$; $\mathcal{U}[24.44, 34.21]$ \\
TOI-5375 b & $\mathcal{U}[9560.00, 9600.00]$ & $\mathcal{U}[1.600, 1.900]$ & $\mathcal{U}[0.00, 30.00]$ & $\mathcal{U}[-79.34, -44.47]$ \\
TOI-569 b & $\mathcal{U}[8583.47, 8623.47]$ & $\mathcal{U}[6.540, 6.560]$ & $\mathcal{U}[0.00, 13.15]$ & $\mathcal{U}[65.69, 78.25]$; $\mathcal{U}[67.03, 79.85]$; $\mathcal{U}[68.19, 77.65]$ \\
TOI-587 b & $\mathcal{U}[8578.63, 8618.63]$ & $\mathcal{U}[8.040, 8.050]$ & $\mathcal{U}[0.00, 9.32]$ & $\mathcal{U}[-8.52, 1.79]$ \\
TOI-629 b & $\mathcal{U}[9163.81, 9203.81]$ & $\mathcal{U}[8.700, 8.730]$ & $\mathcal{U}[0.00, 16.41]$ & $\mathcal{U}[-13.76, -4.56]$; $\mathcal{U}[-5.57, 3.65]$ \\
TOI-681 b & $\mathcal{U}[8798.77, 8838.77]$ & $\mathcal{U}[15.680, 15.860]$ & $\mathcal{U}[0.00, 20.00]$ & $\mathcal{U}[16.16, 28.23]$; $\mathcal{U}[17.89, 29.88]$ \\
TOI-694 b & $\mathcal{U}[8769.75, 8809.75]$ & $\mathcal{U}[47.950, 48.150]$ & $\mathcal{U}[0.00, 100.00]$ & $\mathcal{U}[20.98, 29.36]$; $\mathcal{U}[15.71, 27.98]$; $\mathcal{U}[18.23, 29.82]$ \\
TOI-746 b & $\mathcal{U}[8802.71, 8842.71]$ & $\mathcal{U}[10.800, 11.000]$ & $\mathcal{U}[0.00, 14.69]$ & $\mathcal{U}[24.82, 40.49]$; $\mathcal{U}[24.79, 40.28]$ \\
TOI-811 b & $\mathcal{U}[8773.86, 8813.86]$ & $\mathcal{U}[25.060, 25.260]$ & $\mathcal{U}[0.00, 6.31]$ & $\mathcal{U}[29.93, 37.23]$ \\
TOI-852 b & $\mathcal{U}[8711.87, 8751.87]$ & $\mathcal{U}[4.930, 4.950]$ & $\mathcal{U}[0.00, 26.86]$ & $\mathcal{U}[-5.83, 5.22]$; $\mathcal{U}[-22.64, -11.55]$ \\
WASP-128 b & $\mathcal{U}[7095.17, 7135.17]$ & $\mathcal{U}[2.208, 2.210]$ & $\mathcal{U}[0.00, 10.81]$ & $\mathcal{U}[8.91, 20.58]$; $\mathcal{U}[8.77, 20.47]$ \\
WASP-30 b & $\mathcal{U}[5075.12, 5115.12]$ & $\mathcal{U}[4.140, 4.160]$ & $\mathcal{U}[0.00, 12.00]$ & $\mathcal{U}[0.80, 15.08]$ \\
\enddata
\tablecomments{(Continued) TOI-2490 b: FEROS;  TOI-2521 b: CHIRON;  TOI-2533 b: TRES;  TOI-2543 b: CHIRON, CORALIE;  TOI-263 b: ESPRESSO; TOI-4603 b: PARAS, TRES;  TOI-503 b: TRES, PARAS, ONDREJOV, TAUTENBURG, FIES;  TOI-5375 b: HARPS-N;  TOI-569 b: CHIRON, CORALIE, FEROS;  TOI-587 b: TRES;  TOI-629 b: CORALIE, TRES;  TOI-681 b: CORALIE, FEROS;  TOI-694 b: FEROS, CHIRON, CORALIE;  TOI-746 b: CORALIE, FEROS;  TOI-811 b: CHIRON;  TOI-852 b: TRES, CORALIE;  WASP-128 b: HARPS, CORALIE;  WASP-30 b: CORALIE.}
\end{deluxetable*}

\end{document}